\documentclass[reprint, amsmath, amssymb, aps, superscriptaddress,showkeys]{revtex4-2}

\usepackage{graphicx}
\usepackage{dcolumn}
\usepackage{bm}

\usepackage{orcidlink}
\usepackage[caption=false]{subfig}
\usepackage{qcircuit}
\usepackage{tabularx}
\usepackage{array}
\newcommand{\unit}[1]{\ensuremath{\, \mathrm{#1}}}

\usepackage{braket}
\usepackage{url}
\usepackage{xurl} 

\begin{document}

\preprint{APS/123-QED}

\title{Cross-Platform Benchmarking of Near‑Term Quantum Optimisation Algorithms}

\author{Kieran McDowall}
\email{kieran.mcdowall@stfc.ac.uk}
\affiliation{National Quantum Computing Centre, Rutherford Appleton Laboratory, Harwell Campus, Didcot, Oxfordshire, OX11 0QX.}

\author{Theodoros Kapourniotis}
\affiliation{National Quantum Computing Centre, Rutherford Appleton Laboratory, Harwell Campus, Didcot, Oxfordshire, OX11 0QX.}

\author{Christopher Oliver}
\affiliation{Department of Physics, University of Strathclyde, Glasgow, G4 0NG, United Kingdom.}

\author{Phalgun Lolur}
\affiliation{Capgemini UK PLC, 95 Queen Victoria Street, London, United Kingdom, EC4V 4HN.}
\affiliation{Capgemini Quantum Lab, 147-151 Quai du Président Roosevelt, 92130 Issy-les-Moulineaux, France.}

\author{Konstantinos Georgopoulos}
\affiliation{National Quantum Computing Centre, Rutherford Appleton Laboratory, Harwell Campus, Didcot, Oxfordshire, OX11 0QX.}


\begin{abstract}
Quantum computers show potential for achieving computational advantage over classical computers, with many candidate applications in combinatorial optimisation. 
We present an application level benchmarking framework for near-term quantum optimisation algorithms using a dense Quadratic Unconstrained Binary Optimisation (QUBO) materials science problem as a representative test-case. To solve this problem, we implement two methods, the Variational Quantum Eigensolver (VQE) and Quantum Annealing (QA), on commercially-available gate-based and quantum annealing devices that are accessible via Quantum-Computing-as-a-Service (QCaaS) models. To analyse the performance of these algorithms, we use a toolbox of relevant metrics and compare performance against three classical algorithms. We employ quantum methods to solve fully-connected QUBOs of up to $72$ variables, and find that algorithm performance beyond this is restricted by device connectivity, noise and classical computation time overheads. The applicability of our approach goes beyond the selected configurational analysis test-case, and we anticipate that our approach will be of use for optimisation problems in general.
\end{abstract}


\maketitle

\section{Introduction}
\label{sect:intro}

Combinatorial optimisation problems are ubiquitous across science and industry, motivating quantum optimisation as a highly active field of research and development, and a candidate for near-term quantum advantage \cite{korte2011combinatorial, abbas2024challenges}. With respect to quantum advantage and performance guarantees, algorithms can be categorised into three classes: provably exact algorithms, which guarantee the optimal solution; provably approximate algorithms with guaranteed performance bounds (such as Decoded Quantum Interferometry \cite{jordan2025optimization}); and heuristic algorithms, which offer no performance guarantees. Near-term quantum optimisation algorithms fall into the last category.

For heuristic quantum algorithms, claims of quantum advantage cannot be established through asymptotic complexity arguments and instead require experimental validation. This motivates the need for rigorous benchmarking, in which state of the art quantum and classical methods are compared in a fair and transparent way. Our work addresses this need through the full disclosure of key metrics and methodologies, an aspect that is currently lacking in the field. Cross-platform benchmarking - where a candidate problem believed to be classically hard with practical relevance is solved with competing methods and devices - provides a natural framework for such comparisons. This not only enables the testing of algorithms, but can help track the progress of the underlying quantum hardware that the algorithms are deployed on. As well as informing improvements to the hardware, cross-platform benchmarking can motivate algorithm development where tweaks can be made to improve performance. Our work contributes to ongoing efforts to assess quantum optimisation methods on practically motivated problem instances \cite{pennington2025boosting,kotil2025quantum,vcepaite2025quantum}.



To perform our cross-platform benchmarking we chose the problem of finding the lowest-energy configuration of defective graphene structures, based on the formulation found in Ref.~\cite{camino2023quantum}. We use the graphene defect problem solely as a representative, high density instance on which to test our benchmarking framework. Suppose we are tasked with removing a specific number of atoms from a graphene sheet of $N$ sites. The goal is to find which atoms, when removed, will result in a configuration with the lowest-energy. The configuration of lowest energy will be the one with the maximum number of remaining atom-atom bonds. This is an instance of the densest k-subgraph problem, with many applications beyond configurational analysis~\cite{lanciano_survey_2024}. In this work, we consider only the case of removing $3$ atoms, and thus $k=N-3$.

To solve the problem we first formulate it as a QUBO~\cite{glover2022quantum}. QUBOs are paradigmatic formulations capable of encoding a variety of optimisation problems and known to be nondeterministic polynomial time hard (NP-hard)  to exactly solve in worst case\cite{pardalos1992complexity}. Unless P$=$NP, these problems are intractable for classical computers whereas, for quantum computers, modest speed-ups are possible~\cite{grover_fast_1996,gilliam2021grover}, while they remain unlikely to be tractable~\cite{aaronson2010bqp,aharonov2008adiabatic} in the worst case. Nonetheless, it is believed that the correlations arising in real world problems are complex enough to generate hard instances and at the same time provide some structure to be exploited by a quantum computer to potentially solve the problem faster or achieve better solutions (e.g.~\cite{boulebnane2024solving,jordan2025optimization}).

Quantum algorithms for optimisation problems exist for both gate-based quantum computers and quantum annealers. In this work, we apply two well studied practical approaches, variational quantum algorithms (VQAs) and quantum annealing algorithms (QAAs). Despite the simplicity of these methods, and despite efforts in investigating VQAs and QAAs in real quantum hardware, there are significant challenges in using these approaches to achieve quantum advantage in optimisation. These include difficulties such as the approximate nature of these algorithms, issues that inhibit their performance such as barren plateaus \cite{mcclean2018barren} and non-adiabatic effects in QAAs \cite{hauke2020perspectives}. There are also more practical issues around, for example, noise and decoherence \cite{cerezo2021variational} and optimisation of controls. As a result, there are many open questions that must be addressed before quantum advantage may be achieved. Valuable insights for dealing with these issues can be attained by experimental implementation of these algorithms on real near-term quantum devices.

Motivated by these challenges, in this work we present a systematic testing of performance of VQA and QAA methods on commercially-available quantum hardware: a gate-based QPU (\textit{ibm\_fez}) and a quantum annealer (D-Wave Advantage 6.4). We use the graphene defect problem as our test-case. The simplicity of this model and its use in a previous pedagogical study \cite{camino2023quantum} removes a level of complexity in understanding our analysis, making it more accessible. However, most of our analysis applies to optimisation problems in general.

Our main contributions are:
\begin{itemize} 
   \item We introduce a benchmarking framework for quantum optimisation that enables fair and consistent comparison of gate-based variational algorithms, quantum annealing and classical algorithms on the same QUBO instances.
   
   \item We apply this benchmarking framework in practice to the graphene defect problem on QCaaS hardware, using this hardware to solve fully connected QUBOs up to 72 variables.

    \item We explore the scaling behaviour of the algorithms, finding an efficient classical solution with simulated annealing that scales polynomially with system size $N$. Using simulated annealing, we solved problems with up to 388 variables. We also present some provisional evidence of quantum annealing scaling in polynomial time on the problem sizes that were feasible to run on hardware.

    \item Our work reveals key technical details such as: how higher probability of obtaining the optimal solution can be obtained by using small penalty coefficients, $\lambda$, which keep the QUBO energy range small, in combination with our post-selection, which removes any solutions that violate the constraint (due to using a small $\lambda$); the large overhead embedding poses to quantum annealing and how this can be reduced with a more efficient technique; the large standard deviation error in optimal solution probability due to differing random initial parameters for VQE; how noise affects VQE convergence and increases the difficulty of finding good variational parameters; how device connectivity limits performance on larger problems; and how classical computation overheads limit the number of experimental repeats that can be excecuted on the IBM QPU due to runtime constraints.

\end{itemize}

   





The assumptions and limitations of our work are as follows: we assume a simple energy model, specifically the one studied in Ref.~\cite{camino2023quantum}. We emphasise that our work is a benchmarking study, with the graphene defect problem serving as a representative test-case. For a more detailed discussion of the underlying use-case, its applicability to real-world problems, and comparisons between quantum annealing and density functional theory: we direct the reader to Ref.~\cite{camino2023quantum}. In our work, the problem constraint is encoded into the QUBO formalism using penalty terms, rather than solving a hard-constrained version of the problem. Furthermore, our study was conducted using some of the best commercially available quantum devices at the time of experimentation, which took place in January 2025. As of January 2026 the quantum devices used in this investigation are still accessible. \textit{Ibm\_fez} using the Heron r2 processor still has the joint highest number of qubits of any IBM device. Newer Heron r3 devices have lower two qubit error rates but these error rates are of the same order of magnitude. Readout errors in the Heron r3 devices are of an order of magnitude lower than those of \textit{ibm\_fez}. IBM recently announced the Nighthawk r1 processor with square lattice connectivity instead of heavy hex. This change would reduce SWAP overheads and enable shorter VQE circuits for fully connected problems such as those considered here. D-Wave’s Advantage 6.4 system remains accessible, while newer Advantage2 devices are now available with improved qubit connectivity. Nevertheless, our work remains relevant due to the generality of our benchmarking framework and the algorithms considered. In particular, the algorithmic bottlenecks identified persist despite hardware upgrades. We did not apply error mitigation or other techniques that could improve algorithm performance (as listed in Section \ref{sect:discuss}).

In Section~\ref{sect:rel_work} we present other recent efforts in benchmarking optimisation algorithms and real world implementations. We then formulate our specific problem in Section~\ref{sect:prob_form}. In Section~\ref{sect:methodology_for_bench} we describe our methodology for the cross-platform benchmarking including the chosen metrics, whose results we present in Section~\ref{sect:results}. Finally, in Section~\ref{sect:discuss} we discuss open questions and future directions. Additional details of our methodology are included in the Appendices.

\section{Related Work}
\label{sect:rel_work}

Progress has been made in solving quantum chemistry problems with classical hardware, by exploiting domain knowledge about the structure of specific problems \cite{mcardle2020quantum}. While a standard approach to solving electronic structure problems on classical computers is density functional theory (DFT) \cite{lim2011dft}, we instead use the QUBO formulation of the problem given in Ref.~\cite{camino2023quantum}. Some of the best classical methods for solving QUBOs include linear programming (e.g. \cite{bliek1u2014solving}), greedy algorithms (e.g. \cite{vince2002framework}) or heuristics (e.g. \cite{van1987simulated}). 

More specifically, Ref.~\cite{camino2023quantum} demonstrates how QAA can be applied to solve a version of the graphene defect use-case formulated as a Densest k-Subgraph (DkS) problem (see Appendix~\ref{app:complexity} for more detail on its formulation and complexity). This problem is NP-hard for general $k$ and bipartite graphs of degree up to $3$~\cite{feige1997densest}.
In~\cite{camino2023quantum}, an $18$ atom lattice is encoded as a fully-connected QUBO and executed on quantum annealers. In our work we execute this on larger lattice sizes, using quantum annealers as well as gate-based quantum computers and classical methods. Another related piece of work~\cite{carnevali_vacancies_2020} studies a slightly different formulation of the problem where the goal is to minimise the number of dangling bonds (i.e. connections between vacancies and carbon atoms in the lattice), again executed on a quantum annealer. The same DkS problem as in~\cite{camino2023quantum} has also been investigated in a biology-inspired test-case~\cite{calude_quantum_2020} on randomised bipartite graphs, where the topology of the D-Wave device is identified as the main limitation for applying it on quantum hardware.

Ref.~\cite{abbas2024challenges} provides a comprehensive review of the current state of quantum optimisation, highlighting the challenges and future prospects of the field. The review outlines best practices for making fair comparisons when solving optimisation problems with different methods. The authors showcase recent publications where optimisation problems with varying numbers of qubits and problem connectivity (known as density) are solved using VQAs, comparing each solution's quality through the approximation ratio metric. The most common problem connectivity involves 3-regular graphs (where each node is connected to three others), which aligns well with quantum hardware layouts. Studies that explore higher qubit counts typically maintain this low problem connectivity. Increasing connectivity for larger qubit systems requires additional SWAP gates, leading to increased hardware noise. However, Refs. \cite{dupont2023quantum,harrigan2021quantum,maciejewski2024improving,maciejewski2024design} all consider Sherrington-Kirkpatrick problems with 100\% density, which can be reformulated as QUBO problems. These studies emphasize both the challenge and the importance of choosing standardised metrics to ensure that benchmarks are fair and results are comparable. The chosen set of metrics must fully capture the diverse behaviors of different algorithms. Our work applies the best practices outlined in this review to a particular problem, which stands out in terms of qubit count and high density compared to the other highlighted studies. We achieve a competitive mean approximation ratio with VQE on a QPU relative to the results reported in Table IV of Ref.~\cite{abbas2024challenges}. In terms of problem size, comparison with Table IV shows that our work is competitive with the state of the art in achieving the largest qubit counts for problems solved at 100\% density. While Ref.~\cite{abbas2024challenges} primarily focuses on gate-based algorithms, we also address questions regarding how QAAs can be fairly compared to variational gate-based methods. 

Other studies \cite{bettonte2022quantum,cutugno2022quantum} also compare QAAs and gate-based variational methods for specific use-cases. However, a comprehensive study that includes time comparisons, solution quality, and scaling analysis has yet to be conducted. Ref. \cite{sachdeva2024quantum} compares the Quantum Approximate Optimisation Algorithm (QAOA) to quantum annealing and claims that QAOA, when run on a real gate-based device, can outperform quantum annealing on a 5-regular max-cut graph problem. On the other hand, Ref.~\cite{mcgeoch2024comment} argues that these comparisons could be seen as ambiguous, explaining that the reported results from QAOA were enhanced by post-selection. These studies further highlight the need for transparent and fair benchmarking. In our work, we expose post-selection impacts - both in our metrics and distributions - as well as hyperparameter tuning, embedding overheads, and runtime break downs in a transparent and reproducible manner.

Recently, Ref. \cite{montanez2025evaluating} benchmarks gate-based hardware using the Linear Ramp QAOA (LR-QAOA) method \cite{montanez2025toward}. The advantage of using LR-QAOA over traditional near-term gate-based algorithms is that it does not involve variational parameters. The difficulty of optimising variational parameters, especially on quantum hardware, is demonstrated in our work.

While prior work typically focuses on sparse problems on gate-based devices, we solve dense QUBOs with up to 72 variables. We demonstrate best practices in our cross-platform benchmarking, enabling the evaluation of a gate-based variational algorithm and quantum annealing within a unified framework.

\section{Problem Formulation}
\label{sect:prob_form}
\begin{figure}[hhtbp]
\centering
\includegraphics[width=0.35\textwidth]{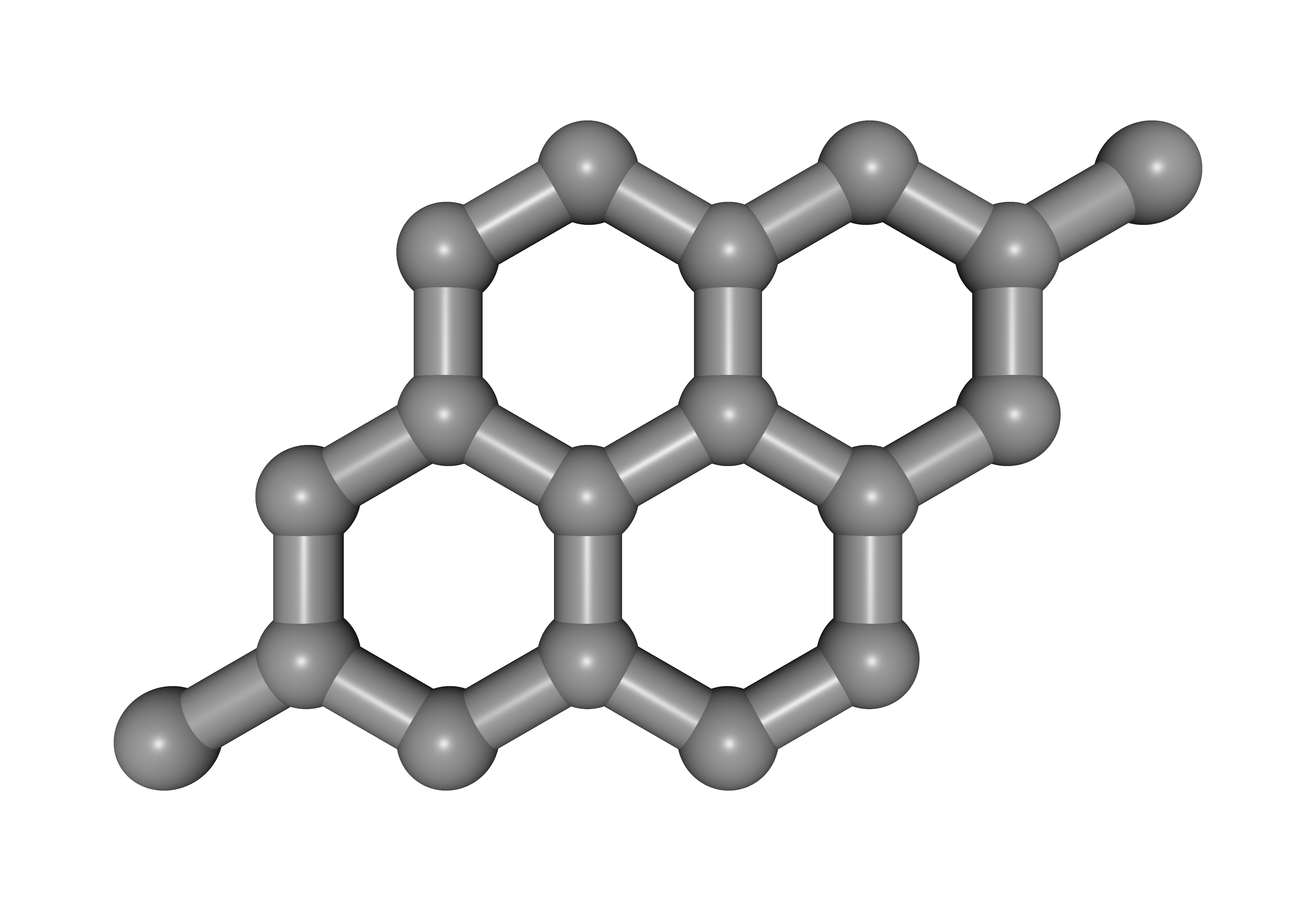}
\raisebox{1.3cm}{\includegraphics[width=0.1\textwidth]{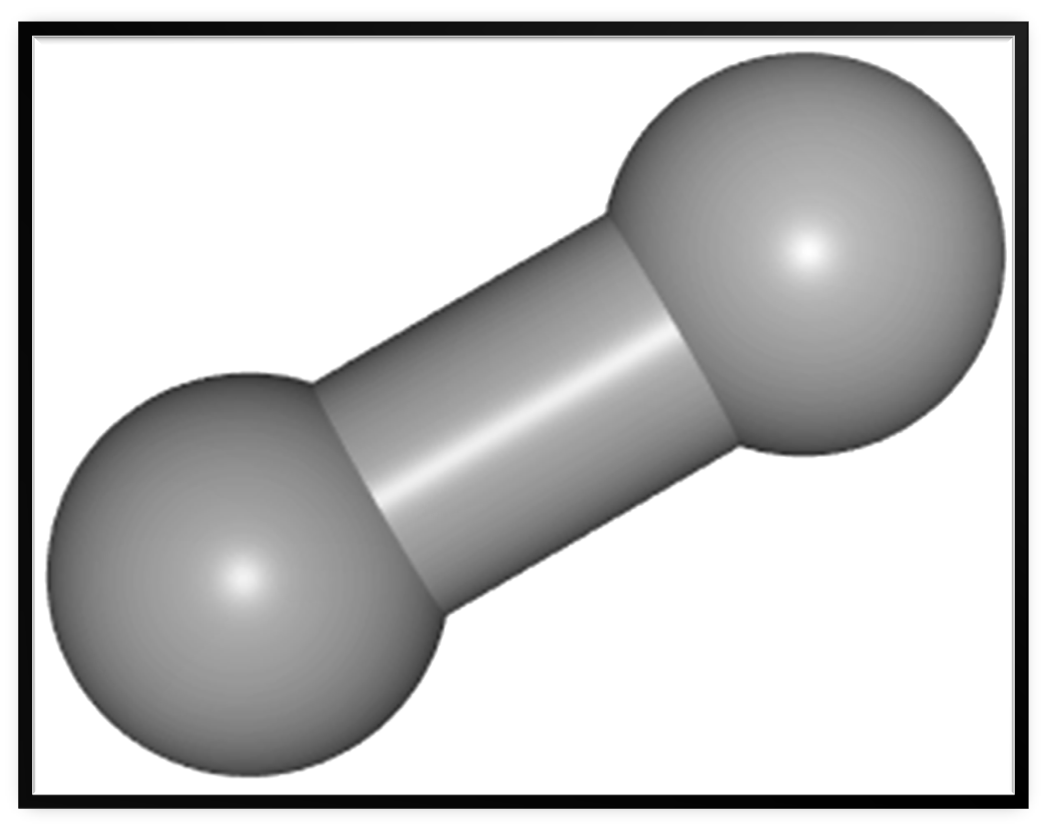}}
\caption{\small{An example of one of the graphene structures explored in this work: a $3\times3$ supercell (which corresponds to an $18 \times 18$ QUBO matrix), where the grey spheres represent carbon atoms. The unit cell is shown in the inset. There is no vacancy in this example. Periodic boundary conditions apply, making this structure a $3$-regular graph (note: the graph that corresponds to the QUBO matrix is fully connected after adding constraints).}} 
\label{fig:cell}
\end{figure}

We use the problem of finding the minimum energy of defective graphene structures as an illustrative example of our benchmarking strategy. This problem is discussed in more detail in Appendix \ref{app:complexity}. We model the system as an $N$-site hexagonal structure, as shown visually in Figure \ref{fig:cell} ($N=18$ in this example). The first graphene structure explored is a $3\times3$ supercell, a unit cell containing two carbon atoms repeated in a $3\times3$ arrangement, which corresponds to an $18$ variable QUBO. We also investigate other $n\times n$ supercells, which translate to QUBOs of varying size. Each (binary) QUBO variable represents a site containing either a carbon atom or a vacancy.

Each of the carbon atoms in Figure \ref{fig:cell} can be removed, creating a vacancy that results in the breaking of carbon-carbon bonds. The breaking of these bonds increases the energy of the system's ground state configuration. The energy increase depends on where the vacancies are introduced. In particular, and following Ref.~\cite{camino2023quantum}, the energy of the system depends only on the number of remaining carbon-carbon bonds. In this model we impose periodic boundary conditions. Referring to \ref{fig:cell}, the periodic boundary conditions can be visualised as connecting atoms on the opposite side of the cell. As a result, the lattice can be represented as a 3-regular graph. The supercell's 2D geometry is maintained for all configurations.

The optimisation problem is, therefore, translated to finding the lowest energy configuration of the system, subject to the constraint of a certain number of vacancies. We work with three vacancies, the highest studied in Ref. \cite{camino2023quantum}, treating this number as a constant throughout our work. Holding the number of vacancies constant at three means the space of feasible solutions grows as ${N\choose 3}$, where $N$ is the number of QUBO variables, which results in polynomial growth. Choosing the number of vacancies to grow as a function of $N$ would cause the feasible search space to grow at a larger rate. We experimented with this and found that it made no significant difference, for this particular QUBO instance, to the time scaling of simulated annealing. Therefore, for simplicity we kept the number of vacancies constant at three.

We note that the introduction of vacancies to graphene structures is analogous to doping materials. Hence, this problem may have relevance when optimising configurations in semiconductors, superconductors, thermoelectrics and optoelectronics \cite{lusk2008nanoengineering,yazyev2006magnetism,zhou2010making}. 

To solve our problem on a quantum device, we consider the QUBO cost function:
\begin{equation} \label{eqn:qubo}
\begin{split}
    &H(x) = \boldsymbol{x}^{T}Q\boldsymbol{x} \\& = \sum_{i}Q_{i,i}\boldsymbol{x}_{i}+\sum_{i}\sum_{j>i}Q_{i,j}\boldsymbol{x}_{i}\boldsymbol{x}_{j},\hspace{0.1cm} \boldsymbol{x}_{i}\in \{0,1\},
\end{split}
\end{equation} 
where $\boldsymbol{x}$ is an $N$-component vector containing the binary QUBO variables that represent the different configurations, and $Q$ is the cost matrix encoding the cost function of the particular problem instance. We aim to solve the minimization problem:
\begin{equation} \label{eqn:obj_funct}
    \min_{\boldsymbol{x}} \left( \sum_i Q_{i,i} \boldsymbol{x}_i + \sum_i \sum_{j,i} Q_{i>j} \boldsymbol{x}_i \boldsymbol{x}_j \right).
\end{equation}

In order to map our problem to the above QUBO, we represent the presence of a vacancy or an atom in each site $i$ by $x_i$ being $0$ or $1$ accordingly. The hexagonal graphene structure (with boundary conditions) can be represented by a graph with adjacency matrix $A$ and dimension $N\times N$. Since our objective is to find the maximum number of carbon-carbon bonds left after the atoms are replaced by vacancies, we want to reward (by decreasing the energy of the configuration) only the edges with a $1$ in both of their vertices. Therefore, our goal is achieved by choosing a cost matrix that is strictly upper diagonal and contains $-\kappa A_{ij}$ in its non-zero elements. Parameter $\kappa$ is the bond energy, such that adding a single bond lowers the system energy by $\kappa$.

To specify that we must have a certain number of vacancies we can add a constraint term:
\begin{equation} \label{eqn:constr}
    \lambda \left( \sum_i x_{i}-N_{C} \right)^{2},
\end{equation}
to our cost function, where $N_C$ is the number of carbon atoms we want to have in our cell and $\lambda$ is a constraint coefficient. Vectors $\boldsymbol{x}$ which do not have the number of carbon atoms equal to $N_C = N-N_{\text{vacancies}}$ are penalised in terms of their associated cost value. Bringing all this together, we get:
\begin{equation}
\begin{aligned}
&H(x)  = \underbrace{-\kappa \sum_i^{N} \sum_{j>i}^{N}A_{i,j} \boldsymbol{x}_i \boldsymbol{x}_j}_{\text{objective}} \\
           & \quad + \lambda \underbrace{\left( \sum_i^{N} \left(1 - 2 N_C\right) \boldsymbol{x}_i 
           + \sum_i^{N} \sum_{j>i}^{N} 2 \boldsymbol{x}_i \boldsymbol{x}_j \right)}_{\text{constraint}} \\
           & = \lambda \sum_i^{N} \left( 1 - 2 N_C \right) \boldsymbol{x}_i + \sum_{i,j>i}^{N} \left( 2\lambda - \kappa A_{i,j} \right) \boldsymbol{x}_i \boldsymbol{x}_j,
\label{eq:full_cost_func}
\end{aligned}
\end{equation}

The $Q$-matrix can be read off by comparing Equation~\ref{eq:full_cost_func} with Equation \ref{eqn:qubo}. The value of $\lambda$ is chosen through a tuning procedure described in Appendix \ref{app:sim_annealing}, where $\lambda/\kappa$ is defined as the QUBO penalty term coefficient. To find the optimal value of the penalty coefficient, we set $\kappa = 1$ and search for the optimal $\lambda$ for each algorithm. A balance must be struck between enforcing the constraint by increasing $\lambda$ and not making $\lambda$ too large so that it increases the energy range of the QUBO. Consequently, some solutions can violate the constraint and have an incorrect number of vacancies.

When plotting the problem graph associated with our matrix $Q$, we obtain a fully connected graph. This graph is constructed by adding edges between nodes for every $Q_{i,j}\neq 0, j>i$. Note that the graph is fully-connected despite the atoms not being bonded in an all-to-all structure; this is because the constraint term in Equation~\ref{eq:full_cost_func} adds in extra on-site and coupling energy terms. To solve the problem on quantum hardware, the graph must be embedded on physical qubits using classical techniques, as described in Appendices \ref{app:VQE} and \ref{app:annealing}. This is one of the major challenges in general when solving combinatorial optimisation problems using quantum computing approaches.

\section{Methodology for Multi-Algorithm Benchmarking} \label{sect:methodology_for_bench}

\begin{figure}[h!]
\centering
\includegraphics[width=0.48035\textwidth]{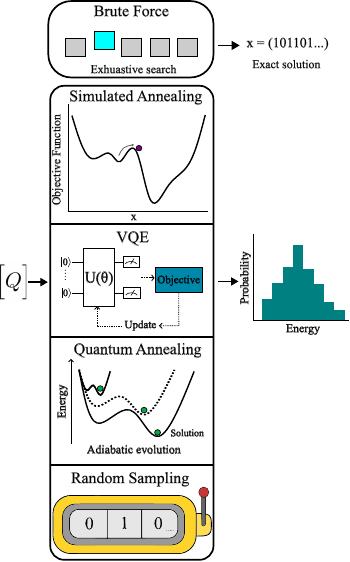}
\caption{\small{
Visual representation of the methods used in this work, as described in the main text.
}}\label{fig:1} 
\end{figure}

Having specified our problem, we discuss the methods and benchmarks we employ. We consider three classical techniques for solving our optimisation problem: brute force, simulated annealing and random sampling. On the quantum side, we use two algorithms: VQE, executed both on a state vector solver and a quantum processing unit (QPU), and quantum annealing implemented on a QPU. Figure \ref{fig:1} provides a schematic overview of the key principles underlying each method. To understand the methodology and solutions from different hardware platforms and algorithms it is important to establish performance metrics which fairly reflect the capabilities of each device and algorithm and that are device and algorithm agnostic. Firstly, the differences between each method are highlighted, along with their limitations. Then, the role of post-selection is discussed and, finally, we outline the chosen metrics and motivation for their use.

\subsection{Algorithms and Parameters}
\label{sect:algor}

Classical brute force simply involves generating all possible solutions to the QUBO problem, calculating their associated energy, and then selecting the optimal ones. Note that we could `hard-encode' the constraint on the total number of vacancies in our calculation by restricting the search space to those configurations that respect the constraint. However, we focus on developing a general framework for the multi-algorithm benchmarking of QUBO optimisation, where the constraint is incorporated as a penalty term as described above. Therefore, in order to ensure a fair comparison, we treat all methods in the same way and do not employ extra steps to enforce the constraint. Brute force is an exact method, so we know the minimum energy returned will be 100\% accurate for our model. The major drawback of brute force is the time it takes to exhaustively produce all possible solutions, which grows exponentially with the size of the configuration space we are searching (e.g.\cite{jalowiecki2021brute}).

Simulated annealing \cite{van1987simulated} is a probabilistic method which draws analogies from the heating and intentional slow cooling of physical systems, such as metals, to avoid structural defects. The cost function is used to compare the current solution against a newly selected one, and improvements on the former are selected. Solutions that do not improve on the current cost function evaluation are also selected with a probability determined by a temperature parameter as a way of combating getting stuck in local sub-optimal minima. The temperature parameter is set to gradually tend to zero, which means fewer worse evaluations will be accepted over time \cite{henderson2003theory}.

Many of the methods that we employ here involve hyperparameters whose values must be set. Our hyperparameter search finds the set of parameters for each method that return the highest probability of sampling the ground state. A hyperparameter search could be performed aiming to optimise performance with respect to another metric such as runtime. Our hyperparameter optimisation is a grid-based search, conducted by defining feasible search spaces and checking different combinations of values. The results of the hyperparameter search for all methods in this section are provided in Appendix \ref{app:hypparam}. Figures~\ref{fig:SA_hyp}, \ref{fig:rand_hyp}, \ref{fig:VQE_hypop}, and \ref{fig:nsup3_minor_embedding} illustrate the robustness of our hyperparameter search: they display not only the optimal parameter settings but also how performance varies across the full set of tested parameters, highlighting the sensitivity of the methods to parameter choices.

Simulated annealing provides a probabilistic final solution with a number of samples equal to the number of times it is repeated. Between repetitions only the random seed changes which affects the starting state. Limitations of simulated annealing include the fact that the temperature parameter range and the rate of its decrease must be carefully set. If the rate of decrease is too fast, the algorithm may get stuck in local minima; if the rate of decrease is too slow, the algorithm could take too long to converge due to bouncing around the energy landscape. The simulated annealing hyperparameters we optimise are: (i) the temperature range and (ii) the number of iterations (sweeps).

The final classical method we consider is random sampling. It randomly selects a particular binary vector out of all possible bit strings $\boldsymbol{x}$ of $N$ bits (like in brute force, we do not exclude from our search space the solutions that have an incorrect number of vacancies) and evaluates its energy with respect to the cost function. 
This means that each binary vector has probability $1/2^{N}$ of being sampled. This is, of course, not uniform with respect to energy as multiple bitstrings can have the same energy. Random sampling provides a useful baseline for solving QUBOs with probabilistic methods, as it uses no technique that exploits the problem structure to improve the probability of finding good solutions.

Turning to our quantum methods, VQE prepares a parameterised trial wavefunction, or \textit{ansatz}, \(\ket{\psi(\mathbf{\theta} )}\), where \(\mathbf{\theta}\) is a set of parameters that can be tuned by changing circuit parameters, to approximate the ground state of a quantum system. The quantum system's Hamiltonian, $\hat{H}$, is set to correspond to the classical cost function Equation \ref{eqn:qubo} upon mapping of the binary variables to Ising spins and subsequent quantisation. Measurements are taken over multiple executions of the same circuit, to estimate the energy expectation value. Each execution of the same circuit and subsequent measurement we call a \textit{shot}. Motivated by the variational principle, a classical optimiser then adjusts circuit parameters to minimize the energy expectation value, \(\bra{\psi(\mathbf{\theta} )}\hat{H}\ket{\psi(\mathbf{\theta} )}\), of the Hamiltonian $\hat{H}$ with respect to the variational quantum state $\ket{\psi(\mathbf{\theta} )}$. 

Each cycle of updating the circuit parameters and estimating energy expectation is called an \textit{iteration}. The algorithm keeps iterating this process until either the convergence criteria are met based on a tolerance parameter or the maximum number of iterations is reached. This approach of using a quantum resource to prepare and manipulate a quantum state within a classical loop is known as hybrid quantum computing. After convergence, the final circuit can be sampled (again for the same number of shots as each previous iteration) to obtain a probability distribution over the classical bitstring representations \cite{fedorov2022vqe}. The number of solutions obtained with VQE is equal to the number of shots, which is set to ensure shot noise remains negligible. The whole process described here constitutes an \textit{experiment}. As in our other methods, there are a number of hyperparameters whose values must be chosen. For VQE, these are (i) the choice of ansatz, (ii) the tolerance of the classical optimiser, (iii) the maximum number of iterations for the classical optimiser, and (iv) the CVaR $\alpha$ parameter (described shortly). Note that in this work, we use both IBM quantum hardware and a noiseless classical simulator. 

Some of the known limitations of VQE (and VQAs in general) are their iterative nature and their execution time \cite{tilly2022variational}. The iterative nature means that the classical (optimiser) part of the algorithm is susceptible to getting stuck in local minima and not converging on the global minima. This susceptibility is heightened when we consider VQE in the presence of noise and when considering systems of many qubits. We know that the addition of noise can cause the variance of the cost function to decrease \cite{wang2021noise} which means it becomes harder for optimal parameters to be found. Ref. \cite{ragone2023unified} discusses vanishing variance in loss functions and defines barren plateaus, where the variance vanishes exponentially with the system size. Due to the limited connectivity of current gate-based hardware, many SWAP gates may be needed for QUBO problems with high density. This can result in long circuits, the execution of which is limited by the gate times of the specific device. Long circuits also cause the effects of noise and decoherence to accumulate. This issue is particularly problematic for QAOA which is further described in Appendix \ref{app:VQE:QAOA} and Ref.~\cite{harrigan2021quantum}.

We use VQE with the conditional value at a risk (CVaR) \cite{barkoutsos2020improving} as the objective function of the classical optimiser. In this case, instead of using the usual expectation value as the objective function, which is the average energy of all the measured bit strings, only a fraction of the best bit strings are used to calculate the average energy. This fraction is determined by the parameter $\alpha$, a hyperparameter we tune in our experiments, with $\alpha=1$ corresponding to standard VQE without CVaR. Using CVaR can help convergence on the optimal parameters in fewer iterations \cite{barkoutsos2020improving, barron2024provable}. 

The convergence criteria, set for VQE with the COBYLA optimiser (from the SciPy library \cite{virtanen2020scipy}), aim for objective values to stop varying within a certain tolerance and within a maximum number of iterations dependent on the problem size. The COBYLA tolerance in the SciPy library  is defined as a lower bound on the size of the trust region \cite{yuan2000review}. We note here that setting stricter tolerance or convergence criteria could improve the VQE's solution quality in exchange with sacrificing its total runtime. A tolerance informed through hyperparameter search was found to consistently allow VQE and COBYLA to meet the convergence criteria before the maximum iterations were achieved. The number of maximum iterations must be selected based on the time feasibility of running many experiments. We show our convergence plots in Appendix \ref{app:VQE}.

Finally, quantum annealing generates a solution to the given problem by adiabatically evolving a given Hamiltonian to the quantum (target) Hamiltonian that encodes the problem, in an attempt to reach the ground state of the system. Under the adiabatic theorem, if the system starts in the ground state of the initial Hamiltonian (which is an easy to prepare state), it will track the instantaneous ground state if the Hamiltonian varies sufficiently slowly with respect to other energy scales \cite{kadowaki1998quantum}. The duration of the annealing (typically in the range of nanoseconds to microseconds on a real QPU) is known as the annealing time. Longer annealing times increase the likelihood the state obeys the adiabatic principle but also worsens the effect of decoherence. Each quantum annealing device has a range of annealing times that can be set, which is a hyperparameter we optimise in our experiments. The final probability distribution is then sampled by making measurements of the final state. The number of anneals we apply and consequent measurements on the final state we call the number of \textit{shots} in quantum annealing, as with VQE. Each shot produces a unique solution. The whole process we call an \textit{experiment}, again similarly to VQE above. Both VQE and QA may sometimes prepare a sub-optimal final state, and we hence repeat experiments for these methods, as we discuss further below.

Due to the limited connectivity of real quantum annealing QPUs, which are usually based on superconducting circuits, variables must be mapped to multiple qubits, forming qubit `chains'. A chain is considered broken if the value of one of the qubits in the chain is inconsistent with the others (they should all be the same value as they represent the same variable). Chains are formed via ferromagnetic coupling among qubits, which requires a balance to be struck between increasing the chain strength, so that the chains are not broken, and having the coupling weak enough so that it does not influence the solutions found in the problem \cite{lee2022determination}. 

The use of qubit chains in current non-fully connected architectures increases the number of qubits used for each variable of the problem by a quadratic factor, as compared to mapping single qubits to variables. This means that using quantum annealers currently available on the market either restricts us to solve problems that are not fully connected, or, as it is in our case, risks poorer performance when we go to larger problem sizes \cite{grant2022benchmarking}. The quantum annealing hyperparameters we optimise are: (i) the chain strength and (ii) the annealing time. 

\subsection{Post-Selection} 
\label{sect:post-selection}

The output of each algorithm described in the previous subsection is a collection of samples from an underlying probability distribution, which represent possible solutions to the QUBO problem, except for the brute force approach that produces a deterministic outcome. Notice that the QUBO problem is unconstrained and only energetically penalises the solutions that do not adhere to the constraint encoded in its cost function. We can filter out the solutions that do not satisfy the constraint, i.e. the solutions with the wrong number of vacancies, by classical post-processing.

A natural question that arises is: why were these constraint violating configurations sampled at all? Should we not have a large enough penalty coefficient, $\lambda$, so that these configurations are never sampled? The answer is as discussed before - finding the optimal $\lambda$ is a balance between making sure these solutions are never sampled and trying to not increase the QUBO energy range too much by penalising the cost of these constraint violation configurations. Often the latter is preferred. Removal of these configurations is achievable by sorting through the bit string solutions and counting the number of present vacancies (represented by zeros). If an incorrect number of vacancies is present in the solution, then the solution is removed. For all the explored methods, the post-selection improves the probability of finding the optimal solution for the original configurational analysis problem. 

To perform post-selection for our use-case with a single constraint, a constant number of bitstrings must be checked, equal to the number of samples (shots), which have $N$ entries each. Therefore, the scaling of this method is linear in $N$. The time required for post-selection at the $N=18$ QUBO problem size, was found to be approximately $0.01\unit{s}$ for $10,000$ samples, which is the maximum number of samples produced as output in any of our experiments. However, for larger problems, the number of samples may need to scale with $N$ to achieve non-zero probability of the optimal solution. As $N$ increases, this probability decreases due to increased noise from using more qubits, which we have confirmed experimentally.

The ultimate output from all of these methods is a probability distribution over classical bitstrings or, equivalently, over the configuration energies we sample from, as described above. The classical brute force approach produces the exact solution, but we can associate this with a delta function probability distribution, peaked at the optimal solution. The properties of the full distributions are revealing, and we analyse several of these shortly. Furthermore, this unified output conveniently allows various metrics or summary statistics to be defined that are applicable for all methods. These metrics are important for quick comparisons between different methods and for summarising the important behaviours. We next discuss the metrics considered in this work, which we choose to quantify the performance in terms of both the time to find the solution and the quality of the solution found.

\subsection{Performance Metrics} 
\label{sect:metrics}

Defining good benchmarking strategies for measuring performance of optimisation algorithms has been studied in both quantum and classical computing contexts~\cite{beiranvand2017best}. In general, these strategies fall under two distinct categories. The first category involves allocating a fixed amount of resources, such as time or energy, and measuring the quality of solution obtained, using a well-defined metric for it. The second involves measuring the amount of a resource, using an appropriate metric, that is necessary to acquire a solution that achieves a pre-specified threshold (or target) for the quality of solution.

It is therefore clear from the above that, in order to proceed, we need to define metrics of two different types: ones that measure the quality of solution, and metrics that measure the amount of the resource devoted to obtaining a solution. In our case, we select \emph{optimal solution probability} and \emph{approximation ratio} as metrics for the quality of solution, and \emph{user runtime} and \emph{QPU time} for measuring the resources to solution. 

\textbf{Optimal Solution Probability,} $\boldsymbol{P_s}$ -- The proportion of times, $N_{\text{ground state}}$, the ground state occurs in all $N_{\text{solutions}}$ solutions:
\begin{equation*}
P_s = \frac{N_{\text{ground state}}}{N_{\text{solutions}}}.
\end{equation*}
By `solutions' in the above definition we mean a collection of output samples either before or after post-selection, and this will be specified when the metric is used. This metric is independent of the detailed structure of the probability distribution and describes only the likelihood of successfully finding the solution to the problem.

\textbf{Approximation Ratio,} $\boldsymbol{AR}$ -- The approximation ratio, is defined as:
\begin{equation*} \label{eqn:AR}
    \text{AR} = \frac{E-E_{\text{max}}}{E_{\text{min}}-E_{\text{max}}},
\end{equation*} 
where $E$ is the energy expectation value of the solutions, $E_{\text{min}}$ and $E_{\text{max}}$ are the theoretically calculated minimum and maximum energy values of the cost function over all bit strings that satisfy the constraints, as in Ref.~\cite{moussa2022unsupervised}.

In our case we calculate AR using the solutions after post-selection. The reason we do this is because before post-selection we can have energies outside the range $[E_{\text{min}},E_{\text{max}}]$. These $E_{\text{min}}$ and $E_{\text{max}}$ values are calculated using constrained brute-force techniques. The AR places the average energy found within the context of the whole energy landscape of the problem. We note that there is some variation in the name and precise definition used in the literature for this metric. When exploring new large problems, the theoretical minimum and maximum values of energy may not be readily available without having to solve using an exact method. An option for larger problems is to use, in place of the AR, the average energy and the minimum energy obtained experimentally by a given method.

\textbf{User Runtime} -- The time experienced by the user when running the experiment:
\begin{equation*} \label{eqn:Runtime}
    \text{User Runtime} = T_{\text{encoding}} + T_{\text{latency}}  + T_{\text{device}}
\end{equation*}
where $T_{\text{encoding}}$ is the classical-to-quantum encoding time, $T_{\text{latency}}$ is the job processing time (the time it takes to send the job to the desired device, also known as the latency) and $T_{\text{device}}$ the device runtime, which is the CPU or QPU time (or both for VQE on the QPU). In the case of VQE, $T_{\text{encoding}}$ includes transpilation of the circuit to native gates and optimisation of the physical circuit, whereas, for quantum annealing, $T_{\text{encoding}}$ is just the embedding time. Note that we do not include the time spent in device queues, and that our separation of quantum and classical time is consistent with Ref. \cite{koch2025quantum}.

For classical brute force, random sampling, simulated annealing and VQE with the state vector solver, $T_{\text{encoding}} = 0$. For classical brute force, random sampling and simulated annealing $T_{\text{latency}}$ is also equal to zero as code is run locally on HPC, such that there is no network latency. The post-selection time, $T_{\text{post-selection}}$, is another interesting time metric to show for full visibility. As our post-selection time is approximately equal for each method we do not include it in the user runtime, as discussed in the previous sub-section.

\textbf{QPU Time} -- The time required on the QPU to reach a solution to the problem:
\begin{equation*} \label{eqn:QPU}
    \text{QPU Time} = T_{\text{device,Q}} 
\end{equation*}
where $T_{\text{device,Q}}$ is the QPU runtime, specific to running on real quantum hardware.

For D-Wave devices this is the `QPU access time', which includes the QPU programming time \cite{QPUtimeDwave}. Similarly, for IBM devices this is defined as the time a QPU is committed to complete a job \cite{QPUtimeIBM}, which includes compiling time (converting the transpiled quantum circuit into pulse sequences). Our reported QPU time for the IBM device is calculated by summing the `quantum seconds' of each `job' (in our case a job corresponds to a single \textit{iteration} of VQE). An alternative method for calculating the QPU is described in Ref. \cite{koch2025quantum}. This metric is important for hybrid algorithms as it separates $T_{\text{device}}$ into CPU and QPU time.

Having discussed the selected metrics, we also note a number of other metrics that are used in the wider literature. Ref. \cite{cutugno2022quantum}, considers the number of violated constraints. For our QUBO there is only one constraint specified - the number of vacancies. The number of times it is violated could be reflected in a `validity' metric: if many solutions violate the constraint then the validity returned will be low. We instead choose to encode information about constraint violation by post-selecting the data to remove solutions that do not obey the constraint, as described above. A metric specific to the quality of the quantum annealing process in current devices is the percentage of broken chains. We report this value in Appendix \ref{app:annealing}, Figure \ref{fig:b_chain}.

Another metric often used is the time-to-solution (TTS), where this is the time taken for a method to find the optimal solution once with a desired accuracy (often $99\%$). TTS, as defined in Ref. \cite{zhou2020quantum, zeng2024performance}, is:

\begin{equation*} \label{eqn:TTS}
    \text{TTS} = T\frac{\ln{(1-p_d)}}{\ln{(1-p_{\text{GS}})}}.
\end{equation*}

Where $p_d$ is the desired accuracy ($99\%$). The TTS can be calculated from our metrics realising $T$ is our user runtime and $p_{\text{GS}}=P_s$.

Ref. \cite{bauza2024scaling} uses $\text{TT}_\epsilon$ for approximate optimisation, where $\text{TT}_\epsilon$ is the time to reach an energy within a fraction $\epsilon$ of the ground state energy. $\text{TT}_\epsilon$ is a useful metric for large problem sizes where the ground state may not be found.

Finally, we report error bars on all of our metrics to enable proper statistical interpretation. We have two distinct sources of error in our methods. For our optimal solution probability metric, we have the standard error, SE, which takes into account shot noise when sampling from the output distribution:
\begin{equation}
    \text{SE} = \sqrt{\frac{P_s(1 - P_s)}{N_{\text{solutions}}}}.
\end{equation}

Here we create our collection of solutions by aggregating shots across many experiments. This means that the total number of solutions for quantum annealing and VQE is equal to $N_{\text{solutions}} = N_\text{experiments}\times N_\text{shots}$, with $N_{\text{experiments}}$ being the number of times the experiment is repeated and $N_{\text{shots}}$ the number of shots per experiment. While in simulated annealing and random sampling it is equal to the number of times we run the algorithm. For the definitions of shots and experiments see Section~\ref{sect:algor}. This is necessary with quantum annealing as the D-Wave devices have a maximum QPU access time, and in order to increase the number of shots to a sufficiently large number, one needs to run the experiment again\cite{dwaveTiming}. 

We also compute the standard deviation, $\sigma$:
\begin{equation}
    \sigma = \sqrt{\frac{ \sum_{i=1}^{N_{\text{experiments}}} (x_i - \mu)^2} {N_{\text{experiments}}} },
\end{equation}
where $x_i$ is a value obtained from the experiment and $\mu$ is the experiment mean. The resulting error bar takes into account errors due to imperfect preparation of the ground state. We therefore repeat `experiments': the entire process of state preparation and generating $N_{\text{shots}}$ samples several times ($N_{\text{experiments}}$). For each experiment, we compute the values of our quality of solution metrics, and then report the mean and standard deviation as above to capture this variation in state preparation. Note that we also use $\sigma$ for our time metrics. Finally, we note that we could simply aggregate all shots across all experiments into one big distribution and not discuss two sources of error. We choose to separate out the two error sources to achieve more detailed insight into these methods, and to allow us to meaningfully characterise the runtime. As $\sigma$ is the more dominant error source (notice SE will be small as $N_{\text{solutions}}$ is large) we report this as our error bars in the main text and report the SE in Appendix \ref{app:sim_annealing}, Table \ref{tab:perform_withSE}.

\section{Results} 
\label{sect:results}

\begin{figure*}[ht!]
    \centering
    \begin{minipage}[b]{0.4\linewidth}
        \centering
        \includegraphics[width=\linewidth]{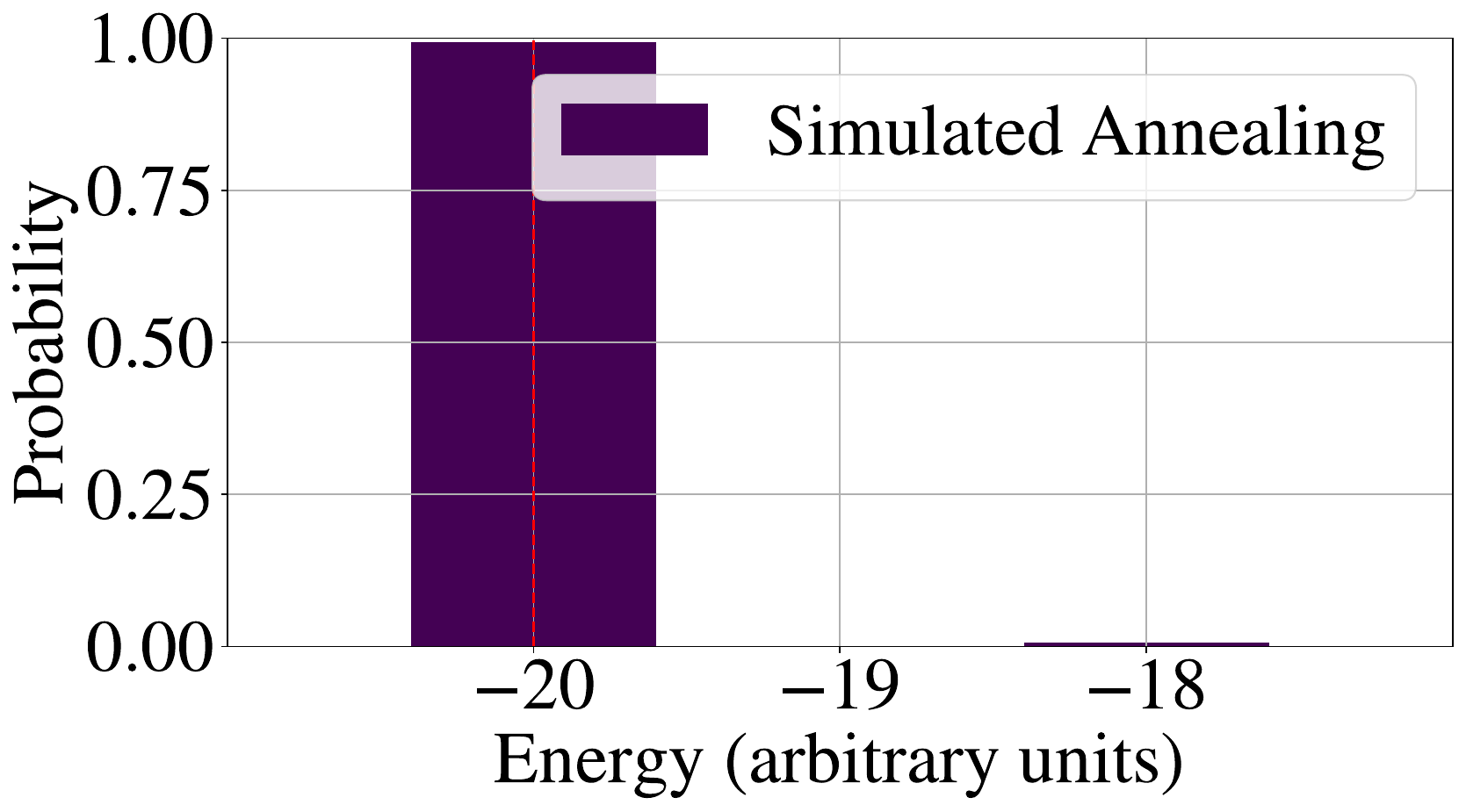}
    \end{minipage}
    \begin{minipage}[b]{0.4\linewidth}
        \centering
        \includegraphics[width=\linewidth]{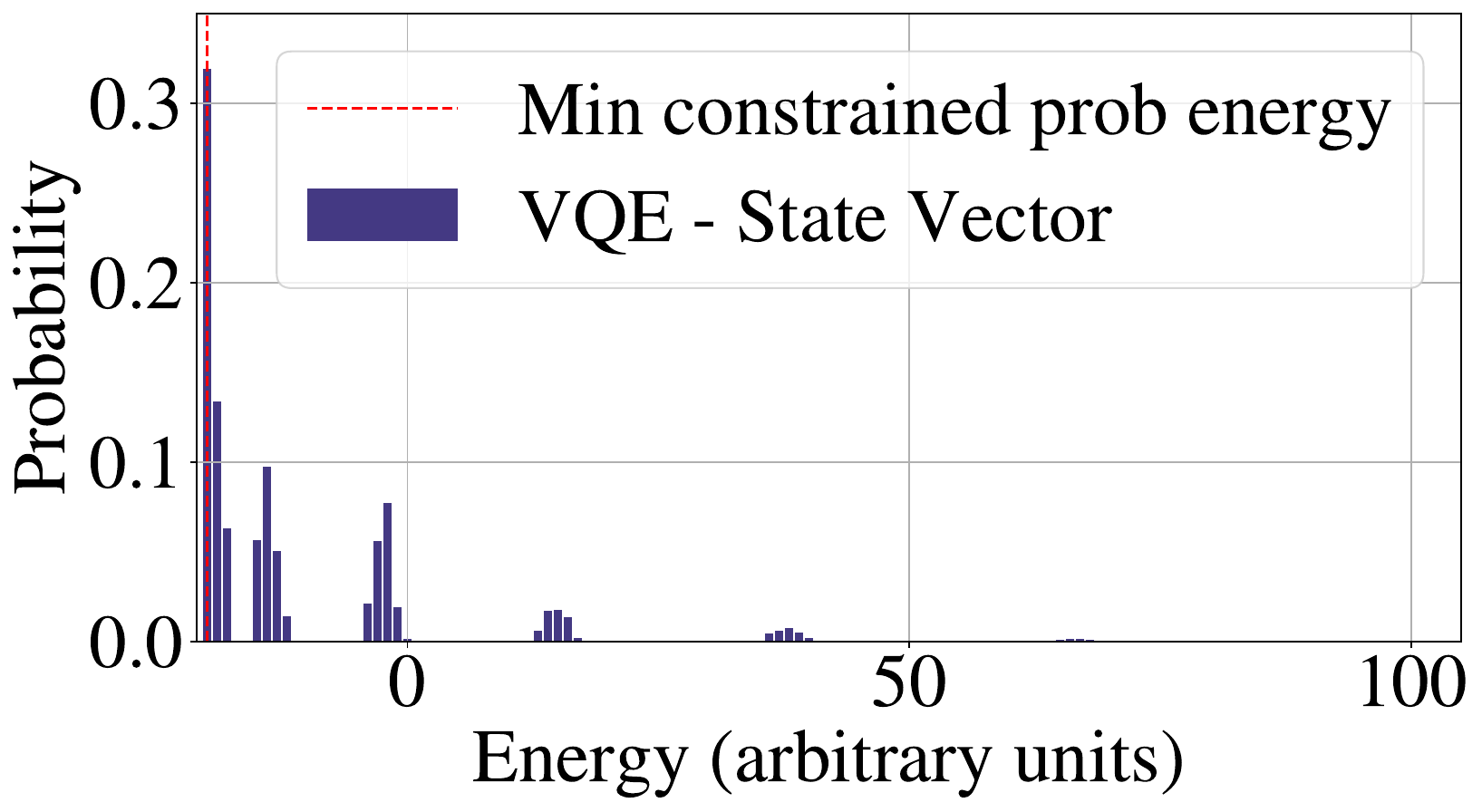}
    \end{minipage}
    \begin{minipage}[b]{0.4\linewidth}
        \centering
        \includegraphics[width=\linewidth]{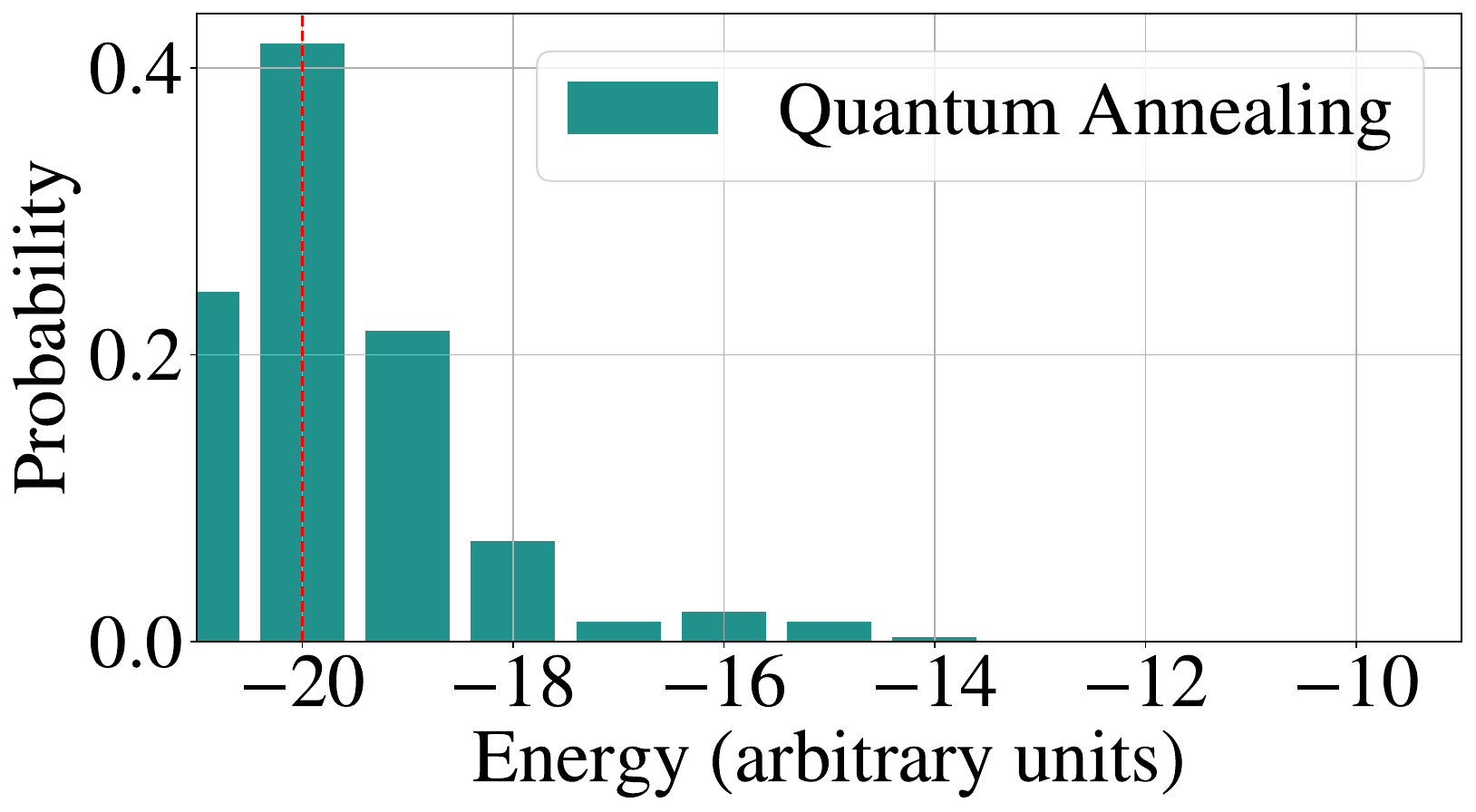}
    \end{minipage}
    \begin{minipage}[b]{0.4\linewidth}
        \centering
        \includegraphics[width=\linewidth]{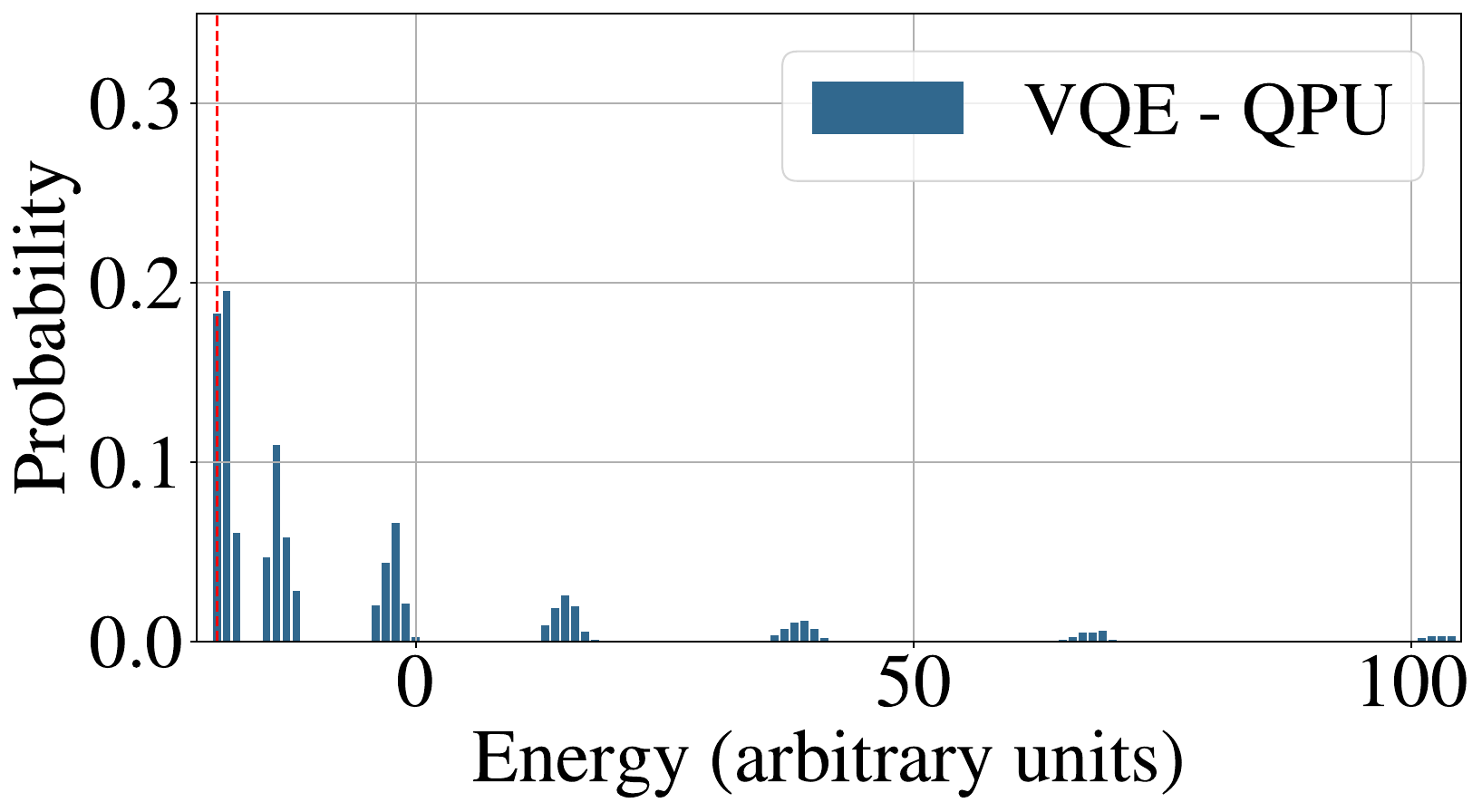}
    \end{minipage}

    \begin{minipage}[b]{0.40\linewidth}
        \centering
        \includegraphics[width=\linewidth]{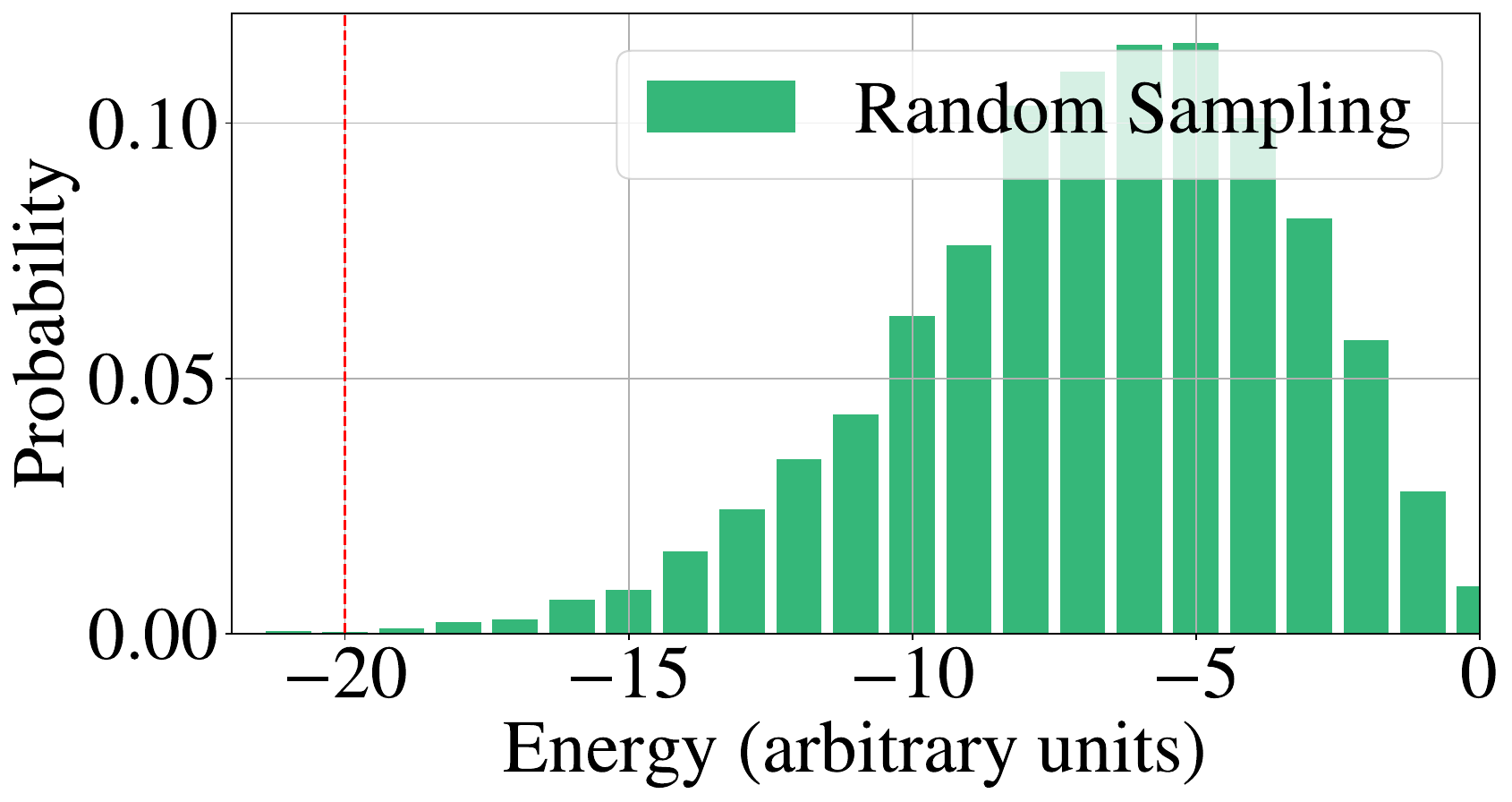}
    \end{minipage}

    \caption{\small{The distributions as a function of energy obtained from each method on the 18 variable problem, without post-selection. Quantum annealing was performed on D-Wave's 5612 physical qubit Advantage System 6.4. The VQE results were obtained using a noiseless state vector solver and the real 156 qubit \textit{ibm\_fez} QPU. Experiments were repeated 10 times to produce accumulated distributions which have been normalised. The minimum constrained problem energy represented by the red dashed line is $-20$ in our units. Note that the distribution for the brute force method is omitted, as it is a trivial delta function distribution centred on the exact solution. The hyperparameters used for simulated annealing were: $\lambda=3$, the temperature range was set at $\beta=[0.1,10]$ with 1000 iterations (sweeps) and number of repeats $=1000$. Quantum annealing with minor embedding: $\lambda=1$, chain strength $=3$, the annealing time $=1400\unit{ns}$ and the number of shots $=1000$. For VQE: $\lambda=3$, with the RealAmplitudes ansatz, using CVaR $\alpha=0.4$, COBYLA with $\text{tol}=1$, and shots $=10000$. For random sampling $\lambda=5$ and 1000 samples were used.}}\label{fig:distributions} 
\end{figure*}

We first solve the $18$ variable QUBO and analyse our methods performance using the metrics defined in Section \ref{sect:metrics}. Figure \ref{fig:distributions} shows the energy probability distributions obtained by each method before post-selection.

\begin{figure}[htb]
\centering
\includegraphics[width=0.38\textwidth]{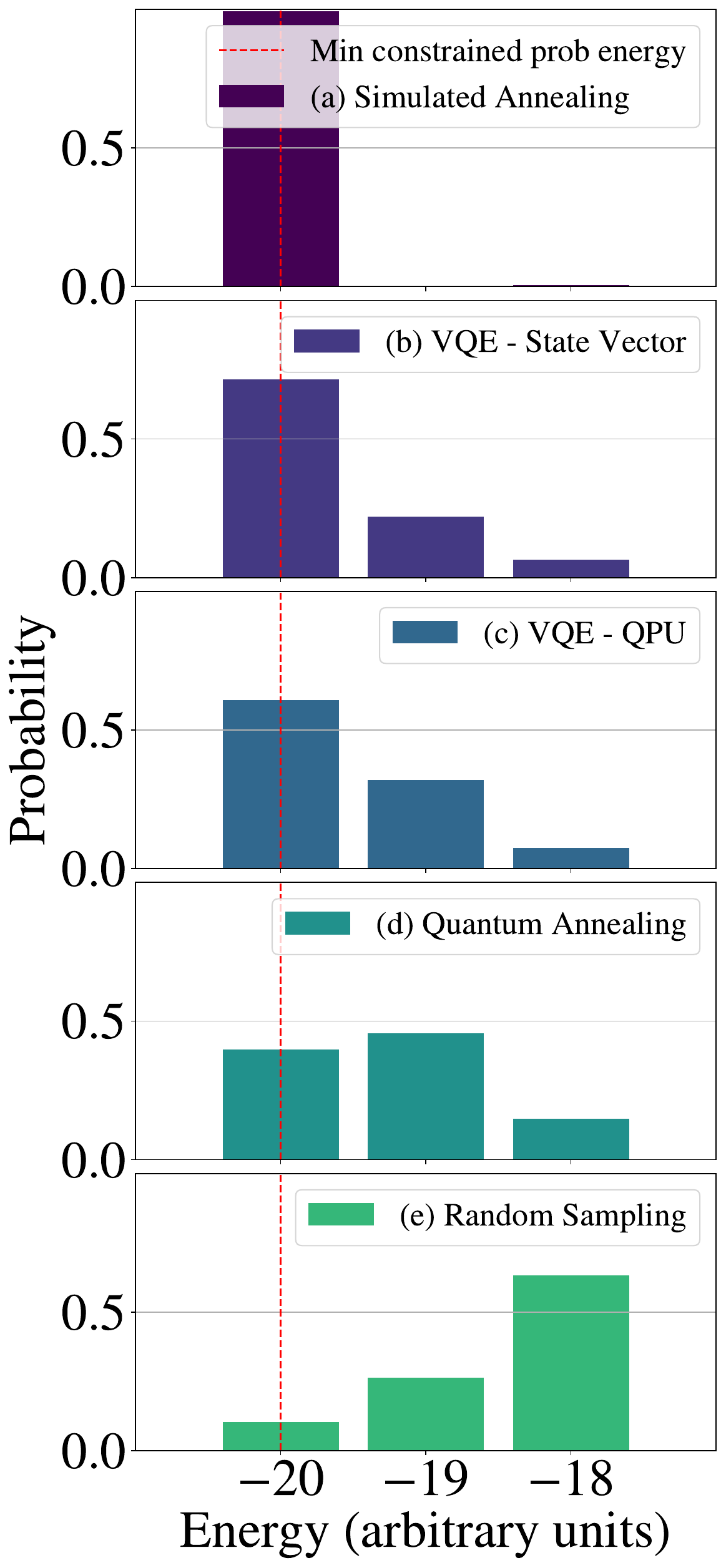}
\caption{\small{The post-selected distributions obtained for each method on the 18 variable problem. Experiments were repeated 10 times to produce accumulated distributions which have been normalised. The minimum constrained problem energy represented by the red dashed line is $-20$ in our units. The hyperparameters used are as stated in the caption of Figure \ref{fig:distributions}.}}\label{fig:post-distr} 
\end{figure}

For VQE on the state vector simulator and quantum annealing, the experiment was carried out $10$ times, the output samples were accumulated and the distributions renormalised. For VQE on the IBM QPU the experiment was carried out $5$ times due to time constraints. We report our metrics averaged across repeat experiments and their error bars as described in the previous section.
Finally, as brute force is an exact method, we take it to have idealised values of our quality of solution metrics, a value of $1$ for $P_s$ and AR. Figure \ref{fig:post-distr} shows the post-selected distributions, in which solutions that violate the constraint on the number of vacancies are excluded. We see that the distributions have more weight over a range of energies close to the minimum.

\begin{table*}[t]
\begin{ruledtabular}
\begin{tabular}{lccccc}
\textbf{Method} & \textbf{$P_s$} & \textbf{$P_s$ Post-Sel.} & \textbf{AR Post-Sel.} & \textbf{User Runtime (s)} & \textbf{QPU Time (s)} \\ 
\colrule
Brute Force & 1 & N/A & N/A & $2.2 \pm 0.03$ & N/A \\
Simulated Annealing & $0.993 \pm 0.003$ & $0.994 \pm 0.003$ & $0.994 \pm 0$ & $0.339 \pm 0.001$ & N/A \\
VQE State Vector & $0.319 \pm 0.1$ & $0.695 \pm 0.2$ & $0.826 \pm 0.01$ & $38.5 \pm 7$ & N/A \\
VQE QPU & $0.183 \pm 0.07$ & $0.595 \pm 0.1$ & $0.761 \pm 0.08$ & $(2.54 \pm 0.4) \times 10^3$ & $941 \pm 200$ \\
Quantum Annealing & $0.189 \pm 0.02$ & $0.398 \pm 0.03$ & $0.626 \pm 0.03$ & $2.54 \pm 0.08$ & $0.399 \pm 0.08$ \\
Random Sampling & $0.0004 \pm 0.0005$ & $0.104 \pm 0.1$ & $0.235 \pm 0.2$ & $0.001 \pm 0$ & N/A \\
\end{tabular}
\end{ruledtabular}
\caption{\label{tab:perform} The performance metric results for our different methods on the 18 variable problem. All metrics are given as mean values, with the standard deviation error included.}
\end{table*}

To produce the quantum annealing results presented, D-Wave's $5612$ physical qubit Advantage System 6.4 was used. The VQE results were produced using the local noiseless IBM state vector solver and the real $156$ qubit device \textit{ibm\_fez}, accessible via the cloud from the IBM Quantum Platform. Not all of the physical qubits from the quantum annealing device and gate-based QPU are used: for further details of the qubit embedding procedure see Appendices \ref{app:VQE} and \ref{app:annealing}. The classical brute force, simulated annealing and random sampling were performed on a machine with 32 CPUs and $123\unit{GB}$ RAM. However, multi-threading was not configured which would better utilize the machine's capabilities. Acceleration through the use of GPUs would also likely have been beneficial.

We now discuss the results for each method and how they scale for larger sizes of the problem in detail. In the following sub-sections we first focus on the results of quantum annealing (Section \ref{sect:results_QA}), and then the results of VQE (Section \ref{sect:results_VQE}). Classical results are discussed throughout both sections.

\subsection{Quantum Annealing Results} \label{sect:results_QA}

The quantum annealing distribution in Figure \ref{fig:distributions} is skewed closer to the optimal solution than random sampling, so it has better general performance in the sense of solution quality, as reflected in the optimal solution probability metrics in Table \ref{tab:perform}. The optimal solution, for the rest of this section, being the optimal energy among the solutions that respect the fixed vacancy constraint in the QUBO. However, solutions below the minimum constrained problem energy are also found. This is a consequence of the small penalty coefficient $\lambda$ which was found from hyperparameter optimisation. If $\lambda$ is large this increases the QUBO energy range and results in a smaller spectral gap in the annealing Hamiltonian. A smaller spectral gap means that more non-adiabatic effects are likely, which can cause erroneous results.

Comparing the quantum annealing distribution in Figure \ref{fig:distributions} with the post-selected distribution in Figure \ref{fig:post-distr}, we see that the post-selected distribution has support over a smaller range of energies because high-energy/less-optimal solutions are invalid and are removed, as expected. The post-selected distribution for quantum annealing is closer to the optimal solution than random sampling but further from it than simulated annealing. This distribution includes energies that are not optimal, which is reflected in the AR metric.

In Table \ref{tab:perform} it can be seen that the QPU time for quantum annealing only makes up a fraction of the user runtime. In Appendix \ref{app:annealing} we dissect the user runtime and show that the majority of this is coming from the embedding procedure (Figure \ref{fig:emb_time}). This motivates us to explore more efficient embedding techniques, which we do in the next section as well as solving larger sizes of QUBO.

\subsubsection{Scaling to larger problem sizes: Quantum Annealing} 
\label{sect:res:qa:largersizes}
\leavevmode

\begin{figure}[!htbp]
    \centering
    \includegraphics[width=0.45\textwidth]{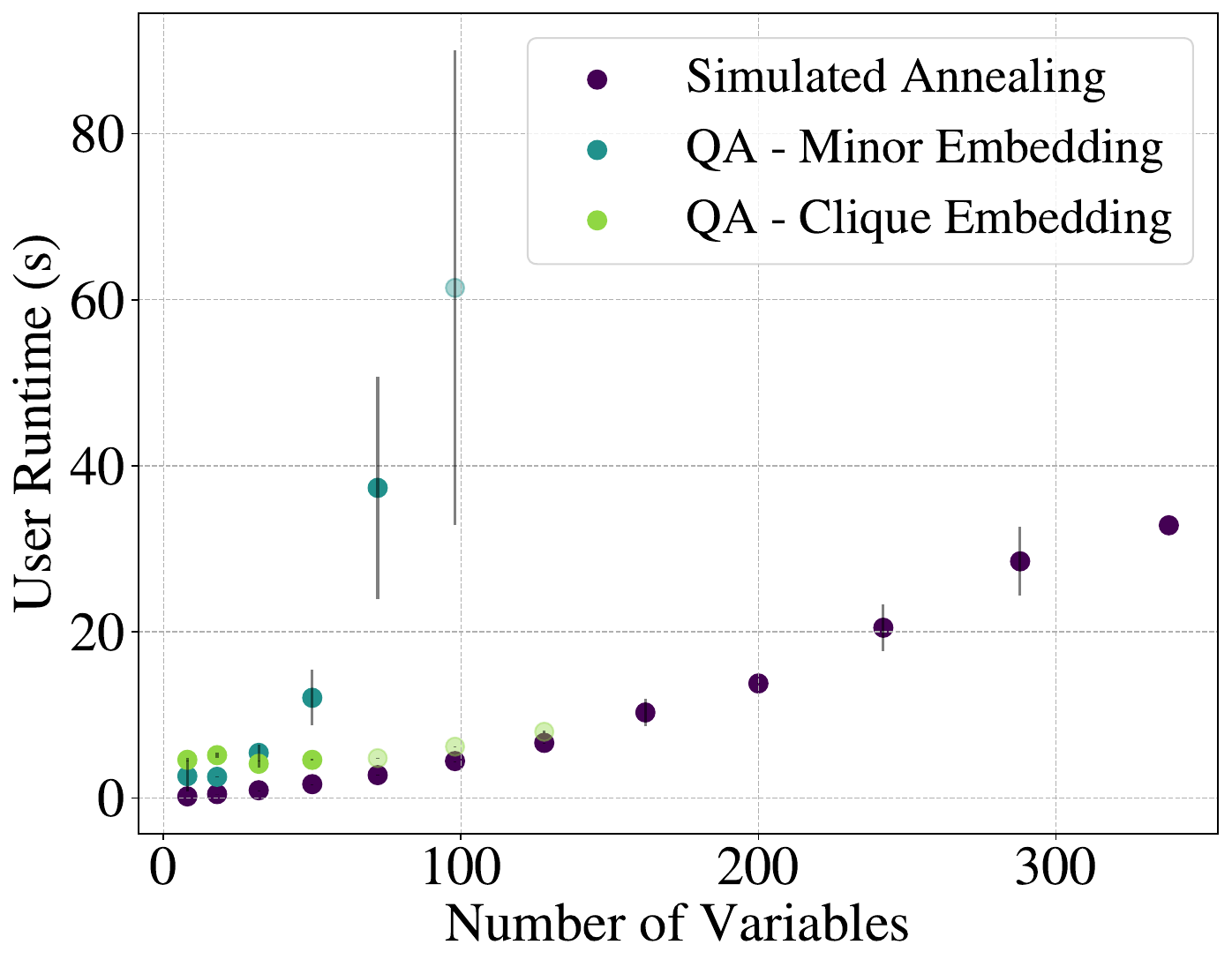}
  \caption{\small{Simulated and quantum annealing user runtime when solving the three vacancy QUBO problem at different sizes up to 338 variables. The results were repeated 10 times, with average values used, and the standard deviation included as error bars. Results with $P_s = 0$ are included as reduced opacity data points. The hyperparameters used for simulated annealing for each problem size were $\lambda=3$, the temperature range was set at $\beta=[0.1,10]$ with 1000 iterations (sweeps) and number of repeats $=1000$. For each problem size the hyperparameters for minor and clique embedding differ and are detailed in Table \ref{tab:QA_hyp_params}. In Figure \ref{fig:scalingTime_log_combined} the same data is plotted with logarithmic axis scales to reveal the polynomial scaling behaviour.}}
  \label{fig:scalingTime_combined}
\end{figure}

\noindent We now discuss how quantum annealing runtime scales as a function of problem size. The time spent by quantum annealing on each problem instance is partly determined by the number of anneals (shots) and the annealing time. The number of anneals is chosen so that our results are statistically significant, and the annealing time is found through hyperparameter optimisation. Figure \ref{fig:scalingTime_combined} displays the user runtime for simulated annealing and quantum annealing (with two different embedding techniques) as a function of the number of QUBO variables, $N$. The quantum annealing embedding technique used for the results in the previous section was minor embedding, D-Wave's default mapping scheme, which is a heuristic method. An alternative embedding technique, `clique embedding' \cite{boothby2016fast}, can be used for QUBO problems where the QUBO graph forms a `clique', which means that it has full connectivity (100\% density). Clique embedding is an algorithm that runs in polynomial time in the worst case\cite{dwaveClique,de2024scaling,boothby2016fast}, which has uniform chain lengths (or close to uniform) \cite{dwaveClique}. We observe near-constant scaling with clique embedding in Figure \ref{fig:emb_time}. At larger problem sizes this may start to grow polynomially.

As discussed previously, the primary contributor to the user runtime of quantum annealing using minor embedding is the embedding time, detailed in Appendix \ref{app:annealing} and explicitly shown in Figure \ref{fig:emb_time}. As problem size increases, the required average qubit chain length also increases, shown in Figure \ref{fig:chain_length}. Consequently, the heuristic minor embedding technique must explore more mapping possibilities, leading to poor scaling. This can also explain the large standard deviation error bars observed in Figure \ref{fig:scalingTime_combined}: there is more variability possible when the minor heuristic embeds larger problems and requires longer qubit chains. A new embedding is found each time the minor heuristic is performed, which can give variable performance in terms of time and quality of solution. In future work, the minor embeddings which yield good performance could be saved and re-used for multiple experiments. 

While considering the runtime scaling, it is important to not lose sight of solution quality. Quantum annealing with minor embedding was able to solve up to the 72 variable QUBO while still returning some non-zero $P_s$. On the other hand, clique embedding was able to solve up to the 50 variable problem while
returning non-zero $P_s$. We include results with $P_s = 0$ as reduced opacity data points in Figure \ref{fig:scalingTime_combined} to give insight into how quantum annealing scales for larger problems. The quality of solution and time metric results are shown in Table \ref{tab:perform_QA}, where we see that, in general, minor embedding returns a higher $P_s$ but with very large standard deviation errors. 

Turning now to the scaling behaviour of classical simulated annealing, we solve the QUBO with up to 338 variables. None of our other methods are performant for such large problems, so we verify these simulated annealing solutions with a modified version of brute force which exploits the structure of the problem. Instead of checking an exponentially-large number of states with brute force, for our particular problem we can search through just the states with the specified number of vacancies. For example, this means that for our 32 variable QUBO where we are looking for solutions with three vacancies that this version of brute force now only has to search through ${32\choose 3} \approx 5\times10^{3}$ configurations. 

We note that we only use this constrained search space for brute force to verify these larger problem sizes. It is fairer practice to compare the brute force method that searches through all configurations with our other methods as this is the space they are also searching. Using this modified constrained version of brute force effectively changes our entire problem from QUBO to a hard-constrained DkS optimisation problem.

We could equally reduce the search space for our variational quantum methods in the constrained case using a space-restricting ansatz \cite{larose2022mixer,bartschi2020grover,bartschi2022short} or quantum Zeno dynamics \cite{herman2023constrained}. We could also consider constrained quantum annealing \cite{hen2016quantum}. Another strategy to reduce the search space is to consider symmetrically equivalent structures only once. With VQAs, for example, this could be done with a symmetry respecting gate-set \cite{arnott2024reverse}.

Returning to our results, we observe polynomial time scaling for simulated annealing. This is not surprising as NP-hard problems have subsets of instances that can be solved in polynomial-time by heuristic algorithms \cite{doerr2022simulated,delahaye2019simulated}. These results highlight the importance of selecting QUBOs which are hard to solve using common classical methods in order to fully stretch quantum computing methods. 

For quantum annealing, the theoretical time scaling is bounded by the adiabatic limit, which depends on how the energy gap $\Delta_{min}$ between the ground state and first excited state of the annealing Hamiltonian closes with system size. This places upper limits on the computational complexity of problems that can be addressed. For first order quantum phase transitions, the annealing rate is bound exponentially with system size. For second order quantum phase transitions, the annealing rate is instead bound polynomially with system size \cite{somma2012quantum}. Therefore, the best scaling we can expect from quantum annealing is polynomial time, which we see experimental evidence of with clique embedding.

\subsection{VQE Results}
\label{sect:results_VQE}

Turning now to the other quantum method we considered, we see from the distributions in Figure \ref{fig:distributions} that both VQE on the state vector solver and on the QPU have small probability of sampling large energies, although the distributions do skew `to the left' compared to naive random sampling.

When comparing VQE on the QPU to the state vector solver we can see that the state vector solver has higher $P_s$. This is due to VQE on the QPU sampling higher energies more often, which is likely a result of noise. Otherwise, their distributions match well. Noise can alter the shape of the cost function landscape, introducing false minima or saddle points. The classical optimiser for VQE on the QPU could be getting stuck in these local minima, failing to converge on the global minima exactly. Post-selection improves $P_s$ for both the state vector solver and the QPU through removal of these higher energies.

A feature of the results in Table \ref{tab:perform} is that VQE has a large standard deviation error, $\sigma$, associated with the $P_s$ metric. As discussed, $\sigma$ captures the inconsistency in convergence with the classical optimiser in VQE. Convergence heavily depends on the initial starting point of the circuit parameters, which is randomly set for each repetition. Sometimes these starting parameters are unfavourable, perhaps close to local minima, and result in VQE struggling to find the global minima. 

Inspection of the time metrics in Table \ref{tab:perform} reveals that the user runtime for VQE state vector simulation is more than $17$ times longer than that of simulated annealing. VQE on the QPU has a user runtime which is over 2 orders of magnitude larger than the state vector simulation. Only $\sim 40\%$ of this time is made up of QPU time. The rest of this user runtime is coming from classical processes, such as: compilation and classical optimiser time.

Strategies to reduce this user runtime could include using a less-optimised transpilation search at the expense of solution quality. Another recent technique introduced by IBM is fractional gates \cite{fractGates}. Fractional gates are essentially new native gates which remove the need for $R_{ZZ}(\theta)$ and $R_{X}(\theta)$ rotation gates to be decomposed into many native gates. This can reduce circuit depth significantly. Our initial experimentation revealed that they can reduce the user runtime by up to a factor of two. Another technique that can be used to reduce gate depth is AI transpilation \cite{kremer2024ai}. As discussed in Section \ref{sect:methodology_for_bench}, less strict convergence criteria would result in less iterations (and shorter user runtime), again sacrificing solution quality.

As we did for QA, we now explore solving different sizes of our QUBO problem with VQE on the QPU.

\subsubsection{Scaling to larger problem sizes: VQE QPU} 
\label{sect:res:VQE:largersizes} 
\leavevmode

\noindent Performing a scaling analysis for VQE on a real QPU proved challenging, primarily due to runtime increases as the problem size is scaled up. Beyond 50 variables, VQE failed to return non-zero $P_s$. However, at the 72 variable problem size, convergence criteria were met and sub-optimal solutions were found (distribution and convergence plots are shown in Figures \ref{fig:VQE_8}, \ref{fig:VQE_18}, \ref{fig:VQE_32}, \ref{fig:VQE_50} and \ref{fig:VQE_72}, as well as the performance metrics for the different problem sizes in Table \ref{tab:perform_VQE_larger}). At large problem sizes, more qubits are required and errors accumulate, causing poor convergence to a state which has no overlap with the ground state. The extended runtime relative to the allocated device access time also meant that multiple experimental runs were not possible. Moreover, thorough hyperparameter optimisation was not possible on the QPU due to these runtime constraints. QPU hyperparameters were chosen based on small problems ran on the state vector solver or extrapolated from those results.  

Due to these complications, we do not report a practical scaling analysis for this problem solved with VQE on quantum hardware. However, we can make some theoretical comments on the expected scaling behaviour of VQE and VQAs in general. Notably, VQE is a heuristic algorithm, and to our knowledge, there is a lack of evidence of provable speed-ups for VQAs on optimisation problems. Ref. \cite{montanaro2024quantum} does claim a provable speed-up for QAOA applied to a specific symmetric optimisation problem and Ref. \cite{boulebnane2024solving} also provides empirical evidence that QAOA offers a modest polynomial speedup over leading classical solvers for random k-SAT, which is a satisfaction problem rather than an optimisation problem (the authors of Ref. \cite{boulebnane2025equivalence} believe their results on the equivalence of QAOA to quantum annealing apply for this k-SAT instance, which implies that quantum annealing should at least match QAOA's performance in terms of AR). Another quantum algorithm which has provable speed-ups for optimisation problems is decoded quantum interferometry (DQI) \cite{jordan2025optimization}, which contains the quantum Fourier transform as a sub-routine and is therefore a fault tolerant algorithm.

Regarding near-term algorithms, it remains to either find structured instances of problems where provable speed-ups exist or to experimentally find possible modest speed-ups. It is also worth noting that using a one-qubit-per-variable encoding makes it impossible to scale to large problems with current hardware due to qubit count limitations. Efforts in multi-variable to qubit encoding are investigated in Refs. \cite{sciorilli2025towards,tan2021qubit}.

\section{Conclusions and Outlook} 
\label{sect:discuss}

We introduced a fair and systematic benchmarking approach for analysing the performance of various algorithms for solving QUBO problems. This approach is applicable to other use-cases and problems in general. We applied it to the configurational analysis problem, determining the energy of defective graphene structures. We defined performance metrics, considering both solution quality and runtime, that are hardware and algorithm agnostic. We then explored the scaling of the problem to larger sizes.

Our results show that simulated annealing was the best-performing method in both solution quality and runtime, exhibiting polynomial time scaling. It can also access large problems of hundreds of variables, in contrast to the quantum methods. We also find some provisional evidence of polynomial scaling with quantum annealing, solving QUBOs up to $72$ variables. Of the two embedding techniques that we tested, clique embedding was found to be optimal in terms of runtime due to its suitability for fully connected problems. We found that using small penalty coefficients, $\lambda$, which keep the QUBO energy range small, in combination with our post-selection, which removes any solutions that violate the constraint (due to using a small $\lambda$) can return higher probability of obtaining the optimal solution for all methods.

We encountered challenges in scaling the VQE method for several reasons, namely lack of convergence within the maximum number of iterations, convergence to states with no overlap with the ground state and limited experimental repetitions due to large IBM runtimes. However, we were able to solve a dense QUBO with 50 variables and obtain sub-optimal solutions for a 72 variable dense QUBO. Using CVaR helped return a higher probability of obtaining optimal solutions with VQE.

Other techniques that could further improve the performance of gate-based VQAs exist, such as:
\begin{itemize}
    \item Warm start \cite{egger2021warm}, where instead of choosing random initial parameters for the ansatz, the starting guess comes from a classical solution to a relaxed (continuous variable) version of the problem.
    \item Concentration of parameters (also referred to as transfer of parameters) \cite{akshay2021parameter}, where, similar to warm start, the starting parameters of the VQA are chosen based off information from other, easier to solve, problems.
    \item Classical surrogate training, in which variational parameters for the QPU are transferred over from those found in approximate classical simulation \cite{herrero2025born}.
    \item Ascending-CVaR objective functions can also be used, where the CVaR $\alpha$ parameter is increased in each iteration \cite{kolotouros2022evolving}.
    \item The use of a Gibbs objective function \cite{li2020quantum} follows similar logic to that of CVaR, where the objective function is formed using the Gibbs exponent.
    \item Recursive QAOA \cite{bravyi2020obstacles}, which iteratively reduces the problem size by fixing variables.
    \item Other techniques include: filtering VQE \cite{amaro2022filtering}, adaptive VQE (ADAPT-VQE) \cite{grimsley2019adaptive}, and the generative quantum eigensolver (GQE) \cite{nakaji2024generative}.
\end{itemize}
  
Techniques also exist for quantum annealing, such as reverse annealing \cite{venturelli2019reverse}, using different annealing schedules and the variational adiabatic algorithm \cite{schiffer2022adiabatic}. The effect of applying error mitigation techniques on both platforms could also be explored further. However, implementing all of these techniques to see improvements in performance was beyond the scope of this work and we leave this as a future research direction.  

We also note that simulation of quantum annealing is possible using path integral Monte Carlo (PIMC) methods \cite{king2023quantum,morrell2024quantumannealing}, which may be of use for verifying small-scale quantum annealing solutions in future studies.

Ideally, we want to find instances of QUBO that are hard for classical techniques but suited to quantum algorithms and hardware. What makes classes of QUBO hard for algorithms in general is an open question. Certainly, solution degeneracy, which dictates the number of local minima in the cost function landscape is a contributing factor \cite{ebadi2022quantum,li2024bit}. 
Through our study, we can make some observations about what makes QUBOs hard to solve on quantum devices: QUBO density, which relates to device connectivity and, for quantum annealing, the total energy range of the problem which corresponds to preferred penalty terms and limits the annealing rate. 
Other fertile areas for future work could focus on constrained optimisation  \cite{larose2022mixer,bartschi2020grover}, optimisation problems with symmetries \cite{montanaro2024quantum} and multi-variable to qubit encoding \cite{sciorilli2025towards,tan2021qubit}.

Quantum hardware is constantly improving, with many companies producing ambitious roadmaps promising more high-quality physical qubits, and early steps towards quantum error correction. We leave finding potential speed-ups to future experimental works and hope that they can follow our framework for cross-platform benchmarking of different algorithms when solving optimisation problems.



Finally, we would like to stress that this paper is not intended to be a complete comparison of gate-based devices and quantum annealing. It is rather a framework for the comparison of `end-to-end' approaches to solving optimisation problems. Our framework compares whole approaches to solving a problem, including the algorithm and hardware platform, and does not focus on the hardware alone. 

\section*{Acknowledgements}

The authors would like to thank Bruno Camino, Stefan Woerner, Andrew King and Robert Cumming for useful discussions, and Vincent Graves for proof-reading the manuscript.

\section*{Code Availability}

The code used to generate the experimental results presented can be found in the public GitLab repository \cite{kmREPO}.

\section*{Author Contributions}

PL conceived, planned and supervised the project. KM carried out the investigation. KM and TK analysed and interpreted the results with support from CO. KG was responsible for overall supervision of the work. KM prepared the manuscript which was then proof-read and revised by all authors.

\newpage

\bibliographystyle{unsrt} 
\bibliography{mybib}

\appendix

\section{Graphene Defect Problem and its Computational Complexity}
\label{app:complexity}

Our configurational analysis use-case is finding the energies of defective graphene structures. More specifically, given a graphene sheet which we represent by a hexagonal carbon lattice of $N$ atoms with boundary conditions, we remove some of the carbon atoms, creating vacancies on the corresponding sites of the lattice. This results in breaking of carbon-carbon bonds, or simply `bonds' hereafter, between the atom we removed and its neighbours in the lattice. The energy of the resulting structure can be determined by the atoms left and their connectivity. Different approaches in determining this energy function and thus the corresponding optimisation problem have been taken in~\cite{carnevali_vacancies_2020} and in~\cite{camino2023quantum}.

We use the same approach as in~\cite{camino2023quantum}. Here the problem is the following: If we are asked to remove a specific number of atoms (i.e. create a fixed number of vacancies), the goal is to find which atoms, when removed, cause the maximum number of bonds remaining in the structure. We can model the above as a graph theoretical problem on a graph $G(V,E)$, with $|V|=N$. Each vertex $v \in V$ represents a site, which can contain an atom, in which case $v \in A$, or a vacancy, in which case $v \in \overline{A}$, where $\overline{A} = V \backslash A$. The edges represent the hexagonal carbon lattice of graphene. The condition of our problem is that the size of $A$, i.e. the number of vacancies, needs to be fixed, $|A|=N-k$. Then the problem is to determine the set $A$, i.e. the position of the vacancies, so that the bonds, which are the edges in $G$ that connect the vertices of $\overline{A}$, are maximised. 

This a very common problem in complexity theory called the Densest $k$ subgraph (DkS)~\cite{lanciano_survey_2024}. Notice that the configuration space, i.e. the number of possible selections of the vertices of the set $A$ of fixed size $N-k$ out of $N$ sites scales with ${N \choose N-k}$. For constant $N-k$ (or $k$) the configuration space is polynomial on $N$ and therefore exhaustive search gives a polynomial time algorithm (note: exhaustive search for the QUBO formulation of the problem is exponential in time as the search space is $2^N$). For arbitrary $k$ and for general graphs the problem is known to be NP-Hard, even for graphs of degree $3$ as ours~\cite{feige1997densest}. The proof is by simply observing that that the maximum clique problem can be reduced to DkS. The best known polynomial approximation algorithm has approximation ratio $N^{1/4+\epsilon}$~\cite{bhaskara_detecting_2010}. In the special case of bipartite graphs, such as our hexagonal carbon lattice with boundary conditions, it has been proven that the DkS problem remains hard~\cite{corneil_clustering_1984}. It is also known that densest problems (more edges in the original graph $G$) are harder in general~\cite{calude_quantum_2020}.

A variety of algorithms to solve this problem, including heuristics, exist, and quantum algorithms have also been used~\cite{camino2023quantum,calude_quantum_2020}. In~\cite{calude_quantum_2020}, based on a biology use-case, randomised bipartite DkS problems with $N = 30$ and $k=15$ are tried on Chimera-architecture D-Wave annealers finding that they cannot handle this problem. We use the more connected D-Wave architecture. For an account of classical algorithms solving this problem see the survey in~\cite{sotirov_solving_2020}. It has been shown that tabu search heuristics achieve better performance than simulated annealing and greedy randomised search~\cite{kincaid1992good}. Also, the variable neighbourhood search heuristic performs better than tabu search for sparse initial graphs, such as ours.

A different approach in modeling the problem is taken in~\cite{carnevali_vacancies_2020}, which requires us to double the number of qubits. Here only the so-called `dangling bonds', which an atom is connected to a vacancy, increase the energy of the structure and therefore its instability. Now, if we are asked to remove a specific number of atoms (a condition not enforced by the authors of~\cite{carnevali_vacancies_2020}), the goal is to find which atoms, when removed, cause the minimum number of dangling bonds. The graph theoretical problem is therefore slightly different. Again we have a graph $G(V,E)$, with $|V|=N$ with vertices representing sites and the edges the carbon lattice, and again $A$ represents the set of vacancies. Again the problem is to find a partition of $V$ into two fixed sized sets $A$ and $\overline{A}$, but the goal now is to minimise the number of elements $e \in E$ for which $e \cap A \neq \emptyset$ and $e \cap \overline{A} \neq \emptyset$ (in other words the edges that connect one member of $A$ with one member of $\overline{A}$, are minimized). This is the well-studied minimum bisection problem (MBP), known to be NP-Hard for general graphs~\cite{garey_simplified_1974, sotirov_graph_2018}. A typical formulation of the problem is to choose $k=\lfloor N/2 \rfloor$. In the special case of $3-$regular graphs, such as our hexagonal carbon lattice with boundary conditions, it has been proven that the MBP problem is hard to even approximate~\cite{berman2002approximation}. We think that this is a very promising version of the problem to be considered in future work.


\section{Hyperparameter Search}
\label{app:hypparam}

\begin{figure*}[htbp] 
    \centering
    \subfloat[\label{fig:SA_hyp:beta}]{%
        \includegraphics[width=0.32\linewidth]{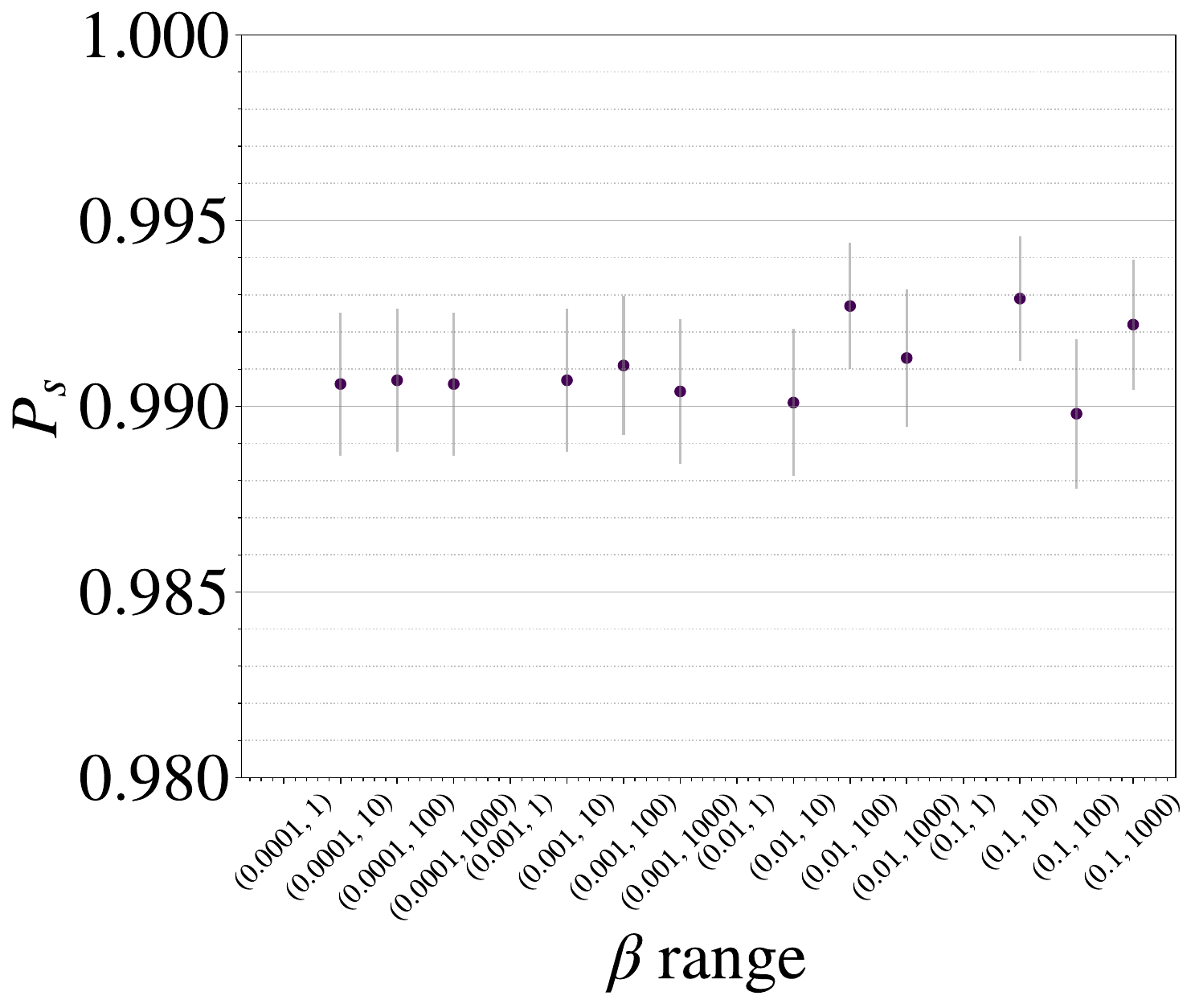}
    }
    \hfill
    \subfloat[\label{fig:SA_hyp:lambda}]{%
        \includegraphics[width=0.32\linewidth]{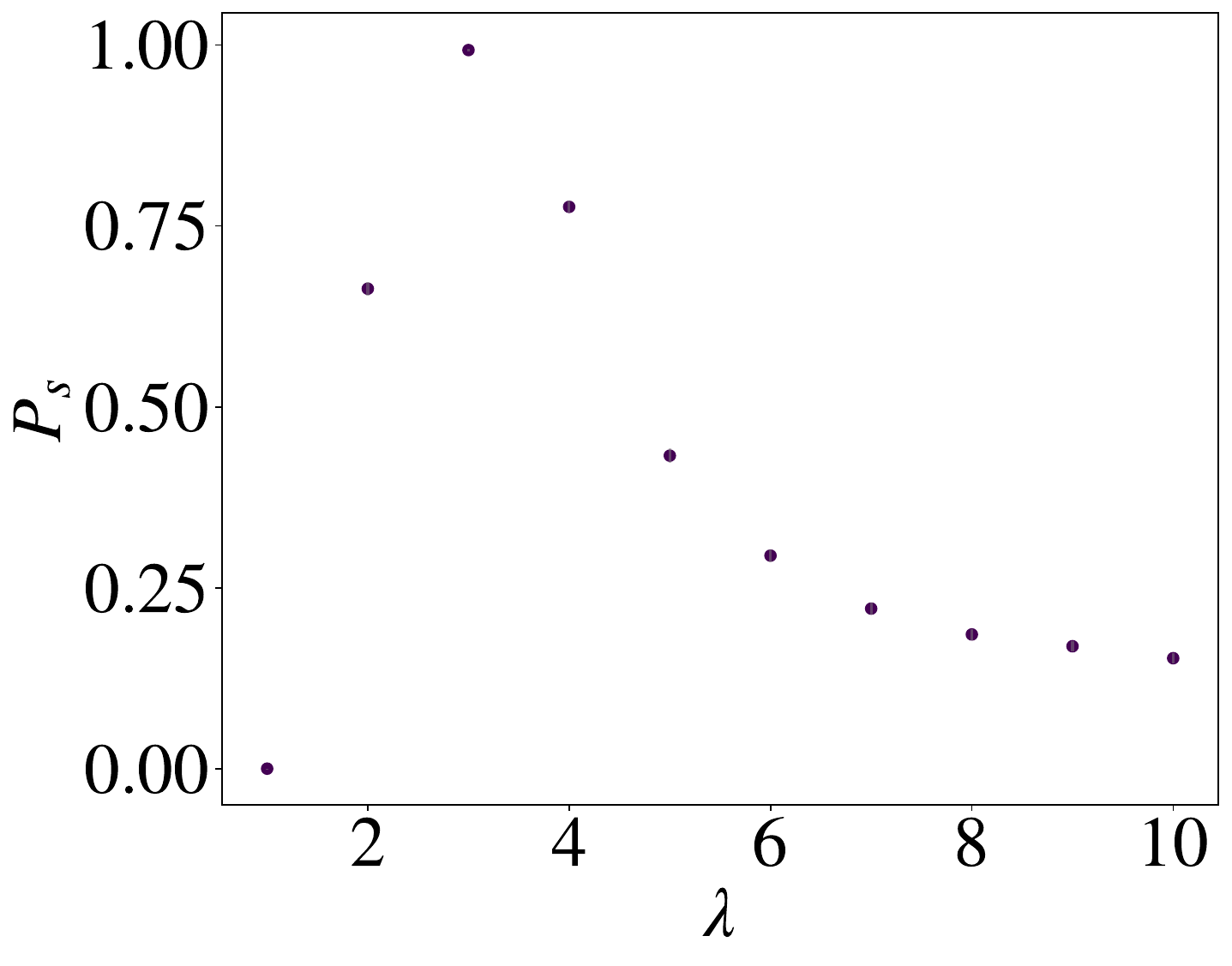}
    }
    \hfill
    \subfloat[\label{fig:SA_hyp:numsweeps}]{%
        \includegraphics[width=0.32\linewidth]{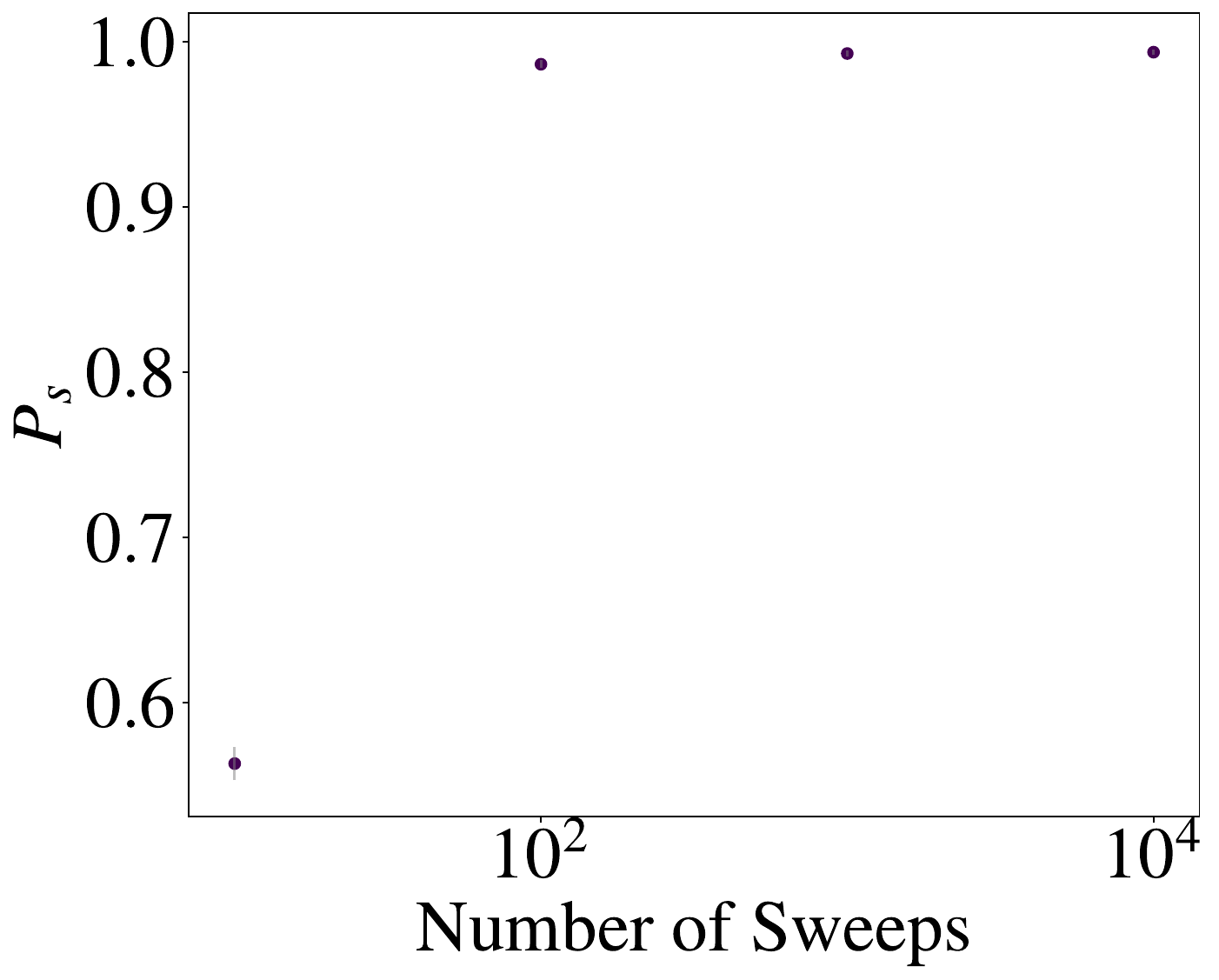}
    }
    \caption{The 18 variable problem hyperparameter search for simulated annealing testing different values of $\beta$, number of sweeps, and penalty coefficient $\lambda$. (a) $\beta$ range optimization, (b) $\lambda$ parameter plot, (c) optimal number of sweeps.}
    \label{fig:SA_hyp}
\end{figure*}

\begin{figure}[htbp]
\includegraphics[width=.8\linewidth]
{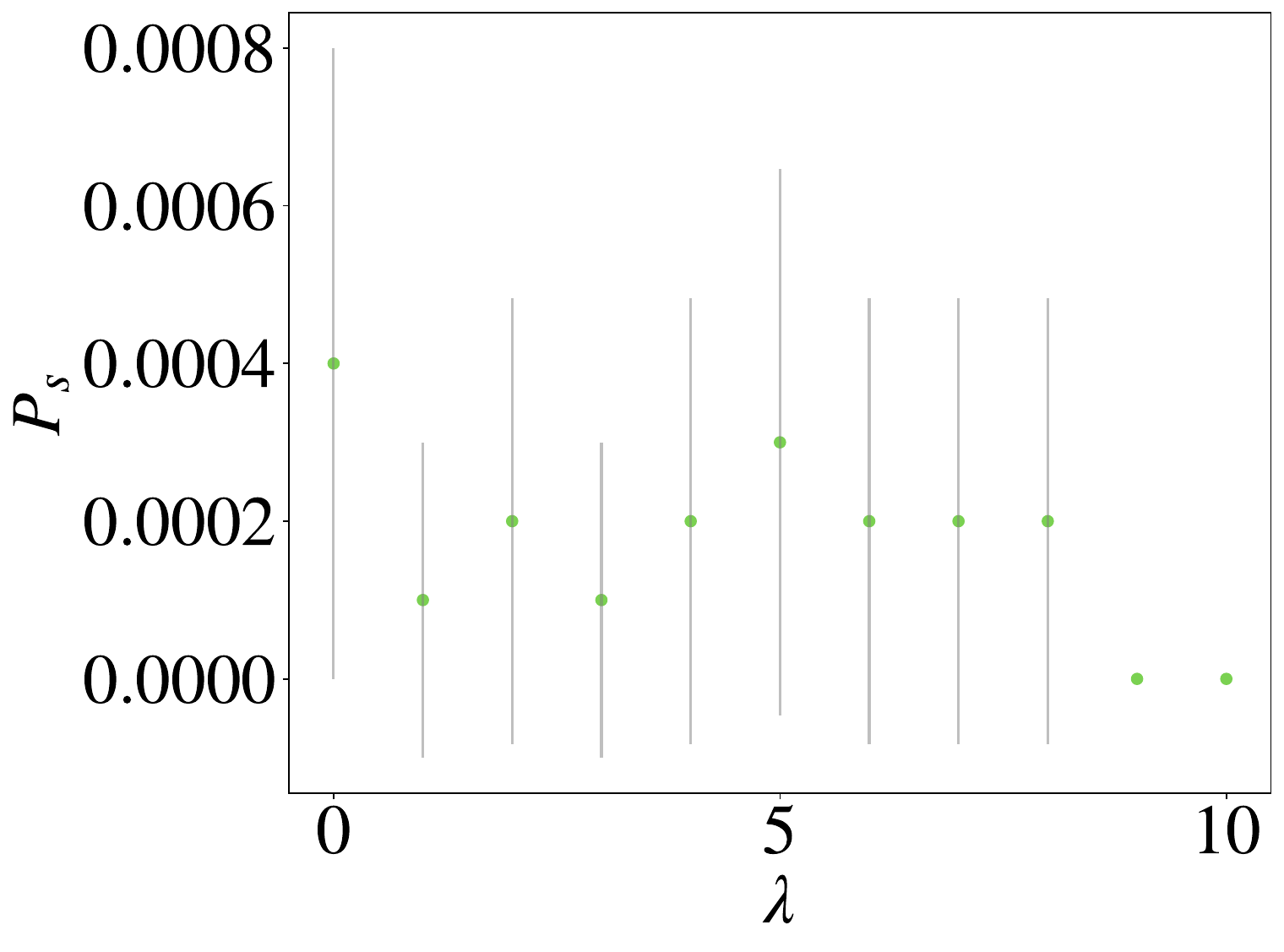}
\centering
\caption{\small{Solving the 18 variable problem with random sampling at different values of the penalty coefficent $\lambda$. Based on these results, $\lambda = 0$ was used for the random sampling results in Section \ref{sect:results}.}} \label{fig:rand_hyp}
\end{figure}

\begin{figure*}[htbp]
    \centering
    \subfloat[\label{fig:ansatz_hyperparameter_l2}]{%
        \includegraphics[width=0.45\linewidth]{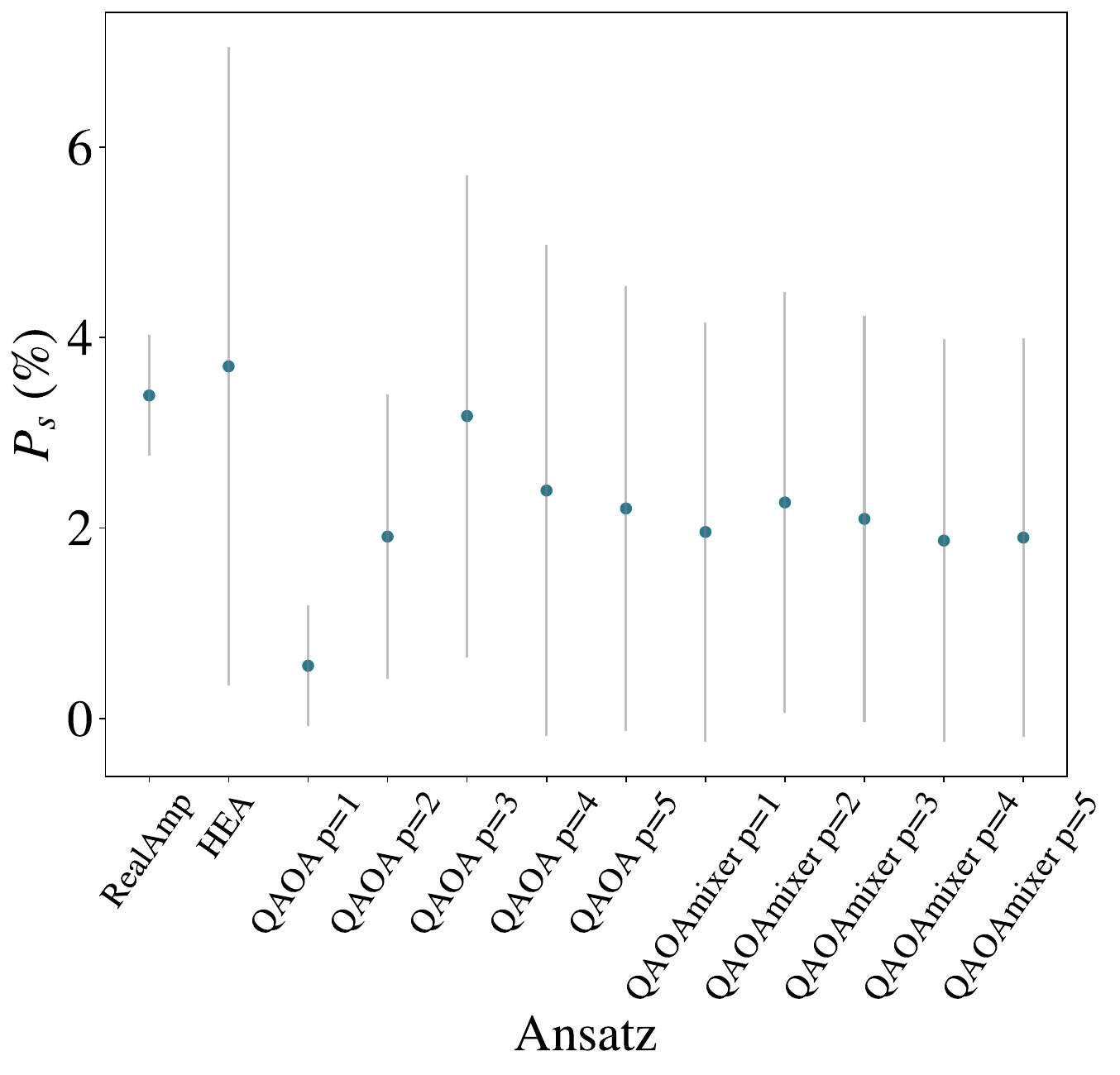}
    } \hfill
    \subfloat[\label{fig:lambda_hyperparameter}]{%
        \includegraphics[width=0.45\linewidth]{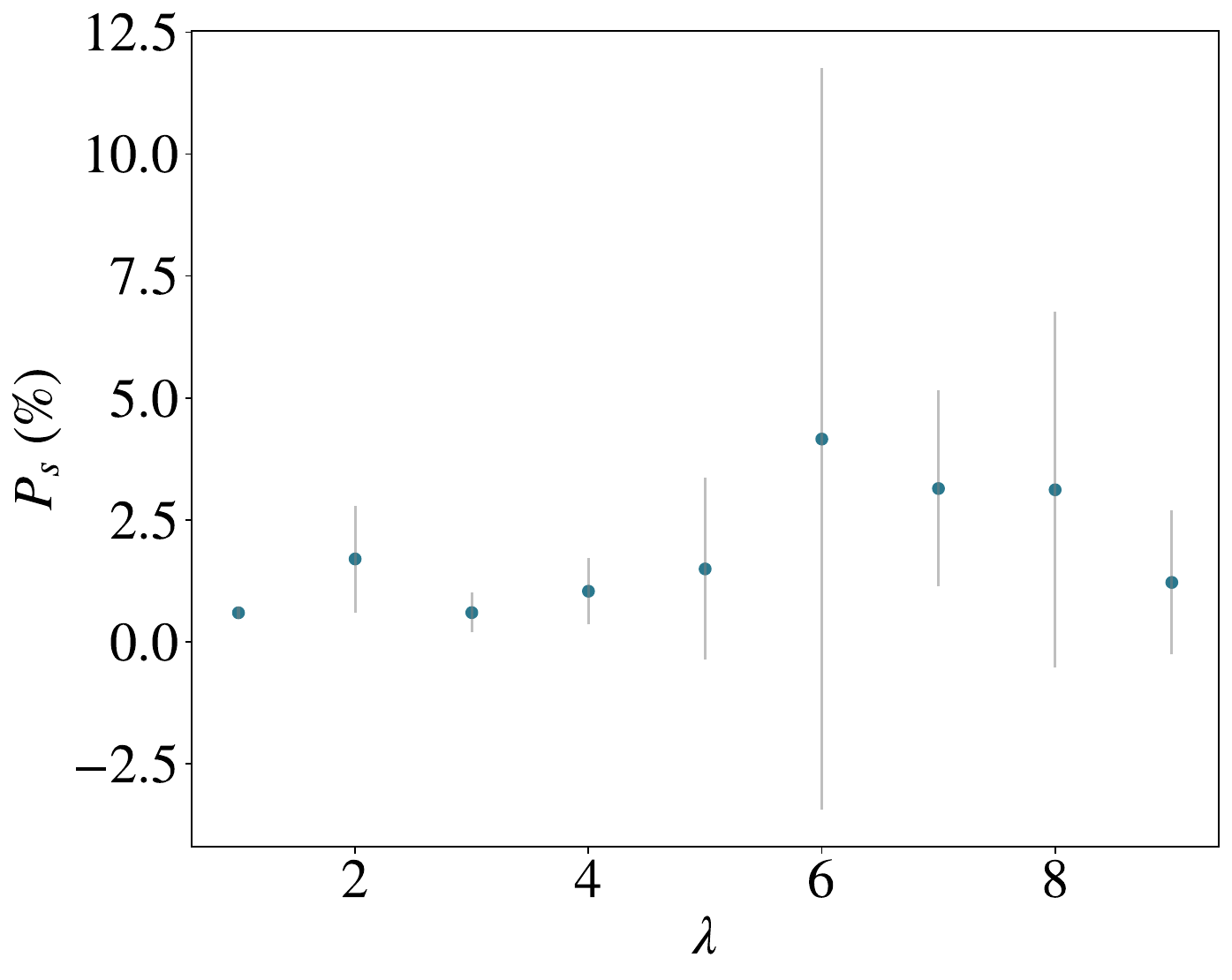}
    } \\
    \subfloat[\label{fig:alpha_hyperparameter}]{%
        \includegraphics[width=0.45\linewidth]{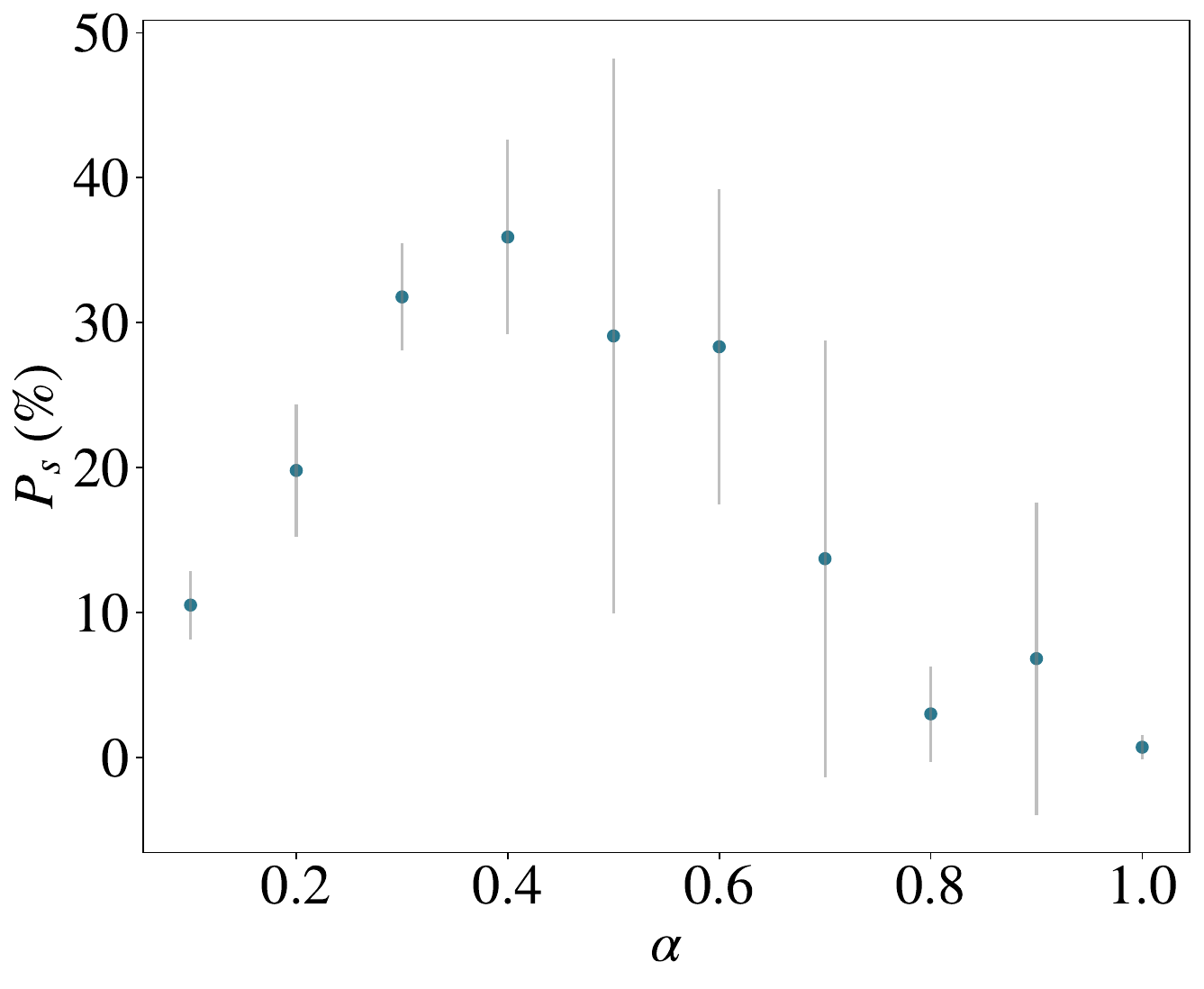}
    } \hfill
    \subfloat[\label{fig:lambda_again_hyperparameter}]{%
        \includegraphics[width=0.45\linewidth]{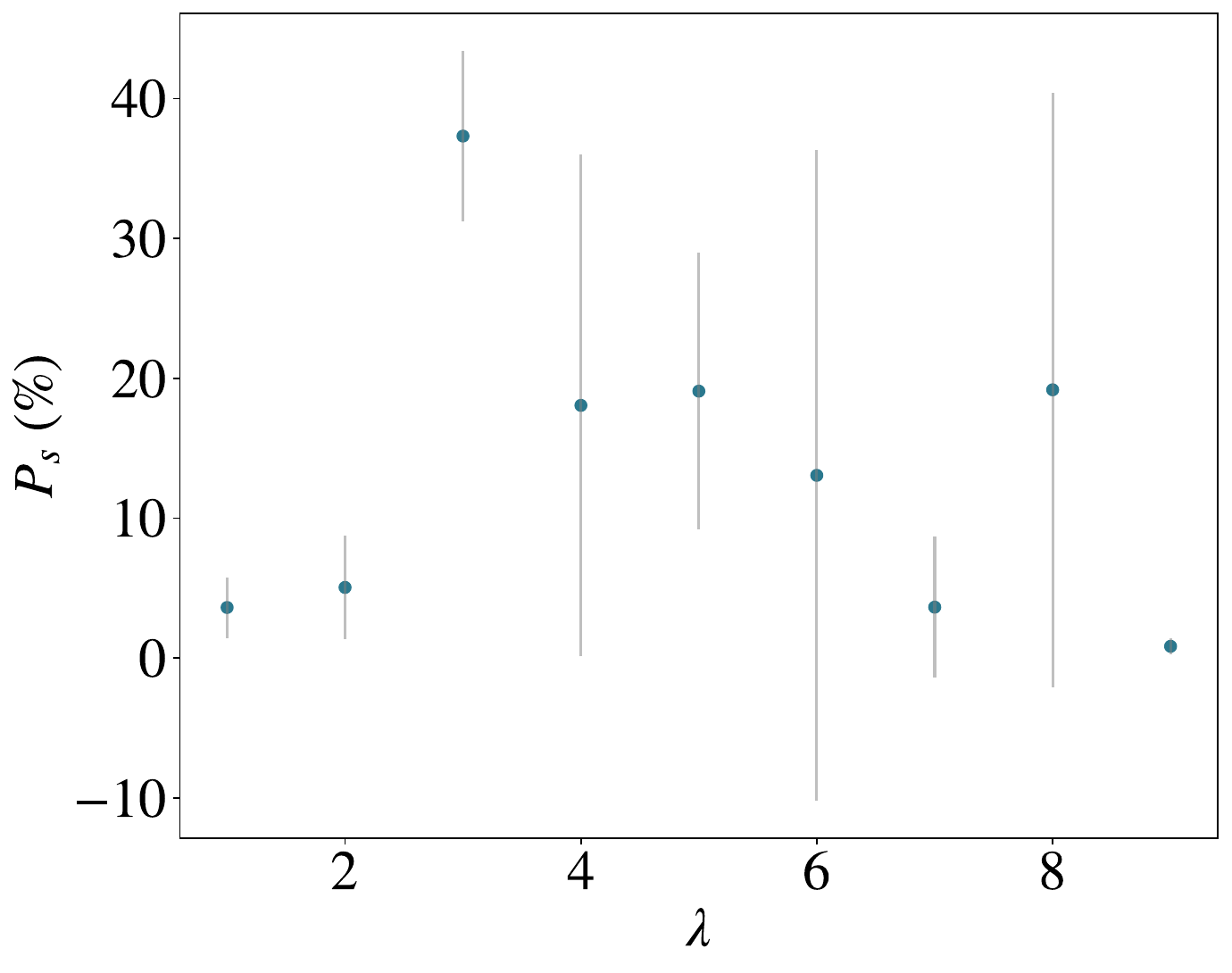}
    } \\
    \subfloat[\label{fig:CVaR_ansatz_l3}]{%
        \includegraphics[width=0.45\linewidth]{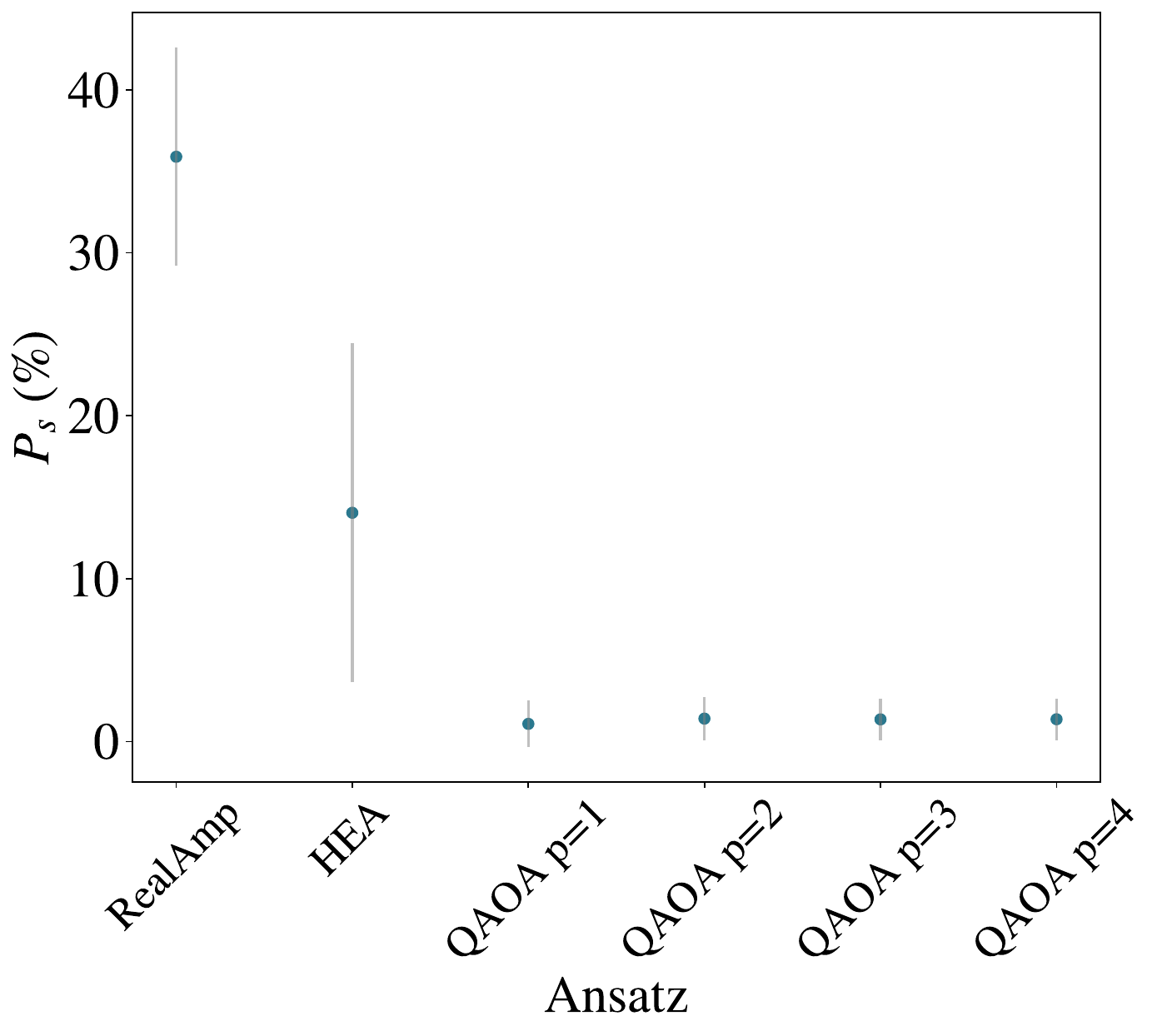}
    }
    \caption{The 18-variable problem hyperparameter search for the VQE state vector solver. (a) Optimal ansatz search with $\lambda=3$ (no CVaR). (b) Optimal $\lambda$ search using RealAmplitudes. (c) Optimal $\alpha$ CVaR parameter search with $\lambda=7$. (d) Optimal $\lambda$ search with CVaR $\alpha=0.4$. (e) Optimal ansatz search using $\alpha=0.4$ and $\lambda=3$. Mean values are shown for 5 repeats (10 for $\alpha$ plot) with standard deviation error bars.}
    \label{fig:VQE_hypop}
\end{figure*}

\begin{figure*}[htbp]
    \centering
    \subfloat[$\lambda = 1$\label{fig:nsup3_lambda1}]{%
        \includegraphics[width=0.45\linewidth]{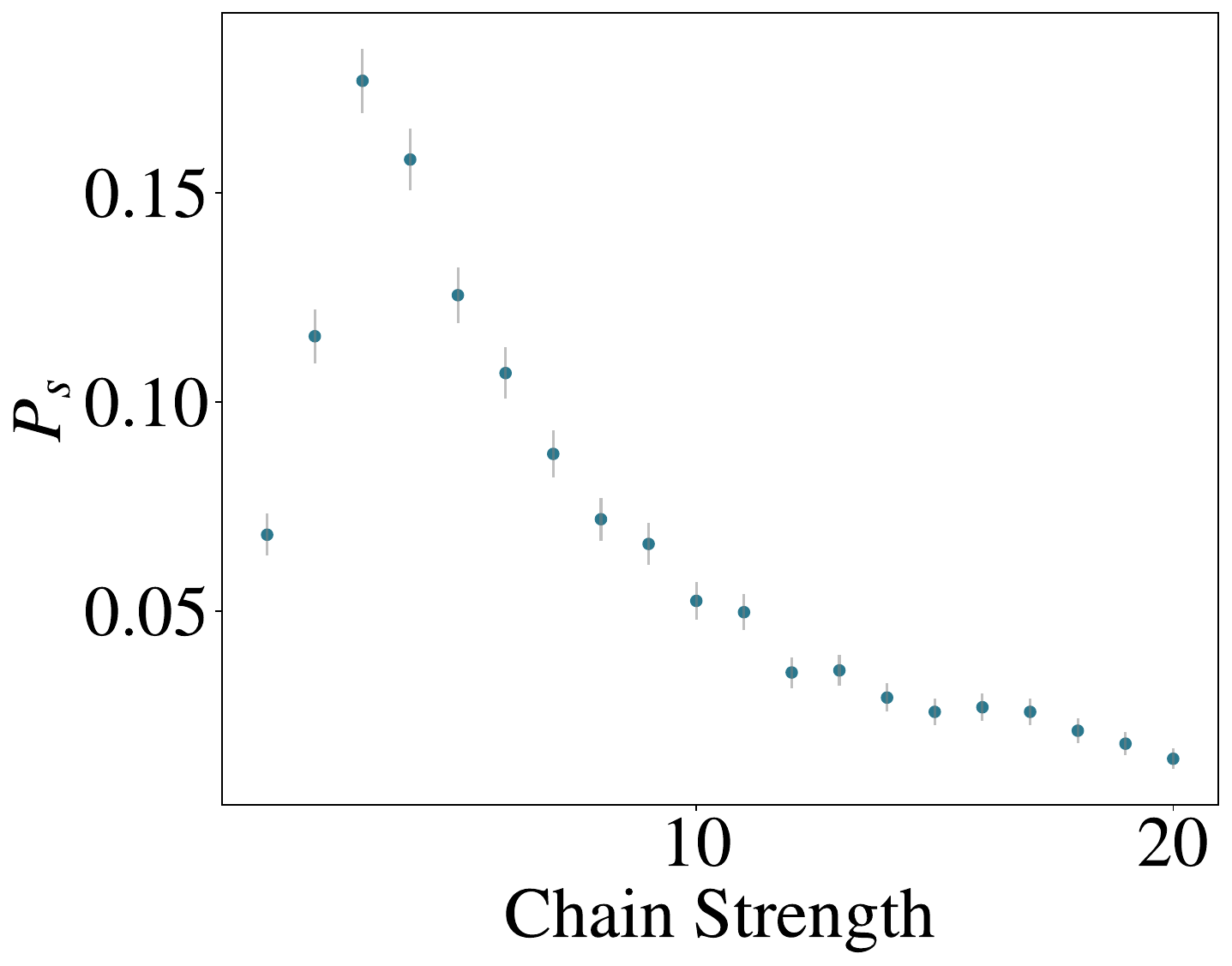}
    } \hfill
    \subfloat[$\lambda = 2$\label{fig:nsup3_lambda2}]{%
        \includegraphics[width=0.45\linewidth]{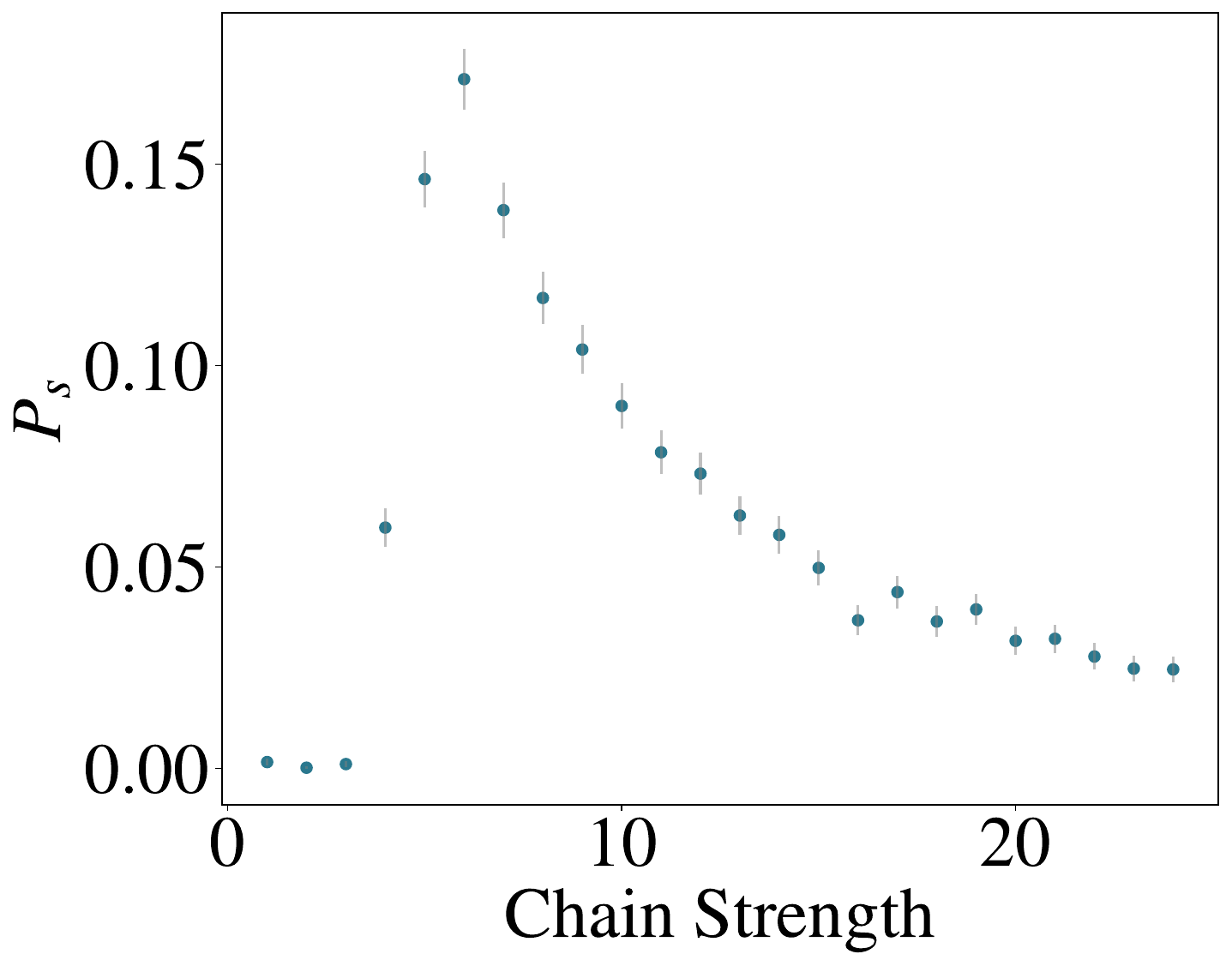}
    } \\
    \subfloat[$\lambda = 3$\label{fig:nsup3_lambda3}]{%
        \includegraphics[width=0.45\linewidth]{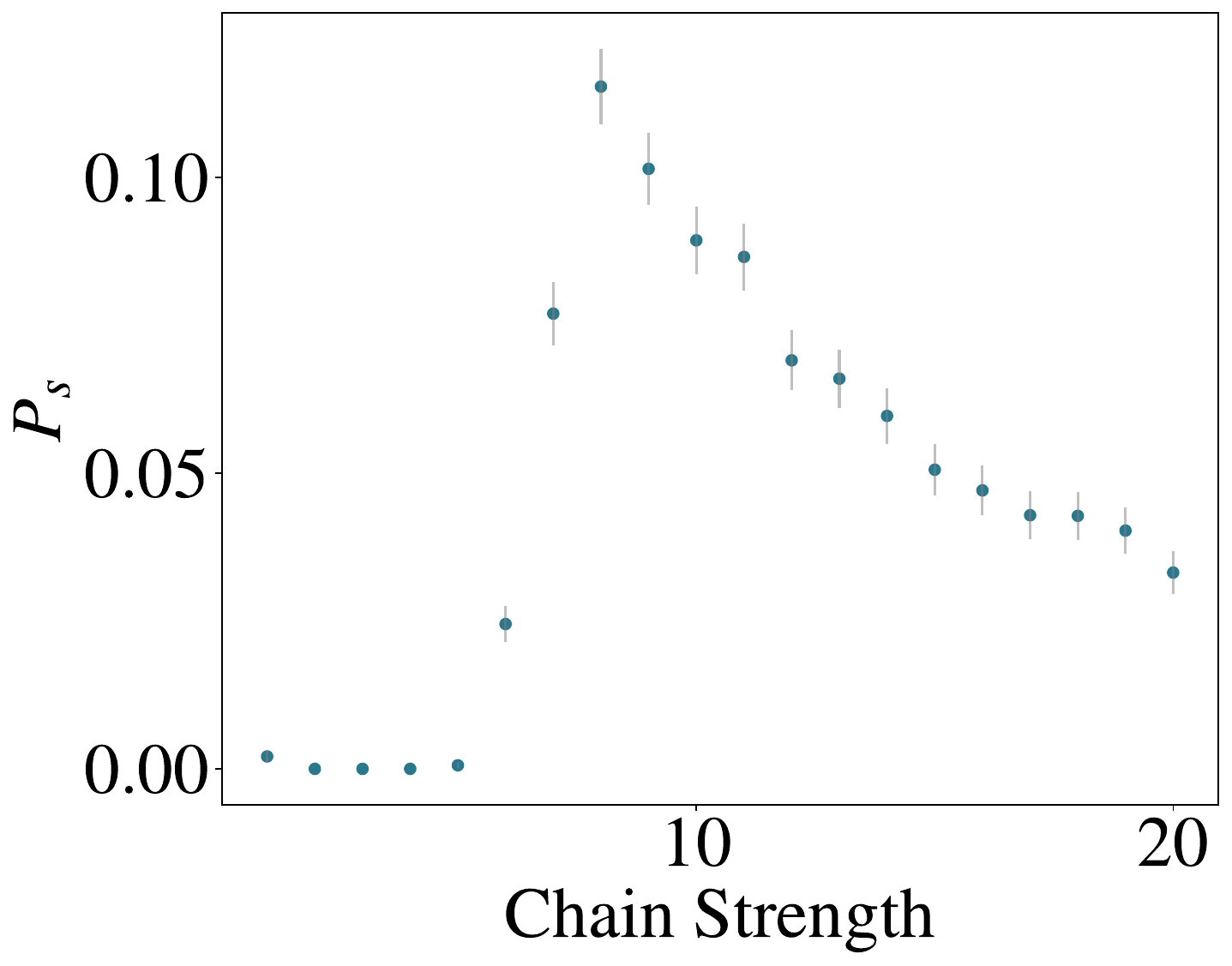}
    } \hfill
    \subfloat[\label{fig:nsup3_anneal_time}]{%
        \includegraphics[width=0.45\linewidth]{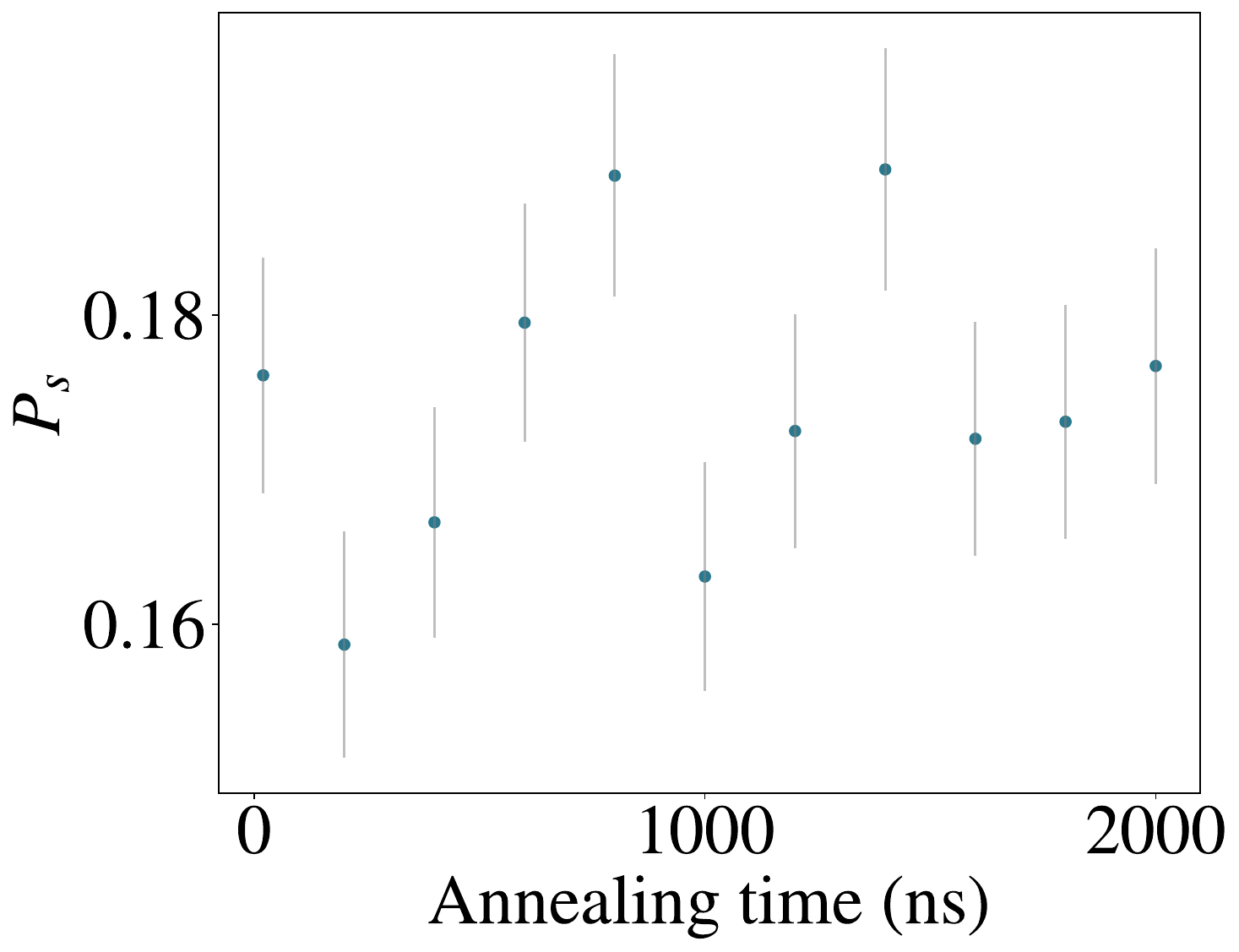}
    }
    \caption{The 18-variable problem hyperparameter search for quantum annealing using minor embedding. The optimal chain strength and $\lambda$ parameters were found by varying the chain strength for $\lambda=[1,3]$. (a)--(c) Chain strength optimization for different $\lambda$ values. (d) Annealing time search using a chain strength of 3 and $\lambda=1$.}
    \label{fig:nsup3_minor_embedding}
\end{figure*}

Our hyperparameter search finds the set of parameters for each algorithm which return the highest probability of sampling the ground state. The hyperparameter optimisation is a grid-based search, conducted by defining feasible search spaces and checking different combinations of values. Other, more rigorous strategies for hyperparameter optimisation exist, such as random search \cite{mcdowall2022finding} and Bayesian optimisation. Use of these methods could yield better hyperparameters. Results are plotted for each algorithm in Figures \ref{fig:SA_hyp}, \ref{fig:rand_hyp}, \ref{fig:VQE_hypop}, \ref{fig:nsup3_minor_embedding}. Hyperparameters for VQE on the QPU were informed by VQE on the state vector solver. Other schemes exist to find the optimal penalty terms. A more systematic approach is explored in Ref. \cite{alessandroni2023alleviating} for finding these terms.

\section{Technical Information and Additional Data for Simulated Annealing} \label{app:sim_annealing}

\begin{table*}[htb!]
\begin{ruledtabular}
\begin{tabular}{lccccc}
\textbf{Method} & \textbf{$P_s$} & \textbf{$P_s$ Post-Sel.} & \textbf{AR Post-Sel.} & \textbf{User Runtime (s)} & \textbf{QPU Time (s)} \\ 
\colrule
Brute Force & 1 & N/A & N/A & $2.2 \pm 0.03$ & N/A \\
Simulated Annealing & $0.993 \pm 0.003, 0.001^*$ & $0.994 \pm 0.003, 0.001^*$ & $0.994 \pm 0, 0.001^*$ & $0.339 \pm 0.001$ & N/A \\ 
VQE State Vector & $0.319 \pm 0.1, 0.001^*$ & $0.695 \pm 0.2, 0.001^*$ & $0.826 \pm 0.01, 0.001^*$ & $38.5 \pm 7$ & N/A \\ 
VQE QPU & $0.183 \pm 0.07, 0.002^*$ & $0.595 \pm 0.1, 0.002^*$ & $0.761 \pm 0.08, 0.002^*$ & $(2.54 \pm 0.4) \times 10^3$ & $941 \pm 200$ \\ 
Quantum Annealing & $0.189 \pm 0.02, 0.004^*$ & $0.398 \pm 0.03, 0.005^*$ & $0.626 \pm 0.03, 0.005^*$ & $2.54 \pm 0.08$ & $0.399 \pm 0.08$ \\ 
Random Sampling & $0.0004 \pm 0.0005, 0.0002^*$ & $0.104 \pm 0.1, 0.003^*$ & $0.235 \pm 0.2, 0.004^*$ & $0.001 \pm 0$ & N/A \\ 
\end{tabular}
\end{ruledtabular}
\caption{\label{tab:perform_withSE} The same table as Table \ref{tab:perform} in the main text with the standard error denoted by $^*$, which arises from shot noise. The standard deviation error is also included.}
\end{table*}

\begin{table*}[htbp]
\begin{ruledtabular}
\begin{tabular}{lccccc}
\textbf{Method} & \textbf{$P_s$} & \textbf{$P_s$ Post-Sel.} & \textbf{AR Post-Sel.} & \textbf{User Runtime (s)} & \textbf{QPU Time (s)} \\ 
\colrule
Brute Force & N/A & N/A & N/A & N/A & N/A \\
Simulated Annealing & $0.894 \pm 0.006$ & $0.903 \pm 0.005$ & $0.903 \pm 0$ & $1.09 \pm 0.004$ & N/A \\ 
VQE MPS & $0.008 \pm 0.01$ & $0.041 \pm 0.05$ & $0.172 \pm 0$ & $343 \pm 40$ & N/A \\ 
VQE QPU & $0.008$ & $0.038$ & $0.188$ & $4.14 \times 10^4$ & $1.88 \times 10^3$ \\ 
Quantum Annealing & $0.047 \pm 0.05$ & $0.094 \pm 0.09$ & $0.295 \pm 0.1$ & $5.42 \pm 1$ & $2.77 \pm 1$ \\ 
Random Sampling & $0 \pm 0$ & $0 \pm 0$ & $0 \pm 0$ & $0.001 \pm 0$ & N/A \\ 
\end{tabular}
\end{ruledtabular}
\caption{\label{tab:perform_32var} The performance metric results for our different methods on the 32 variable problem. The standard deviation error is included. Note: only 1 repeat with VQE on the QPU was possible and a brute force solution was not obtained as the search space is too large. The hyperparameters used for simulated annealing were: $\lambda=3$, $\beta=[0.1,10]$ with 1000 sweeps. Quantum annealing: $\lambda=1$, chain strength $=4$, annealing time $=1600$ ns. For VQE: $\lambda=3$, RealAmplitudes, CVaR $\alpha=0.4$, COBYLA with $\text{tol}=1$. For random sampling $\lambda=5$ and 1000 samples. The MPS solver was used for 32 qubit simulation.}
\end{table*}

D-wave's `SimulatedAnnealingSampler' was used, where all hyperparameters found from the search at the 18 variable problem size were also used for the larger problem sizes. For the rate of decrease of the temperature parameter, a default `geometric' annealing schedule was chosen, and a default number of solution updates or iterations (number of sweeps) was found from the hyperparameter search, which was $1000$ \cite{isakov2015optimised}. The optimal penalty term coefficient $\lambda$ for simulated annealing was found to be $\lambda=3$, which was found via hyperparameter optimisation.

\section{Technical Information and Additional Data for VQE} \label{app:VQE}

\begin{table*}[htbp]
\begin{ruledtabular}
\begin{tabular}{ccccc}
\textbf{QUBO Variables} & \textbf{$P_s$} & \textbf{$P_s$ Post-Selection} & \textbf{User Runtime (min)} & \textbf{QPU Time (min)} \\ \colrule
8  & 0.509            & 0.762           & 63.2           & 17.3           \\
18 & $0.183 \pm 0.067$ & $0.595 \pm 0.12$ & $42.3 \pm 6.5$ & $15.7 \pm 3.57$ \\
32 & 0.0075           & 0.038           & 151            & 31.3           \\
50 & 0.0007           & 0.016           & 137            & 42.4           \\
72 & 0                & 0               & 252            & 98.2           \\ 
\end{tabular}
\end{ruledtabular}
\caption{\label{tab:perform_VQE_larger} Performance metrics for VQE on the QPU at different problem sizes. Only one experiment was conducted for the 8, 32, 50, and 72 variable problem sizes. As seen in Figs.~\ref{fig:VQE_32} and \ref{fig:VQE_50}, the convergence criteria were not met for the larger problems before the maximum iterations. Errors included are the standard deviation from 5 repeats for the 18-variable case only.}
\end{table*}

\begin{figure}[htbp]
    \centering
    \subfloat[\label{fig:fez}]{%
        \includegraphics[width=0.4\linewidth]{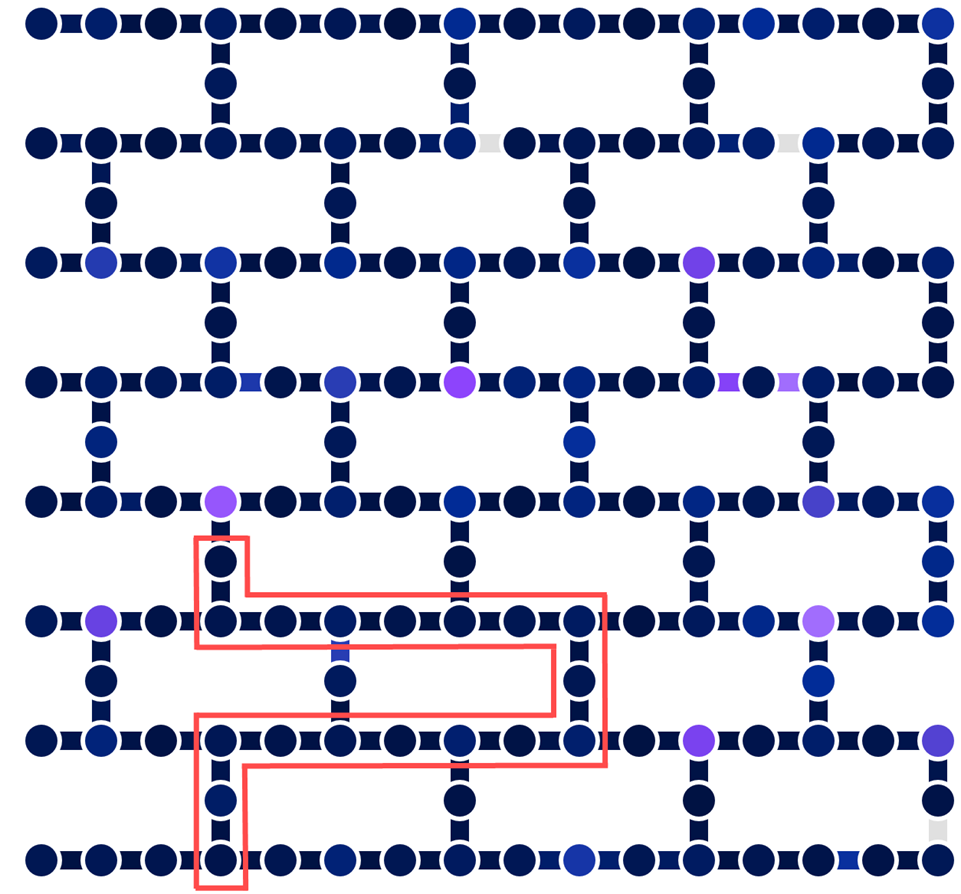}
    }
    \hspace{1cm} 
    \subfloat[\label{fig:d-waveQPU}]{%
        \includegraphics[width=0.4\linewidth]{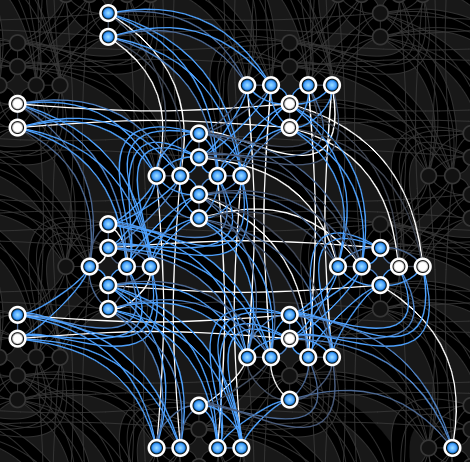}
    }
    \caption{(a) \textit{Ibm\_fez} qubit connectivity. Taken from the IBM Quantum platform. The qubits highlighted in red were used for one of the 18 variable QUBO runs, chosen for their low error rates. (b) D-Wave Advantage\_system 6.4 QPU connectivity and mapping for the problem, generated with the D-Wave problem inspector.}
    \label{fig:layouts}
\end{figure}

The number of shots was set to 10,000 for both the real device and for the state vector solver. To solve the QUBO problem using VQE, each variable is initially mapped to a single physical qubit. Qiskit's circuit transpilation is used when running on the real device, with the optimization\_level=3 \cite{IBMQuantumDocs}. The qubit connectivity plot for \textit{ibm\_fez} is shown in Figure \ref{fig:fez}, where the qubits that were used are highlighted in red. The qubits that were used were chosen based off the error rates of each qubit, which can vary with device calibration. $T_{\text{latency}}$ for VQE on the QPU should be minimal as qiskit serverless was used.

The classical optimiser used was COBYLA for the state vector solver and for the real device. 
For the 18 variable problem an optimiser tolerance of 1 was set as this was the value where successful termination was achieved within the number of specified iterations (250). The chosen tolerance and maximum number of iterations is dependent on the problem size and informed by hyperparameter search. 

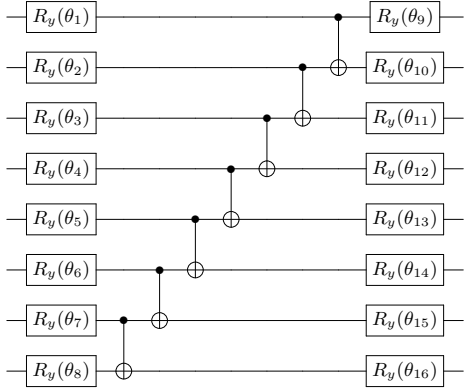
\begin{figure}[!h]
    \centerline{
    \scalebox{0.8}{
        \Qcircuit @C=1em @R=1em {
            & \gate{R_y(\theta_1)} & \qw & \qw & \qw & \qw & \qw & \qw & \ctrl{1} & \gate{R_y(\theta_9)} & \qw \\
            & \gate{R_y(\theta_2)} &\qw & \qw & \qw & \qw & \qw & \ctrl{1} &\targ & \gate{R_y(\theta_{10})} & \qw \\
            & \gate{R_y(\theta_3)} &\qw & \qw & \qw & \qw &\ctrl{1} &\targ &\qw & \gate{R_y(\theta_{11})} & \qw \\
            & \gate{R_y(\theta_4)} &\qw & \qw & \qw & \ctrl{1} &\targ &\qw  &\qw & \gate{R_y(\theta_{12})} & \qw \\
            & \gate{R_y(\theta_5)} &\qw & \qw & \ctrl{1} & \targ &\qw &\qw & \qw & \gate{R_y(\theta_{13})} & \qw \\
            & \gate{R_y(\theta_6)} &\qw & \ctrl{1} & \targ &\qw &\qw & \qw & \qw & \gate{R_y(\theta_{14})} & \qw \\
            & \gate{R_y(\theta_7)} &\ctrl{1} & \targ &\qw &\qw & \qw & \qw & \qw & \gate{R_y(\theta_{15})} & \qw \\
            & \gate{R_y(\theta_8)} &\targ&\qw & \qw & \qw & \qw   &\qw & \qw &\gate{R_y(\theta_{16})} & \qw \\
        }}
    }
    \caption{\small{The real amplitudes ansatz for 8 qubits. 18 qubits were used in Section \ref{sect:metrics} which corresponds to 32 parameters.}}
    \label{fig:ansatz}
\end{figure}

\begin{figure*}[h!]
\centering
\includegraphics[width=0.49\linewidth]{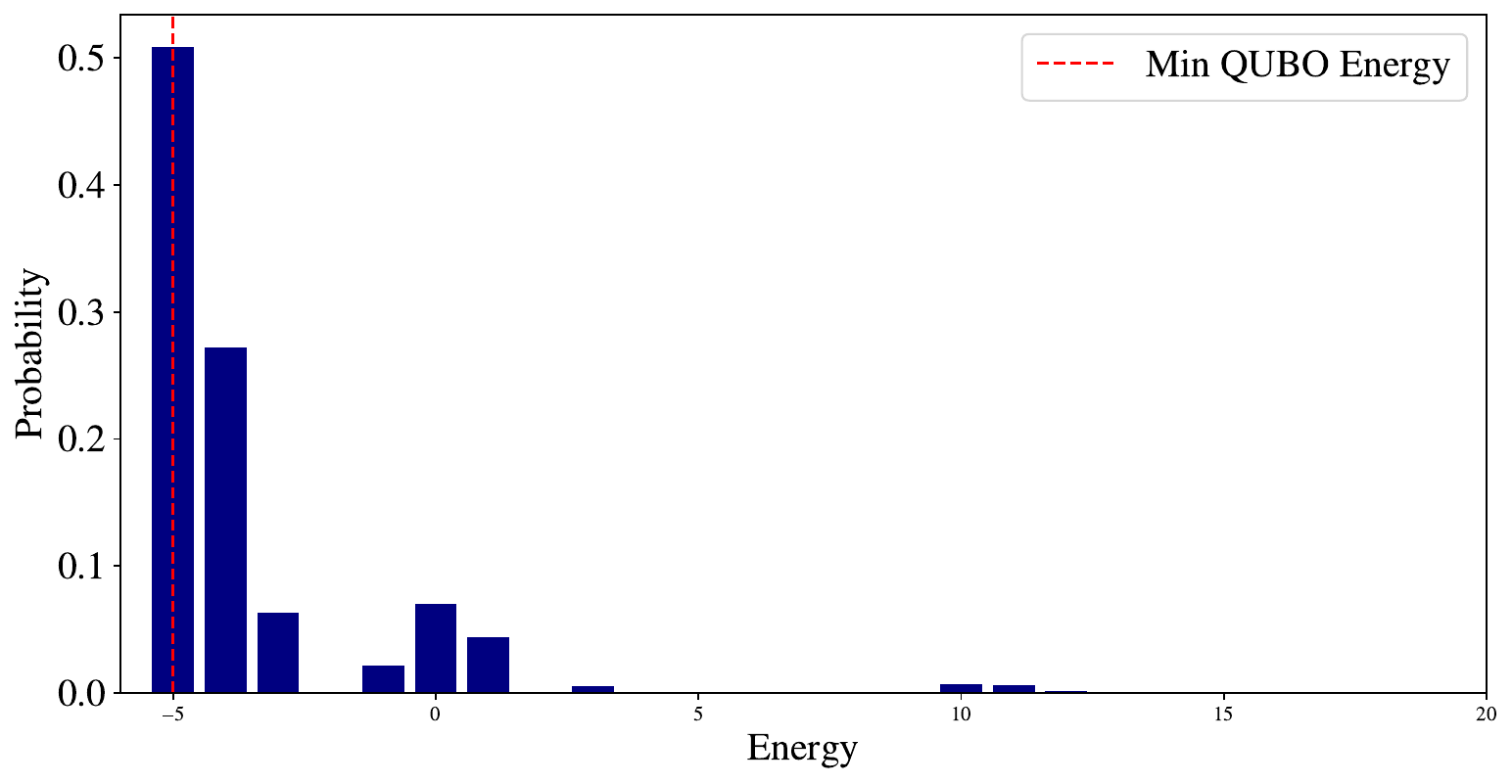}
\includegraphics[width=0.49\linewidth]{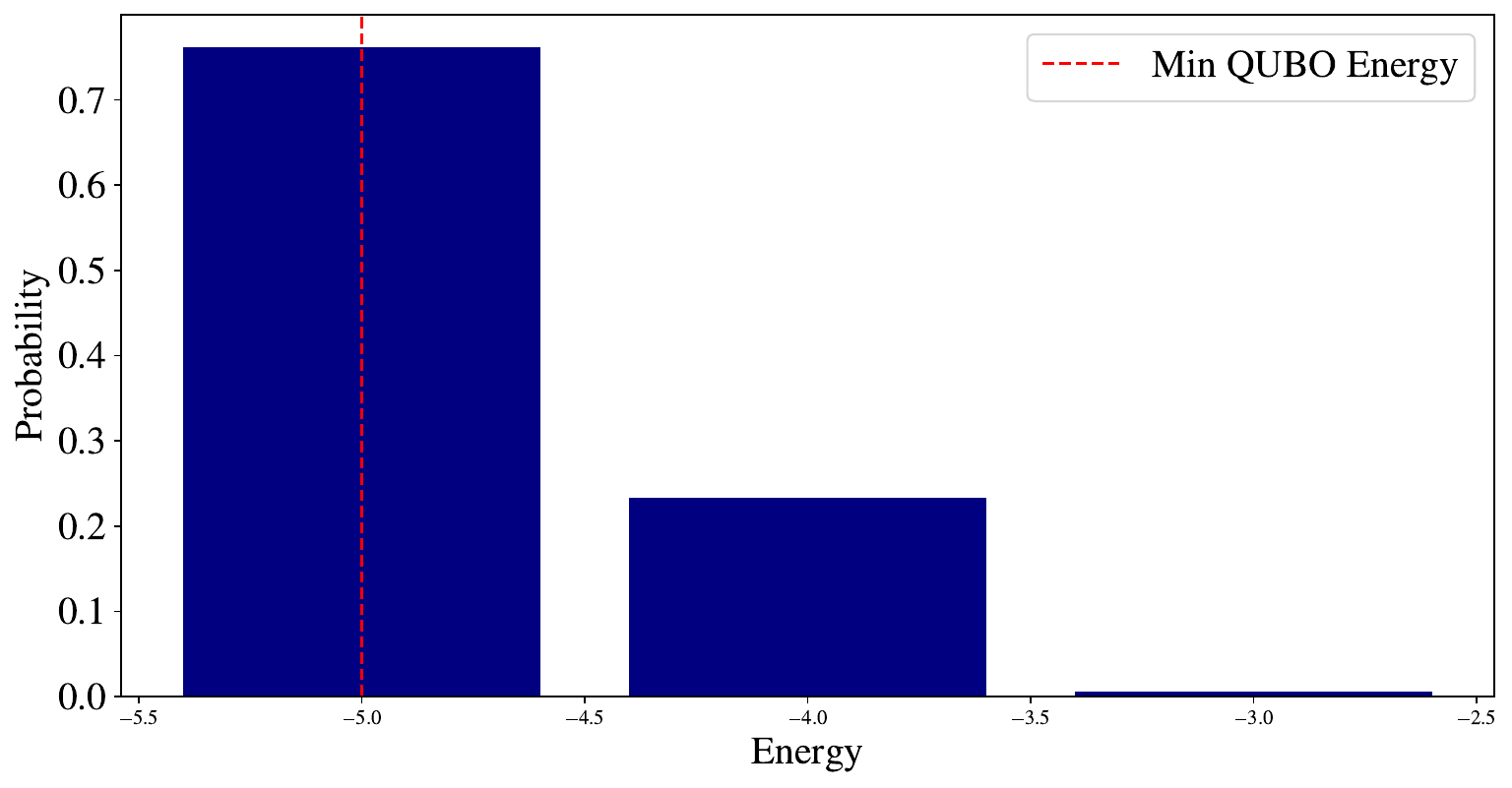}
\includegraphics[width=0.49\linewidth]{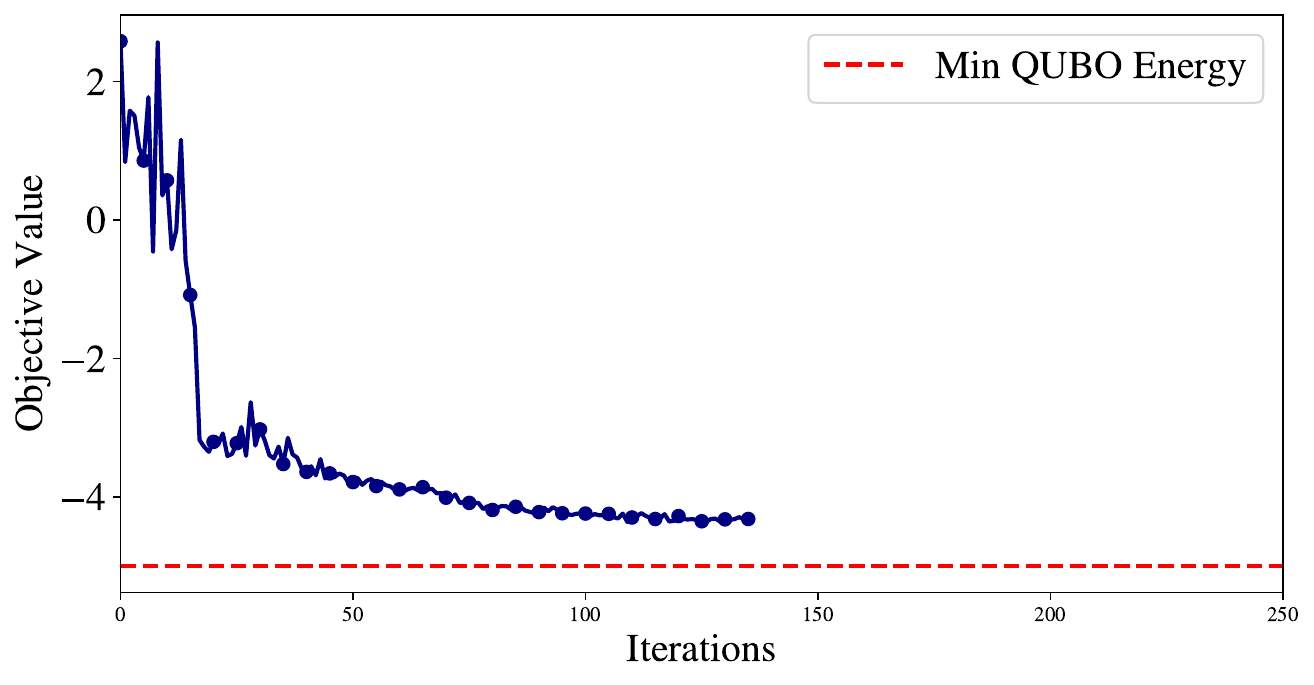}
\caption{\small{8 variable solved with VQE on the QPU. $\lambda = 3$, shots = 10,000, tol = 1e-1, RealAmplitudes ansatz, COBYLA, CVaR with \(\alpha=0.9\), maxiters = 250.}} 
\label{fig:VQE_8}
\end{figure*}

\begin{figure*}[h!]
\centering
\includegraphics[width=0.49\linewidth]{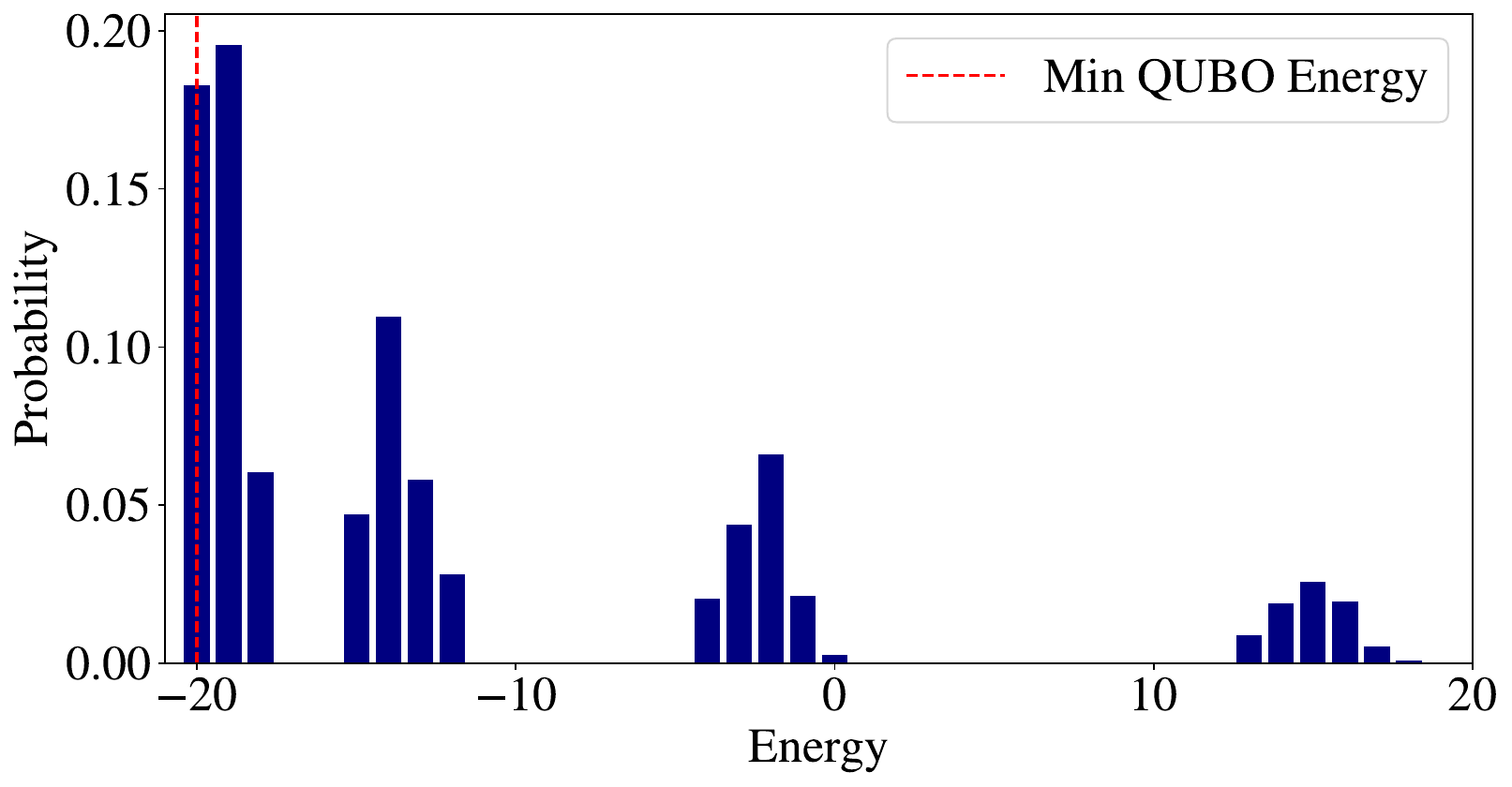}
\includegraphics[width=0.49\linewidth]{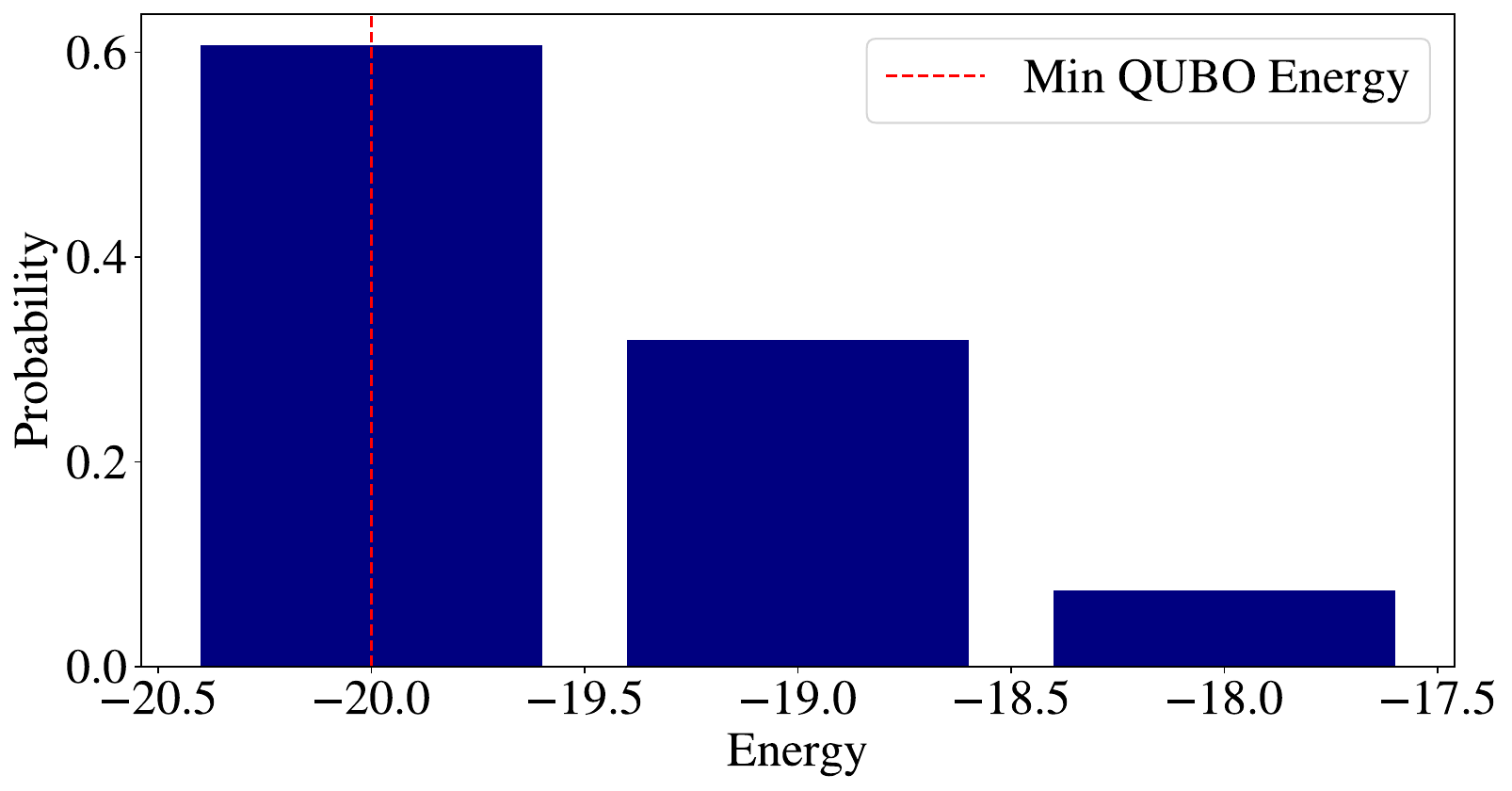}
\includegraphics[width=0.49\linewidth]{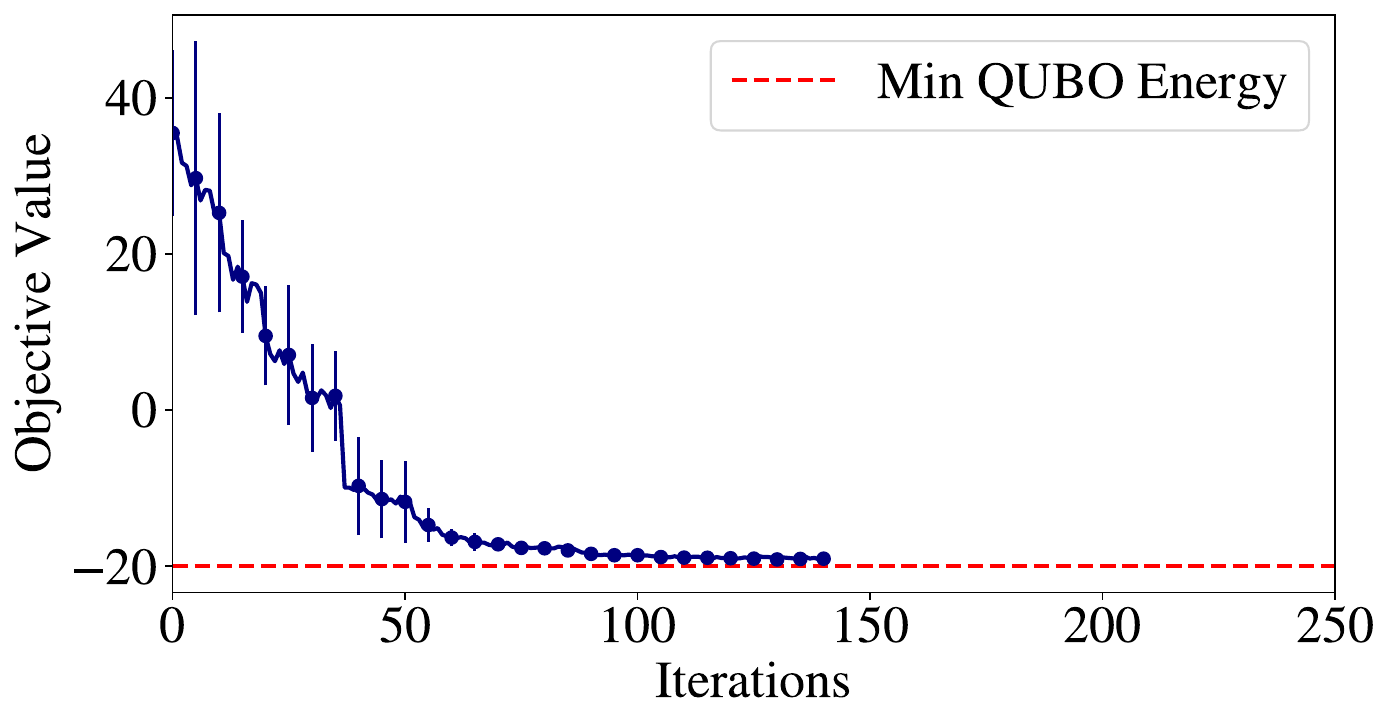}

\caption{\small{18 variable solved with VQE on the QPU. \(\lambda = 3\), shots = 10,000, tol = 1, RealAmplitudes ansatz, COBYLA, CVaR with \(\alpha=0.4\), maxiters = 250. 5 experiments are used for accumulated, renormalised distributions. The convergence plot is truncated at the minimum number of iterations required over the 5 experiments and the error bars are $\sigma$.}} 
\label{fig:VQE_18}
\end{figure*}

\begin{figure*}[h!]
\centering
    \includegraphics[width=0.49\linewidth]{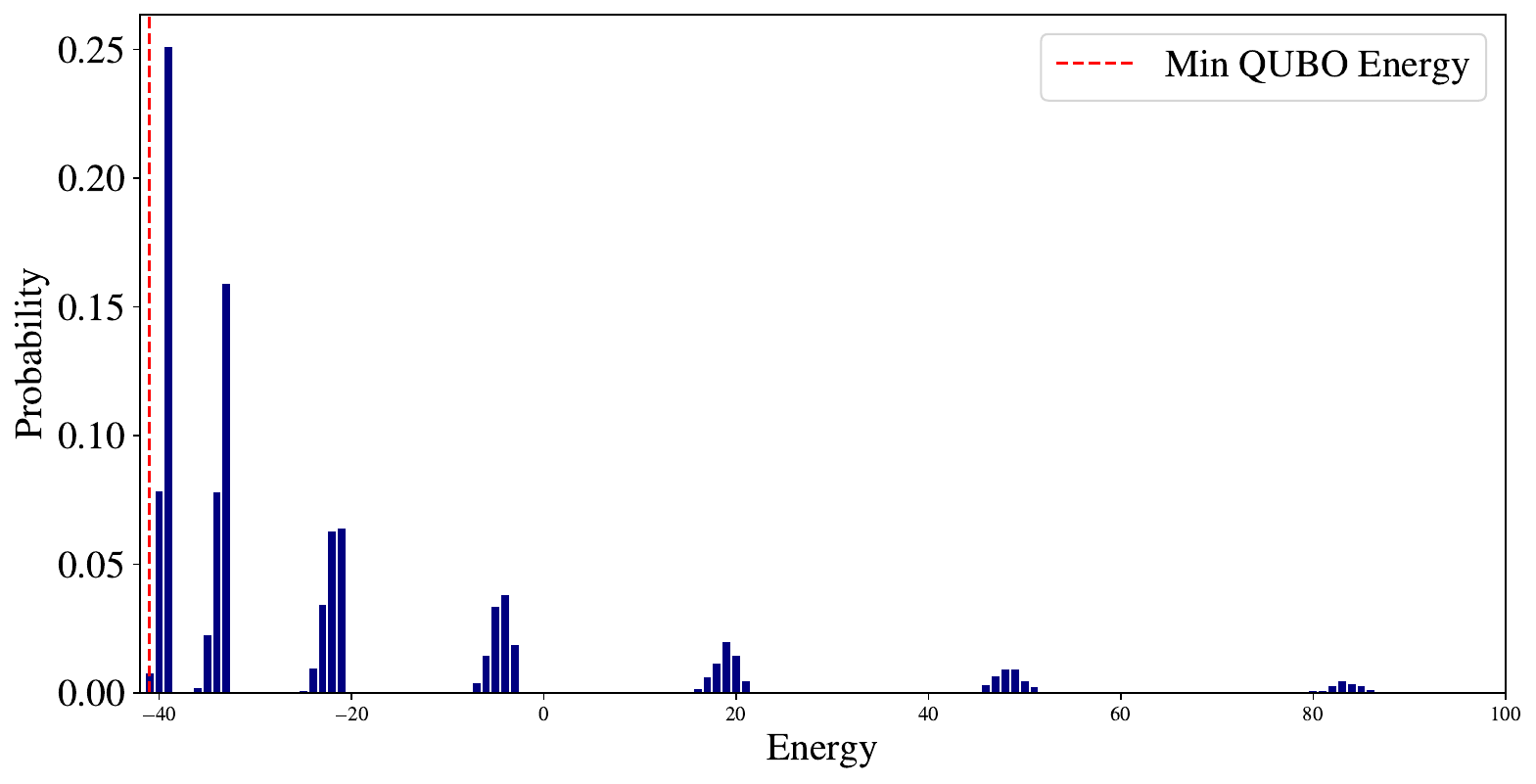}
    \includegraphics[width=0.49\linewidth]{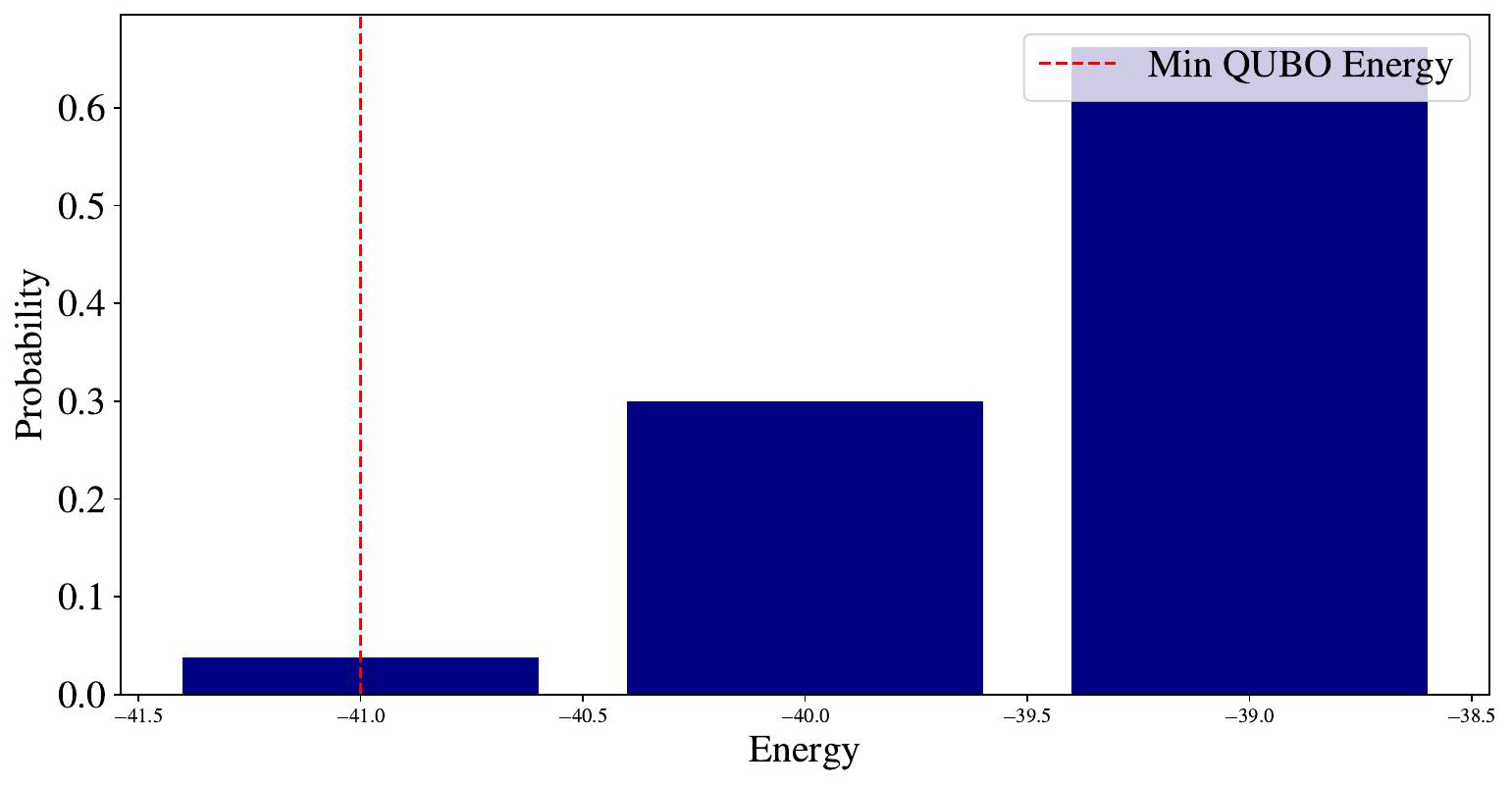}
    \includegraphics[width=0.49\linewidth]{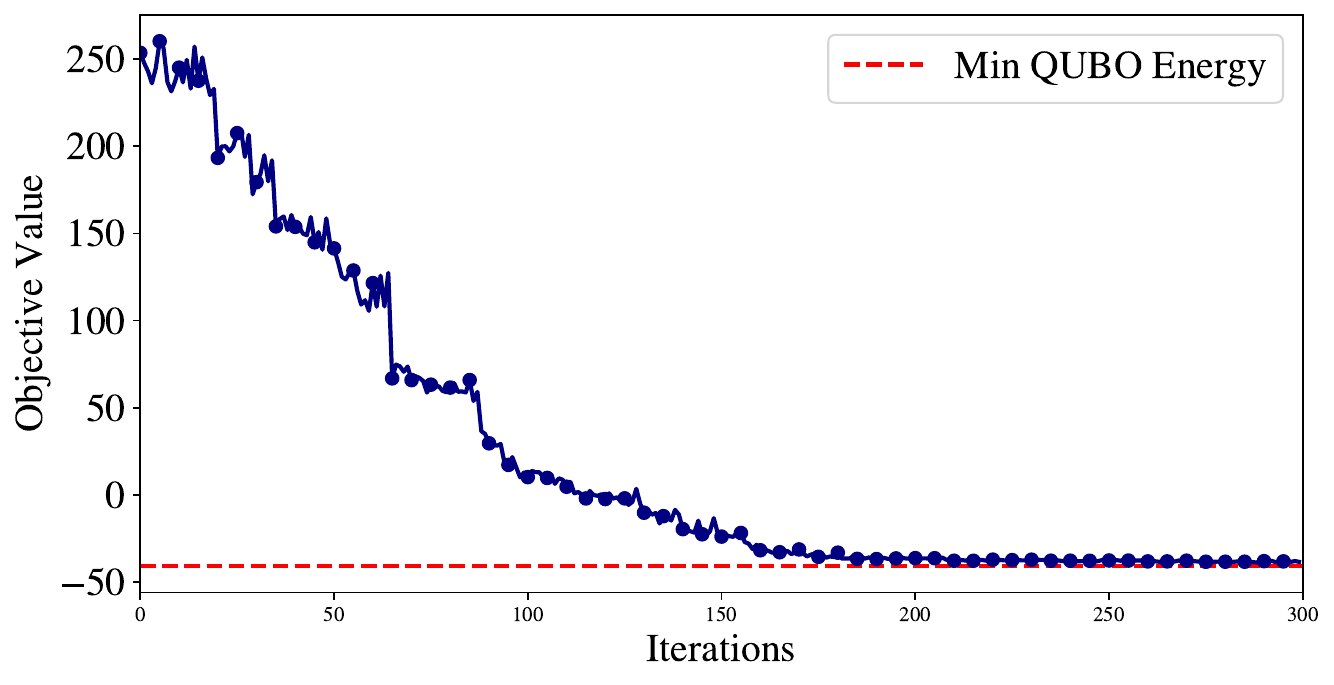}
\caption{\small{32 variable solved with VQE on the QPU. $\lambda=3$ shots = 10,000, tol = 1, RealAmplitudes ansatz, COBYLA, CVaR with \(\alpha=0.4\), maxiters = 300.}} 
\label{fig:VQE_32}
\end{figure*}

\begin{figure*}[h!]
\centering
    \includegraphics[width=0.49\linewidth]{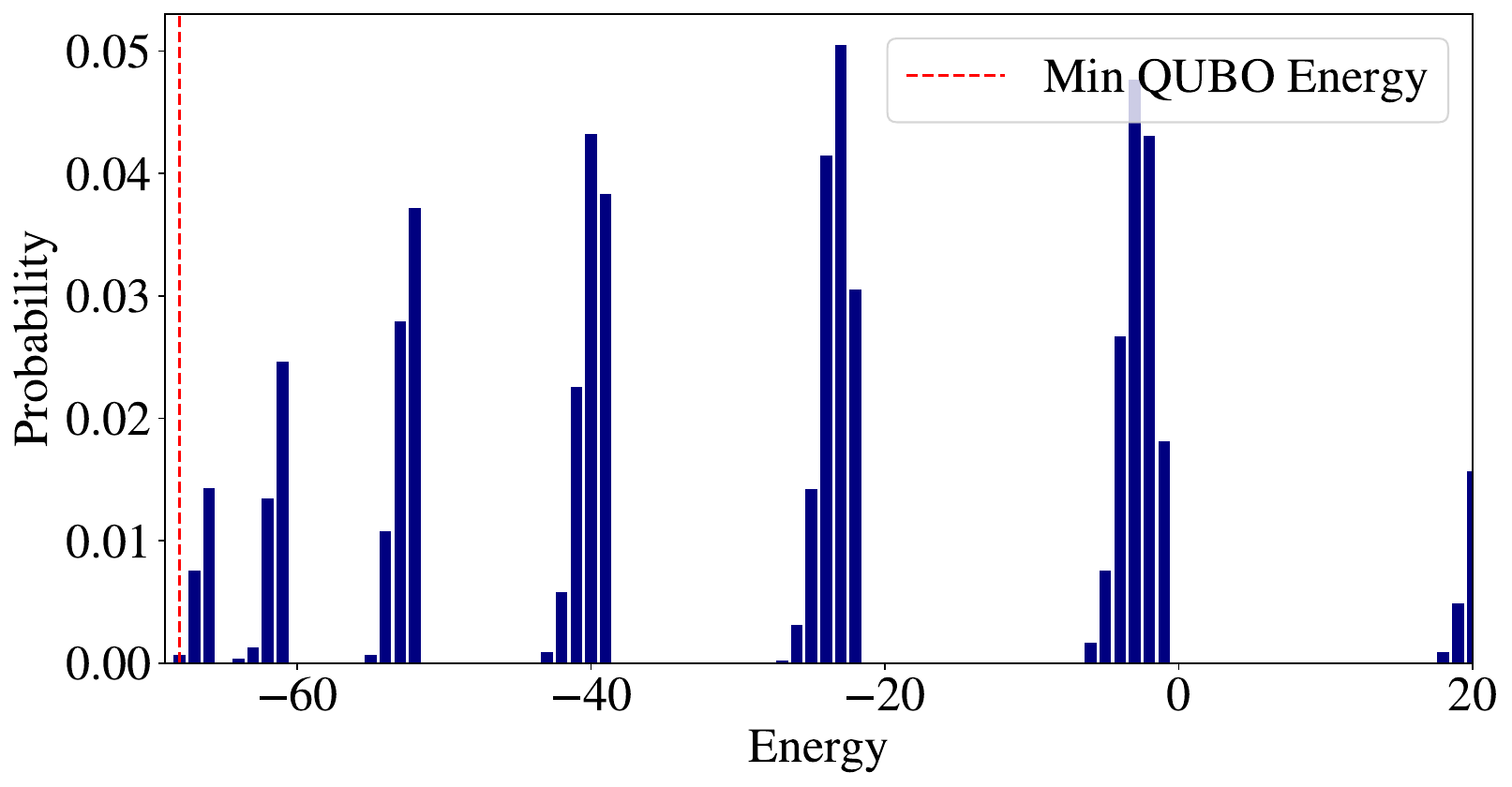}
    \includegraphics[width=0.49\linewidth]{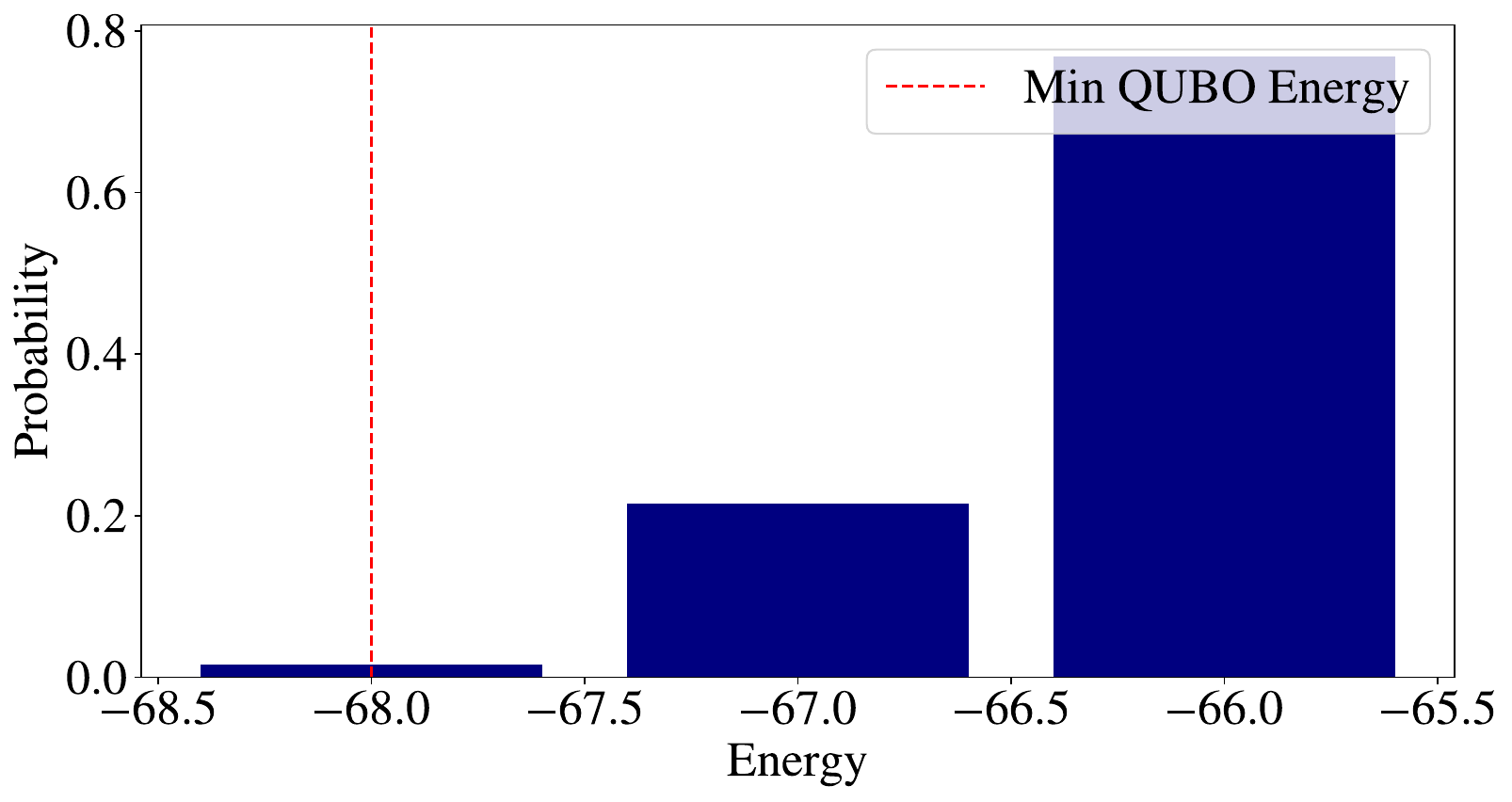}
    \includegraphics[width=0.49\linewidth]{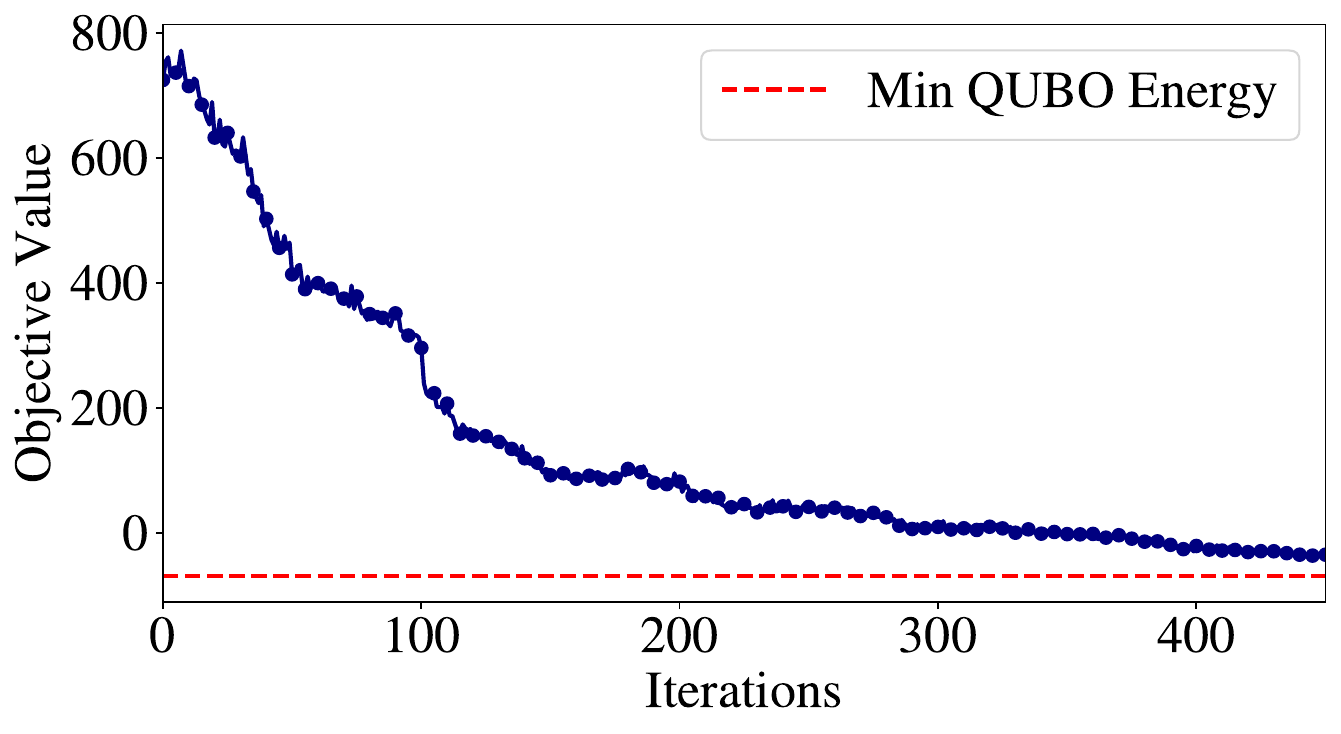}
\caption{\small{50 variable solved with VQE on the QPU. $\lambda=2$, shots = 10,000, tol = 1, RealAmplitudes ansatz, COBYLA, CVaR with \(\alpha=0.4\), maxiters = 450.}} 
\label{fig:VQE_50}
\end{figure*}

\begin{figure*}[h!]
\centering
    \includegraphics[width=0.49\linewidth]{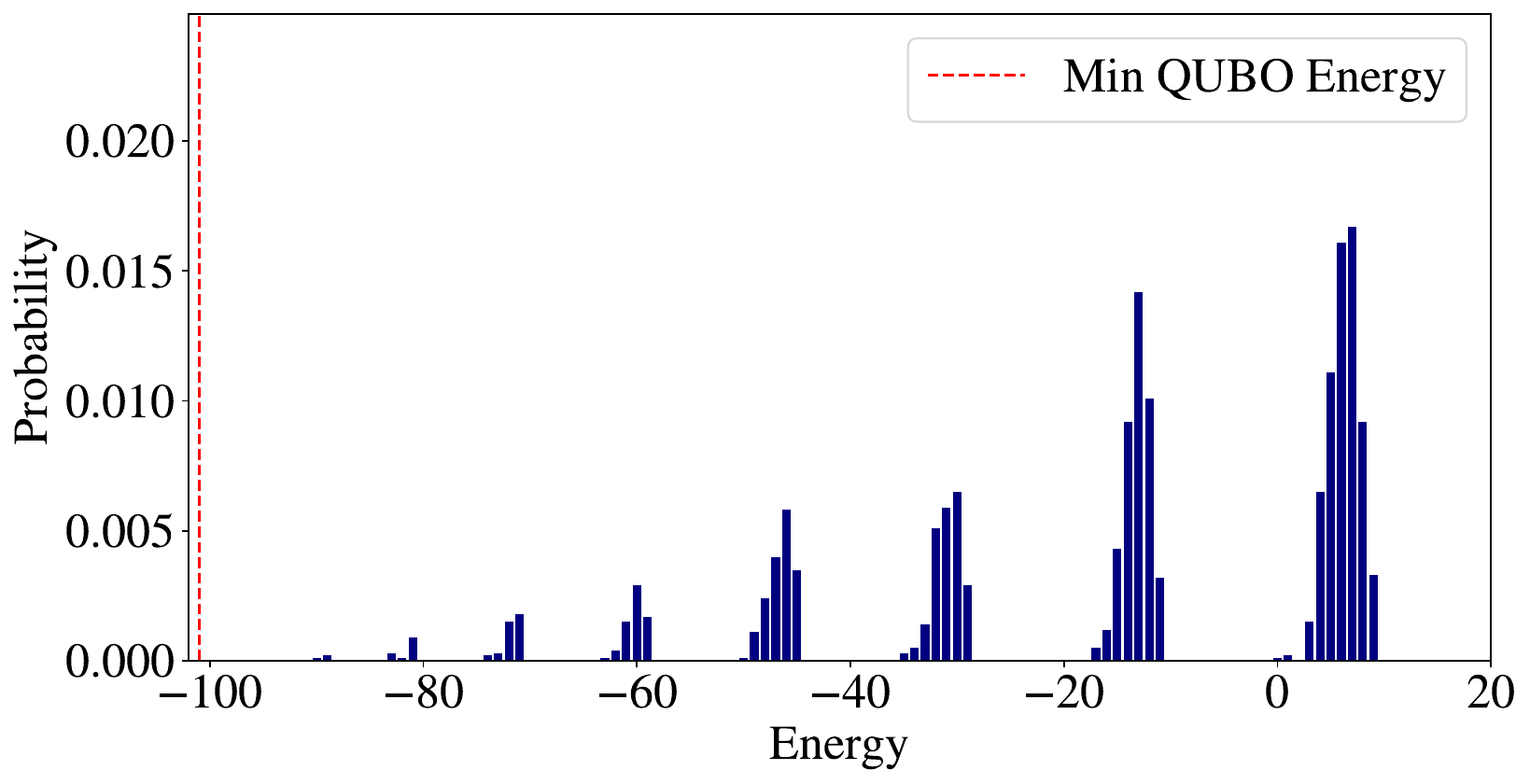}
    \includegraphics[width=0.49\linewidth]{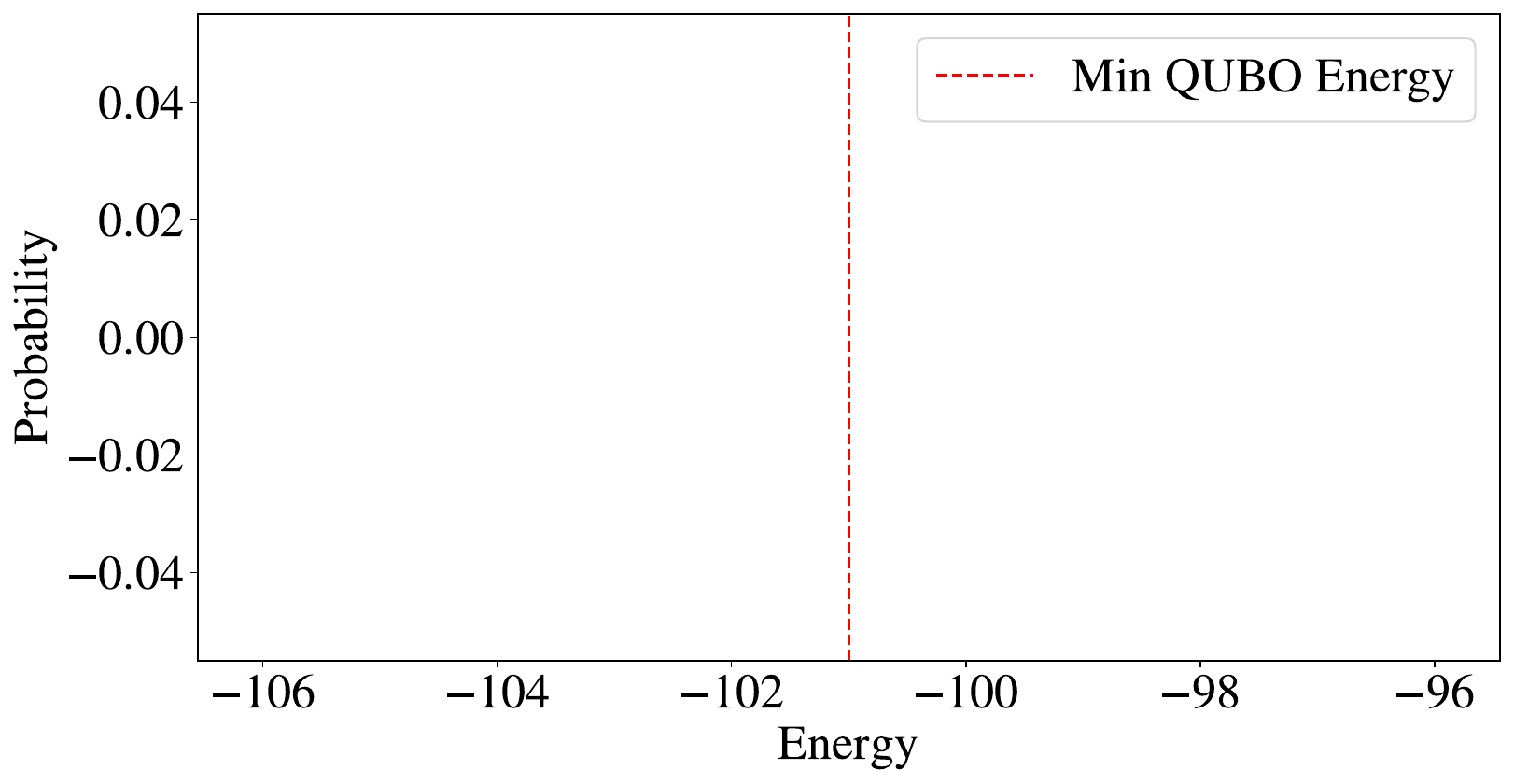}
    \includegraphics[width=0.49\linewidth]{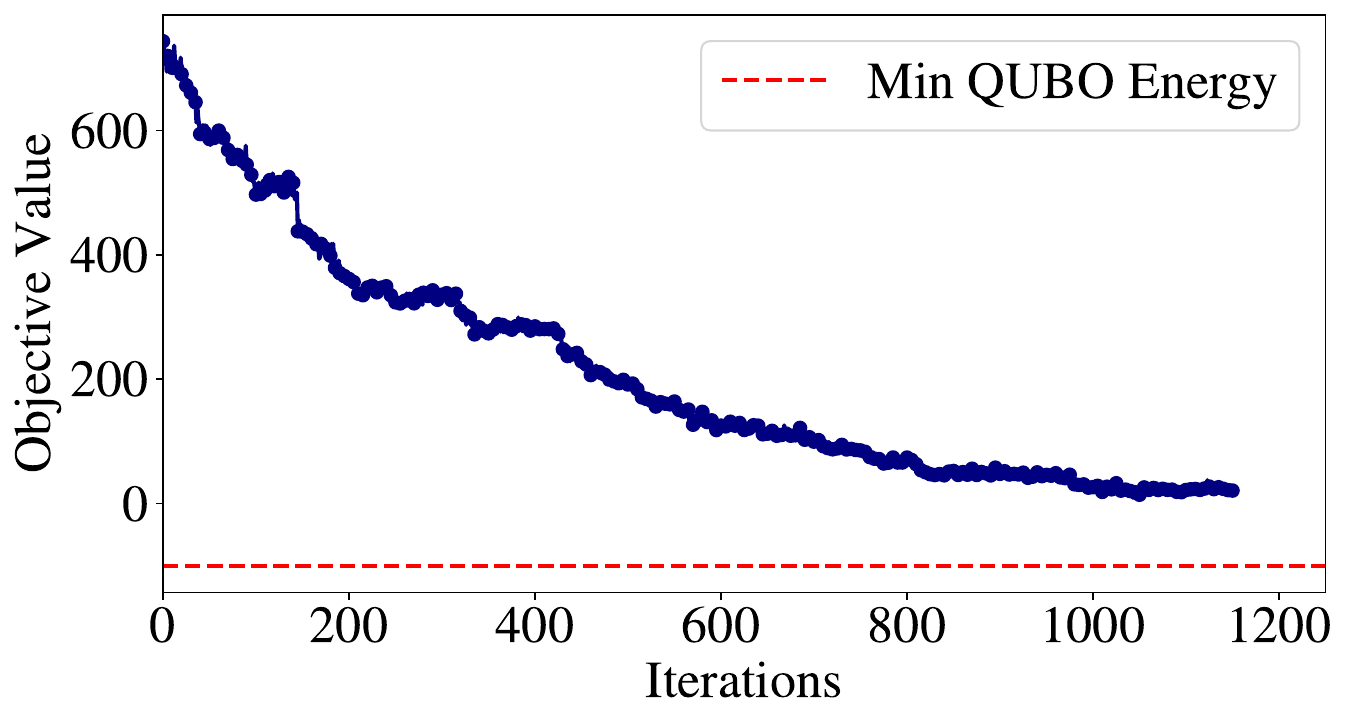}
\caption{\small{72 variable solved with VQE on the QPU. $\lambda=1$, shots = 10,000, tol = 1, RealAmplitudes ansatz, COBYLA, CVaR with \(\alpha=0.4\), maxiters = 1300.}} 
\label{fig:VQE_72}
\end{figure*}

The circuit ansatz used was Qiskit's `RealAmplitudes' \cite{RealAmpl17:online}, shown in Figure \ref{fig:ansatz}, which is commonly used for chemistry applications or classification circuits in machine learning. It is particularly suited for problems with real-valued solutions (since the ansatz only generates real amplitudes, no complex amplitudes). The circuit consists of alternating layers of Y rotations and CX entanglements. For this study, only one alternating layer (repetition) was used, along with the default entanglement pattern `reverse\_linear'. Measurements are made in the computational basis (single qubit Pauli-Z basis), such that a 0 or 1 is measured depending on the overlap with the Pauli-Z eigenstates $\ket{0}$ and $\ket{1}$. For each problem size solved with VQE the distribution, post-selected distribution and convergence plots are shown in Figures \ref{fig:VQE_8}, \ref{fig:VQE_18}, \ref{fig:VQE_32}, \ref{fig:VQE_50} and \ref{fig:VQE_72}.

\subsection{RealAmplitudes vs QAOA Ansatz (Hamiltonian agnostic vs informed)} \label{app:VQE:QAOA}

QAOA is popularly used for optimisation problems as it is Hamiltonian informed. The form of the QAOA circuit ansatz is:

\begin{equation}
    |\psi(\boldsymbol{\beta}, \boldsymbol{\gamma})\rangle = \prod_{k=1}^{p} \left( e^{-i\beta_k H_M} e^{-i\gamma_k H_C} \right) |+\rangle^{\otimes n}
\end{equation}

where, $p$ is the number of layers. $\boldsymbol{\beta} = (\beta_1, \beta_2, \dots, \beta_p)$ and $\boldsymbol{\gamma} = (\gamma_1, \gamma_2, \dots, \gamma_p)$ are variational parameters. $H_C$ is the problem Hamiltonian, which encodes the cost function of the optimisation problem and makes it Hamiltonian informed. $H_M$ is the mixer Hamiltonian, typically chosen as $H_M = \sum_{i=1}^n X_i$, where $X_i$ is the Pauli-X operator acting on the $i$-th qubit. $|+\rangle^{\otimes n}$ is the initial state, which is an equal superposition of all computational basis states \cite{farhi2014quantum}. 

When comparing different ansatz in our hyperparameter search we found that the RealAmplitudes ansatz returned higher values of $P_s$ (results in Figure \ref{fig:CVaR_ansatz_l3}). The simple explanation for this is that the RealAmplitudes ansatz has more parameters and is therefore more expressive allowing it to find a state with good overlap with the ground state. But what happens if we increase the depth of QAOA so that it has the same number of parameters as the RealAmplitudes ansatz? At $p=2$, for the $18$ variable problem size, QAOA has just $4$ independent parameters, but already has a depth of $58$ gates, whereas the RealAmplitudes depth is $20$ gates (these are the abstract circuits before transpilation). This is due to the fact that our problem Hamiltonian $H_C$, is fully connected. Consequently, the QAOA circuit has full connectivity (a fully connected entanglement arrangement), implementation of which on the IBM QPU is possible only with the introduction of SWAP gates. Running and simulating these QAOA circuits at $p=2$ can already take a long amount of time which makes increasing $p$ quite infeasible.


However, Ref. \cite{farhi2022quantum} shows QAOA can have good performance at $p=12$ on a fully connected SK-problem. This is possible due to the concentration of parameters found with QAOA, where parameters at smaller problem sizes can be extrapolated to larger problem sizes. The power of QAOA is that it can perform well with a relatively small number of parameters at low depth which should allow scaling to larger problem sizes. Future work could further explore the practical implementation of QAOA, using concentration of parameters to scale to large, dense QUBOs.


\section{Technical Information and Additional Data for Quantum Annealing} \label{app:annealing}

\begin{figure}[h!]
    \centering
    \includegraphics[width=0.4\textwidth]{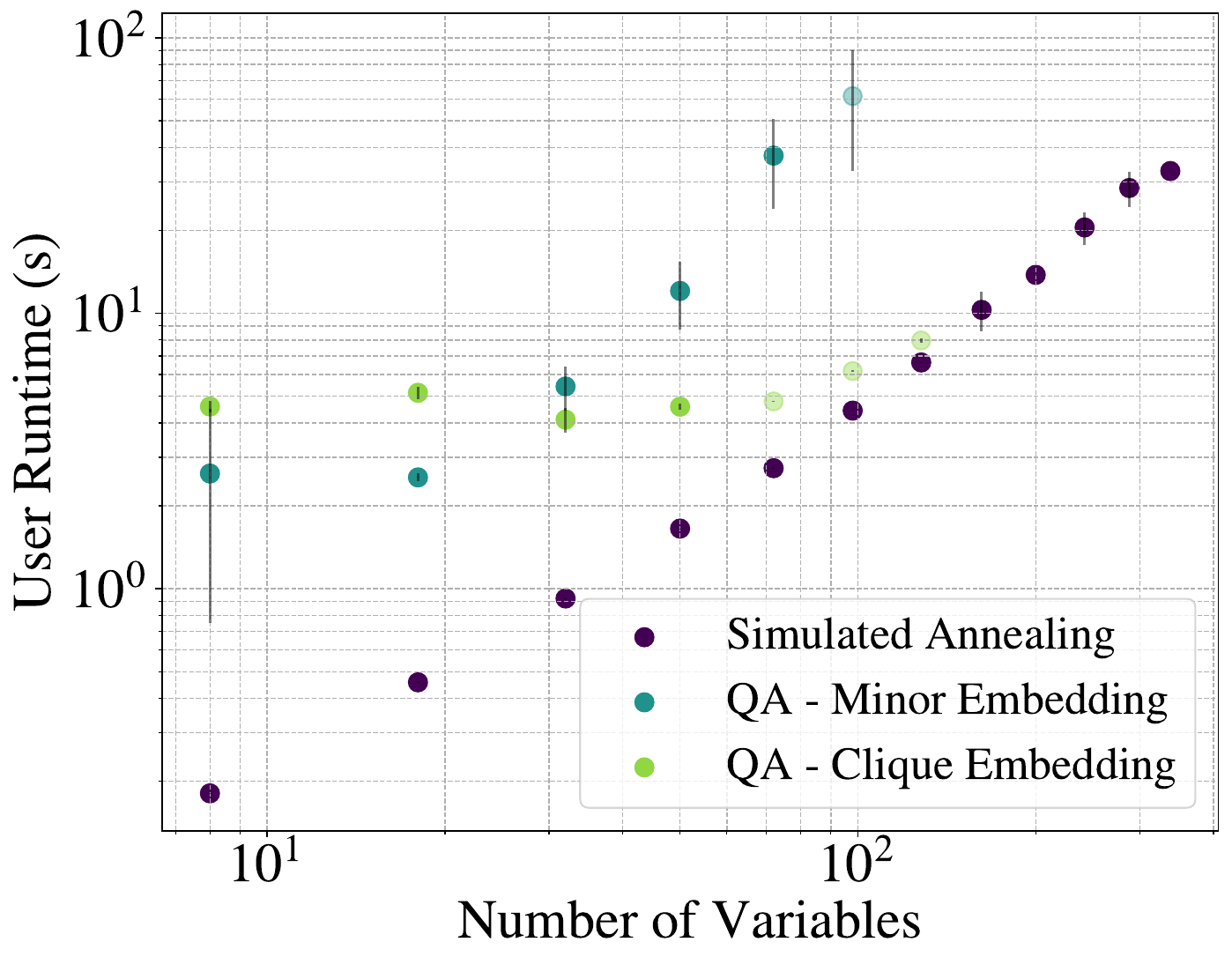}
  \caption{\small{Simulated and quantum annealing user runtime when solving the three vacancy QUBO problem at different sizes up to 338 variables. The same data as in Figure \ref{fig:scalingTime_combined} with logarithmic axis scales. The results were repeated 10 times, with average values used, and the standard deviation included as error bars. The hyperparameters used for each method and each problem size are detailed in Appendices \ref{app:sim_annealing} and \ref{app:annealing}.}}
  \label{fig:scalingTime_log_combined}
\end{figure}

To run a QUBO problem on the quantum annealer, each variable must be encoded to the hardware. Given the limited connectivity of the hardware, multiple physical qubits need to be mapped to a single variable. The mapping procedures used were minor embedding, D-Wave's default mapping scheme, and clique embedding, intended for fully connected problem graphs. The minor embedding mapping to the D-Wave Advantage\_system 6.4 QPU is shown in Figure \ref{fig:d-waveQPU}. The time taken for this embedding procedure on the 18 variable problem was found to be $(0.37\pm0.07)\unit{s}$. The percentage of broken chains for minor embedding on the 18 variable QUBO was $(2.6\pm2.2)$\%. 

\begin{figure*}[htbp]
    \centering
    \subfloat[\label{fig:emb_time}]{%
        \includegraphics[width=0.45\linewidth]{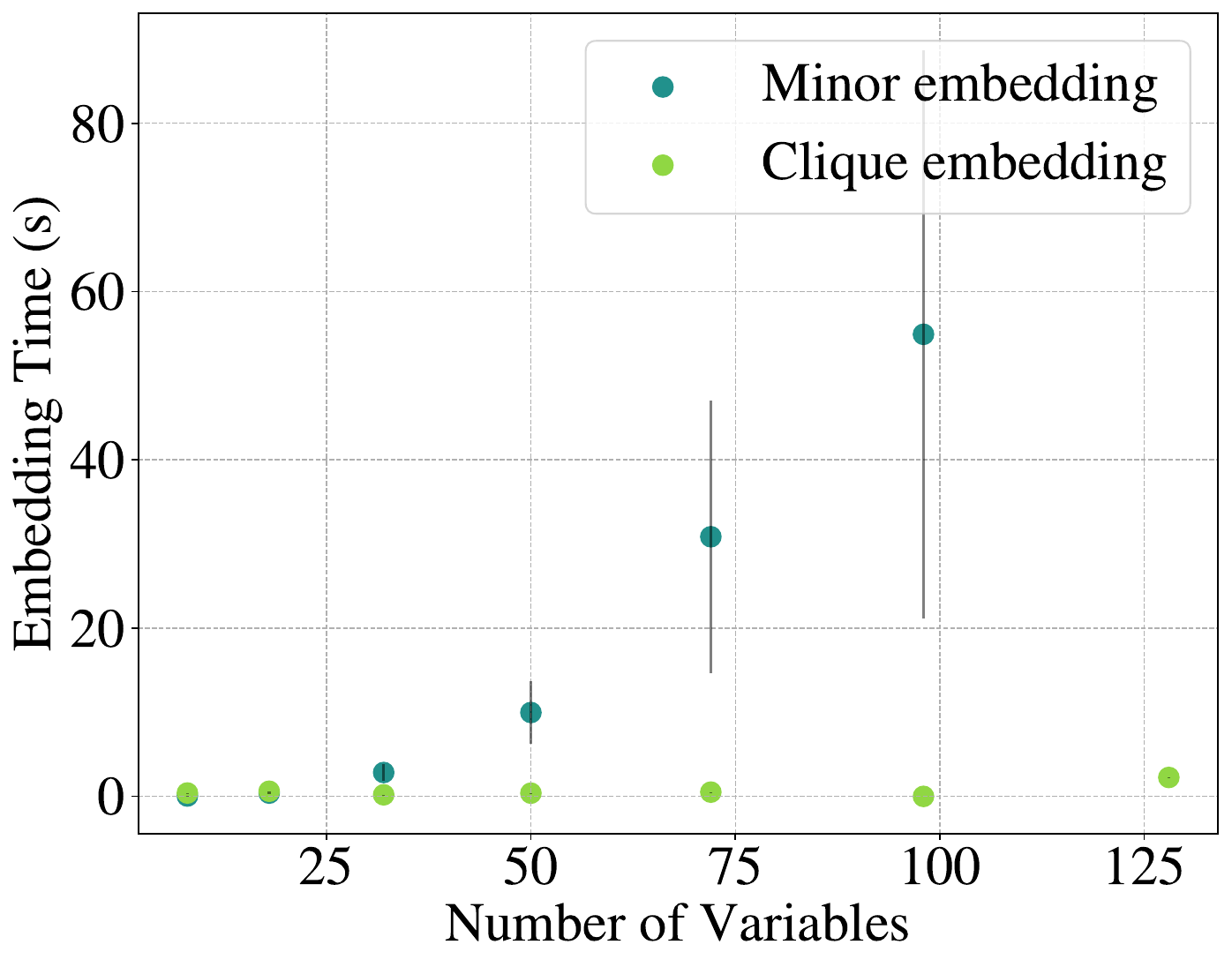}
    } \hfill
    \subfloat[\label{fig:qpu_time}]{%
        \includegraphics[width=0.45\linewidth]{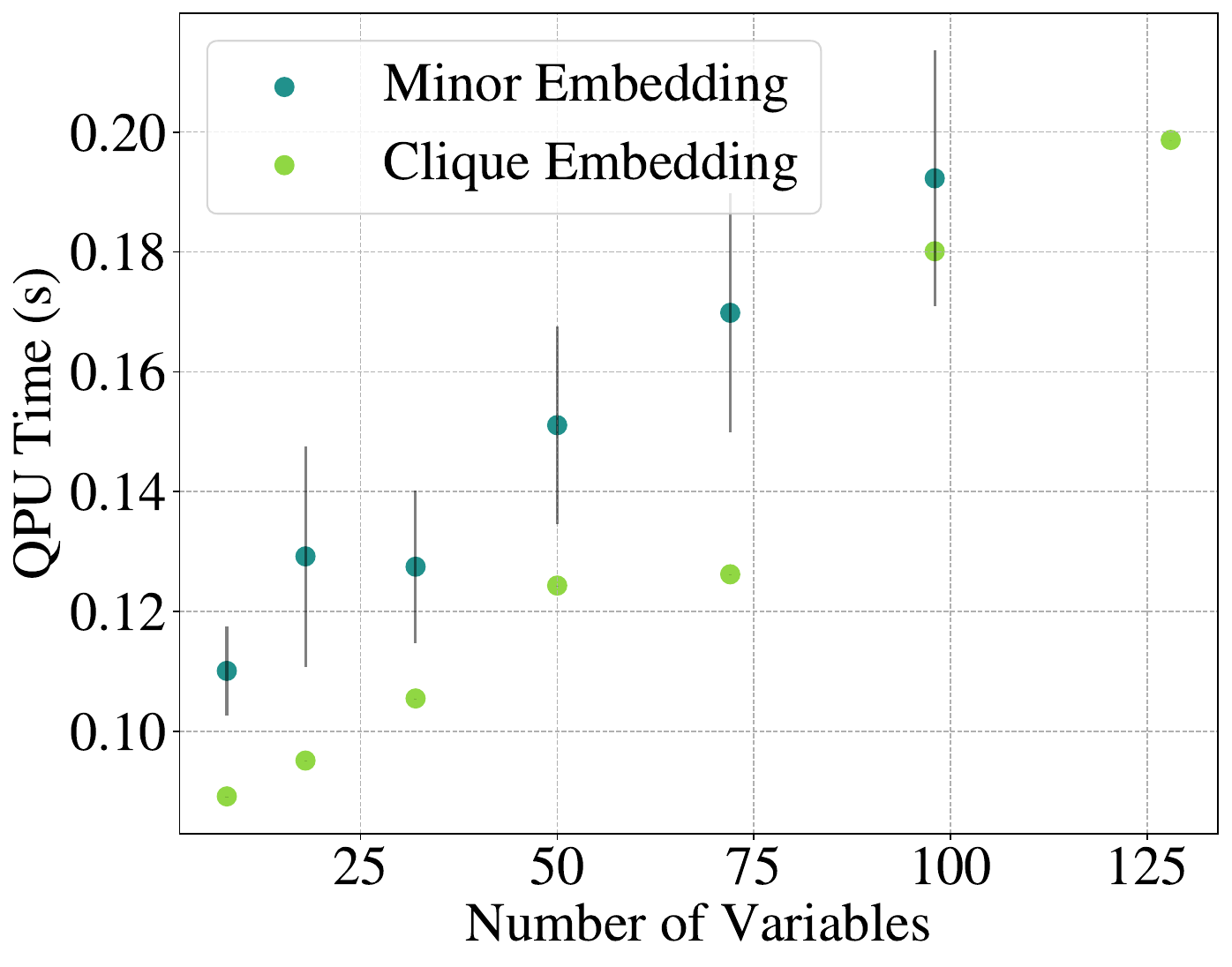}
    } \\
    \vspace{0.3cm} 
    \subfloat[\label{fig:chain_length}]{%
        \includegraphics[width=0.45\linewidth]{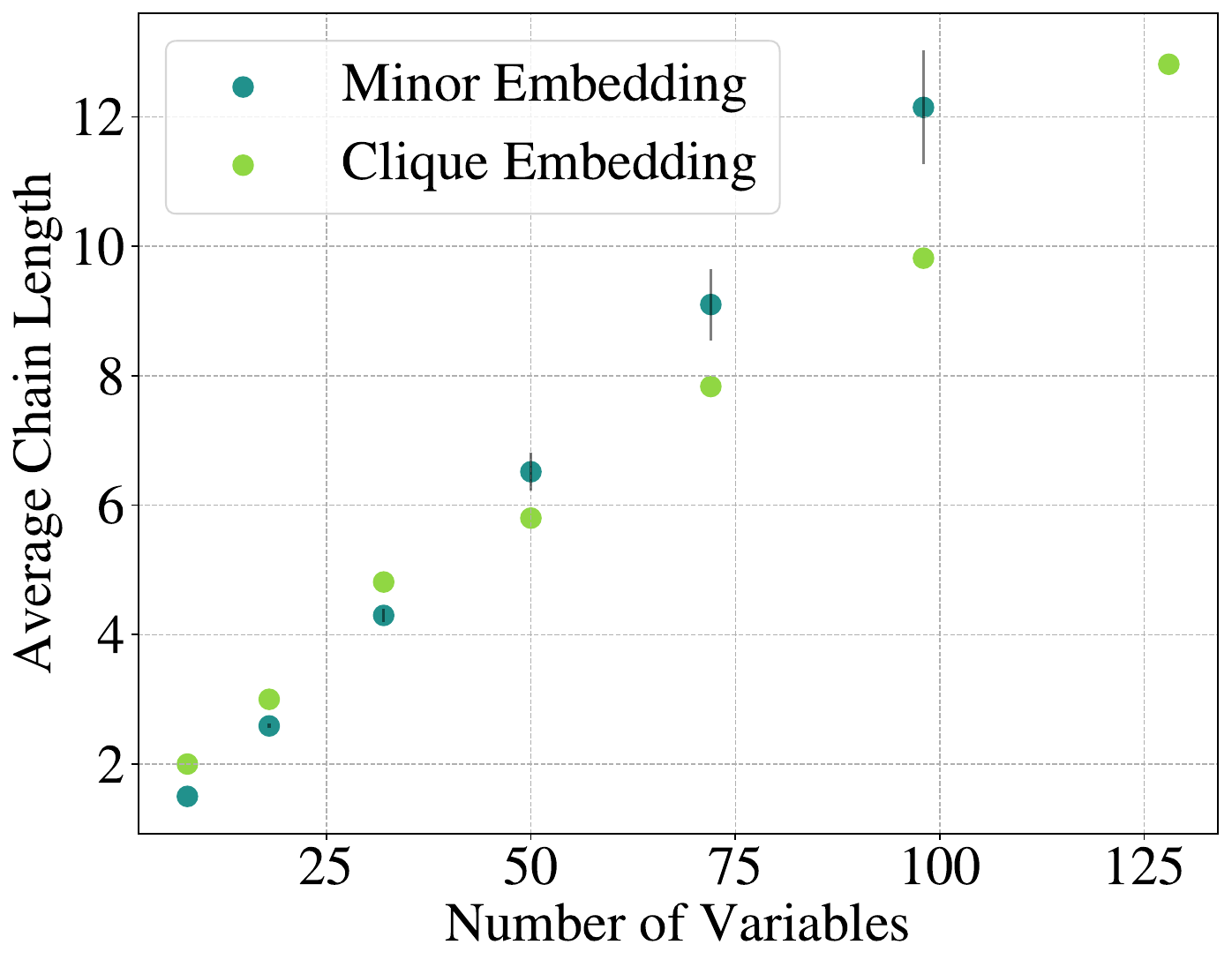}
    } \hfill
    \subfloat[\label{fig:b_chain}]{%
        \includegraphics[width=0.45\linewidth]{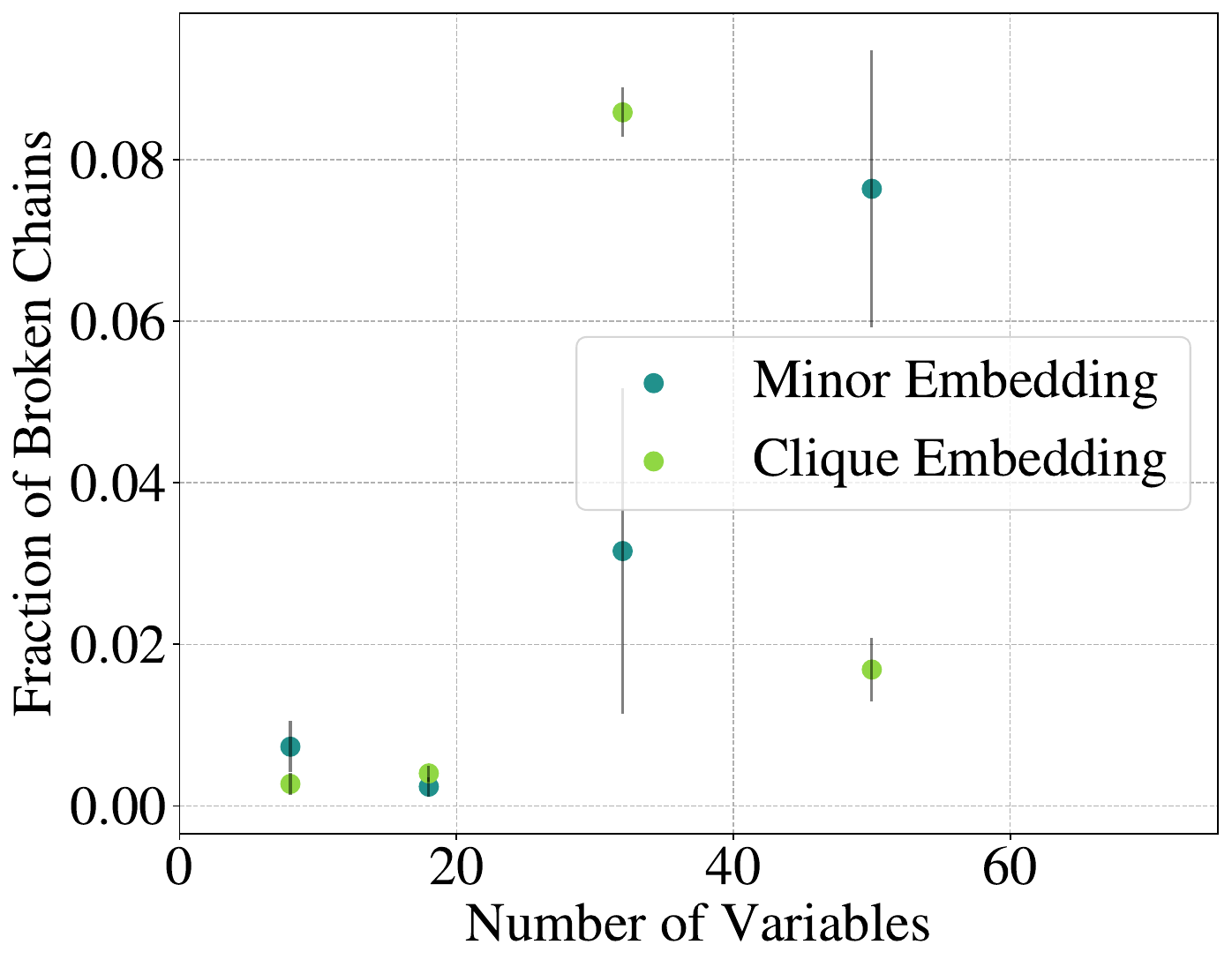}
    }
    \caption{All results were repeated 10 times with average values and standard deviation error bars. (a) Embedding time for the three vacancy QUBO problem up to 128 variables. (b) QPU access time for problem sizes up to 128 variables. (c) Increase in chain length with problem size. (d) Fraction of broken chains for problem sizes up to 72 variables.}
    \label{fig:scaling}
\end{figure*}

Figure \ref{fig:emb_time} shows the embedding time required for all problem sizes solved. Figure \ref{fig:qpu_time} displays the QPU access time that the D-Wave device allocates for solving problem sizes of up to 72 variables, and Figure \ref{fig:chain_length} presents the average chain lengths used. Full results for quantum annealing using both minor and clique embedding are shown in Table \ref{tab:perform_QA} for the larger problem sizes.


\begin{table*}[htbp]
\begin{ruledtabular}
\begin{tabular}{cccccc}
\textbf{QUBO Variables} & \textbf{Embedding} & \textbf{$P_s$} & \textbf{$P_s$ Post-Sel.} & \textbf{Runtime (s)} & \textbf{QPU Time (s)} \\ 
\colrule
8  & Minor & $0.7365 \pm 0.03, 0.004^*$ & $0.8404 \pm 0.02, 0.004^*$ & $2.62 \pm 2$ & $0.063 \pm 0.002$ \\
18 & Minor & $0.1894 \pm 0.02, 0.004^*$ & $0.3980 \pm 0.03, 0.005^*$ & $2.54 \pm 0.08$ & $0.399 \pm 0.08$ \\
32 & Minor & $0.0465 \pm 0.05, 0.002^*$ & $0.0939 \pm 0.09, 0.003^*$ & $5.42 \pm 1$ & $2.765 \pm 1$ \\
50 & Minor & $0.0030 \pm 0.004, 0.0005^*$ & $0.0390 \pm 0.04, 0.002^*$ & $12.06 \pm 3$ & $10.263 \pm 3$ \\
72 & Minor & $0.0002 \pm 0.0006, 0.0001^*$ & $0.0007 \pm 0.002, 0.0003^*$ & $37.34 \pm 10$ & $35.657 \pm 10$ \\
98 & Minor & $0.0000 \pm 0, 0^*$ & $0.0000 \pm 0, 0^*$ & $61.43 \pm 30$ & $59.859 \pm 30$ \\ 
\colrule
8  & Clique & $0.7079 \pm 0.02, 0.005^*$ & $0.8091 \pm 0.01, 0.004^*$ & $4.59 \pm 0.2$ & $2.876 \pm 0.2$ \\
18 & Clique & $0.1580 \pm 0.01, 0.004^*$ & $0.3631 \pm 0.03, 0.005^*$ & $5.15 \pm 0.3$ & $3.209 \pm 0.3$ \\
32 & Clique & $0.0263 \pm 0.006, 0.002^*$ & $0.0608 \pm 0.01, 0.002^*$ & $4.11 \pm 0.4$ & $2.582 \pm 0.01$ \\
50 & Clique & $0.0004 \pm 0.0007, 0.0002^*$ & $0.0238 \pm 0.04, 0.002^*$ & $4.58 \pm 0.1$ & $3.021 \pm 0.1$ \\
72 & Clique & $0.0000 \pm 0, 0^*$ & $0.0000 \pm 0, 0^*$ & $4.79 \pm 0.02$ & $3.167 \pm 0.01$ \\
98 & Clique & $0.0000 \pm 0, 0^*$ & $0.0000 \pm 0, 0^*$ & $6.17 \pm 0.03$ & $4.051 \pm 0.03$ \\
128 & Clique & $0.0000 \pm 0, 0^*$ & $0.0000 \pm 0, 0^*$ & $7.97 \pm 0.1$ & $5.251 \pm 0.08$ \\
\end{tabular}
\end{ruledtabular}
\caption{\label{tab:perform_QA} Performance metric results for quantum annealing with clique and minor embeddings. Errors include standard deviation from 10 repeats and $^*$ the standard error from shot noise. $P_s > 0$ for the 72-variable case with minor embedding holds within standard error bounds.}
\end{table*}

\begin{table*}[htbp]
\begin{ruledtabular}
\begin{tabular}{ccccc}
\textbf{QUBO Variables} & \textbf{Embedding} & \textbf{$\lambda$} & \textbf{Chain Strength} & \textbf{Annealing time (ns)} \\ 
\colrule
8   & Minor  & 3 & 4  & 600  \\
18  & Minor  & 1 & 3  & 1400 \\
32  & Minor  & 1 & 4  & 1600 \\
50  & Minor  & 1 & 6  & 1000 \\
72  & Minor  & 1 & 1  & 2000 \\
98  & Minor  & 1 & 1  & 2000 \\ 
\colrule
8   & Clique & 3 & 4  & 1000 \\
18  & Clique & 1 & 3  & 2000 \\
32  & Clique & 1 & 5  & 600  \\
50  & Clique & 1 & 8  & 2000 \\
72  & Clique & 1 & 13 & 2000 \\
98  & Clique & 1 & 15 & 2000 \\
128 & Clique & 1 & 22 & 2000 \\ 
\end{tabular}
\end{ruledtabular}
\caption{\label{tab:QA_hyp_params} Optimal hyperparameters found for quantum annealing with minor embedding and clique embedding.}
\end{table*}

\end{document}